\documentclass[a4paper,11pt]{article}
\usepackage{jheppub}
\usepackage{mathrsfs}
\usepackage{amsfonts}
\usepackage{setspace}
\usepackage{cellspace}
\usepackage{amsmath,amssymb,bm}
\usepackage[colorlinks=true,linkcolor=blue]{hyperref}
\usepackage{xcolor}
\usepackage{epsfig}
\usepackage{slashed}
\usepackage{caption}
\usepackage{hhline,multirow,tabularx}  
\usepackage{dcolumn}    
\usepackage{url}        
\usepackage{braket}

\title{Scale symmetry breaking, quantum anomalous energy and proton mass decomposition}

\author[a,b]{Xiangdong Ji}
\author[c,d]{Yizhuang Liu}
\author[d]{Andreas Sch\"{a}fer}

\affiliation[a]{Center for Nuclear Femtography, SURA, 1201 New York Ave. NW, Washington, DC 20005, USA}
\affiliation[b]{Department of Physics, University of Maryland, College Park, MD 20742, USA}
\affiliation[c]{Institut  fur Theoretische Physik, Universitat Regensburg, D-93040 Regensburg, Germany}
\affiliation[d]{Institute of Theoretical Physics,
Jagiellonian University, 30-348 Kraków, Poland}

\emailAdd{xji@umd.edu}
\emailAdd{yizhuang.liu@uj.edu.pl}
\emailAdd{Andreas.Schaefer@physik.uni-regensburg.de}

\abstract{We study the anomalous scale symmetry breaking effects on the proton mass in QCD due to quantum fluctuations at ultraviolet scales. We confirm that a novel contribution naturally arises as a part of the proton mass, which we call the quantum anomalous energy (QAE). We discuss the QAE origins in both lattice and dimensional regularizations and demonstrate its role as a scheme-and-scale independent component in the mass decomposition. We further argue that QAE role in the proton mass resembles a dynamical Higgs mechanism, in which the anomalous scale symmetry breaking field generates mass scales through its vacuum condensate, as well as its static and dynamical responses to the valence quarks. We demonstrate some of our points in two simpler but closely related quantum field theories, namely the 1+1 dimensional non-linear sigma model in which QAE is non-perturbative and scheme-independent, and QED where the anomalous energy effect is perturbative calculable.}

\date{\today}

\begin{document}
\maketitle
\flushbottom
\section{Introduction}
In relativistic quantum field theories, a bound state is an irreducible representation of Poincar\'e group characterized by its mass and spin, making it a natural question to ask if and how these two quantities can meaningfully be decomposed into different contributions. In non-relativistic quantum mechanics the mass is simply the sum of the masses of the individual components minus the binding energy which can be unambiguously decomposed into familiar kinematic energy and potential energy. Furthermore, in the macroscopic world, the binding energy effects on mass are entirely negligible and mass
is just the sum of individual parts. In the opposite limit showcased in the electroweak theory, the masses of elementary particles arise entirely from their interactions with the Higgs potential, which acquires a vacuum condensate after the well-known spontaneous gauge symmetry breaking, or the Higgs mechanism~\cite{Peskin:1995ev}.

In quantum chromodynamics (QCD), the fundamental theory of strong interactions, the decomposition of proton mass and spin has been quite interesting for many years. While consensus seems finally to be reached with respect to spin, some controversy seems still ongoing for its mass. The original work by one of us~\cite{Ji:1994av,Ji:1995sv} studied the decomposition of proton mass based on the QCD energy-momentum tensor (EMT) $T^{\mu\nu}$ which was split into trace and traceless parts. It was found in Refs.~\cite{Ji:1994av,Ji:1995sv} that $\frac{1}{4}$ of the proton mass can be attributed to the trace, a statement which reminds of the virial theorem for non-relativistic systems. Moreover, a mass sum rule that contains quark and gluon kinematic energy terms, a quark mass term and an anomalous term has been proposed.  These and related results inspired new variants of the mass sum rule~\cite{Rothe:1995hu,Lorce:2017xzd,Hatta:2018sqd,Metz:2020vxd} and interesting discussions on the subject in the literature~\cite{Ji:2021mtz,
Liu:2021gco, Zahed:2021fxk}. In our view, the controversy is not about mathematical consistency of the different suggestions but on their phenomenological relevance. The latter debate could be ended if, e.g. the quantum anomalous energy (QAE) contribution could be unambiguously extracted from heavy-quarkonium electroproduction off a nucleon at threshold as is tried at Jefferson Lab and is on the physics program for the future Electron-Ion Collider~\cite{Kharzeev:1998bz,Hatta:2018ina,Meziani:2020oks}. While these efforts are ongoing an alternative is provided by lattice QCD calculations. The masses for nucleon and other hadrons have been numerically calculated to such high accuracy~\cite{Durr:2008zz,Fodor:2012gf,Borsanyi:2014jba,Walker-Loud:2018gvh} that the configurations have to contain all major effects contributing to it. Because lattice QCD evidence has thus reached comparable reliability as direct experiments it also suffices to determine the value for QAE in the nucleon on the lattice, similar to what was done in \cite{Abdel-Rehim:2016won,Alexandrou:2017oeh,Yang:2018nqn,He:2021bof}.

Despite this progress in understanding the mass structure of a proton, open questions remains, in particular concerning the anomalous term, see e.g. Ref.~\cite{Metz:2020vxd}. Let us mention that the connection between the trace anomaly and the nucleon mass as well as the related low-energy theorems/effective theories have already been studied in the 70's~\cite{Collins:1976yq,Shifman:1978zn} and that this discussion thus approaches its 50th anniversary. The identification of a piece of QCD Hamiltonian as anomalous contribution
in \cite{Ji:1994av} was based on these
pioneering studies. In a recent paper~\cite{Ji:2021pys}, the authors have not only investigated the anomaly contribution to the QCD energy and argued that the quantum anomalous energy (QAE) is a meaningful part of the proton mass decomposition but it has been suggested that the QAE mechanisms in QCD are very similar to those of Spontaneous Symmetry Breaking (SSB). Results for QED (quantum electrodynamics), QCD and large-$N$ $1+1$ non-linear sigma model were presented in support of this analogy. In the current paper, we provide a more detailed 
arguments and derivations of ~\cite{Ji:2021pys}.

In section \ref{sec:scale}, we will first provide a general review to mass generation and dimensional transmutation in QCD-like theories. We emphasize that the bare coupling constant that leads to the continuum limit is a function of the ratio between the ultraviolet (UV) cutoff and the physical scales. We show how renormalization group equation (RGE) and trace anomaly naturally appear as result of dimensional transmutation.

In section \ref{sec:dec}, based on the principles given in section \ref{sec:scale} we derive the mass sum rule by studying the corresponding Ward identities in detail, paying special attention to regularization and renormalization. We first work in lattice QCD, in which the derivation was first made in Refs.~\cite{Rothe:1995hu,Rothe:1995av}, papers which, unfortunately, seem to have been largely forgotten for many years. We then generalize the derivation to continuum regularization such as dimensional regularization. We show that  while the ``naive'' Hamiltonian $\frac{1}{2}(\vec{E}^2+\vec{B}^2)$ is scheme dependent, the decomposition into traceless and trace part is scheme independent. We identity the operators corresponding to the traceless and trace parts in different schemes and argue that it is the traceless part that can be naturally interpreted as the quark-gluon kinematic energy. We also comment on the renormalization of QCD EMT and compare the results in~\cite{Hatta:2018sqd,Metz:2020vxd} to the more familiar renormalization properties of twist-2 quark and gluon operators~\cite{Peskin:1995ev}.

In section \ref{sec:non}, we explicitly demonstrates the results of the previous section in $1+1$ dimensional $O(N)$ non-linear sigma model in the solvable large $N$ limit that exhibits asymptotically freedom, dimensional transmutation and dynamical mass generation. We show that the quantum anomalous energy contributes half of the mass of the pion-like bosons in such theories, consistent with the  ``virial theorem'' stated above. We study the mass-sum rule in different schemes and show that a proper mass-sum rule does requires a scheme and scale independent anomalous contribution.

In section \ref{sec:qed}, we study the anomalous energy contribution in QED. We show that although there is no dynamical scale generation, there does exists a QAE contribution to the electron pole mass, as well as to the binding energy of a hydrogen atom in the presence of a background field.  In particular, the famous ${\cal O}(\alpha^5)$ Lamb-shift receives a trace anomaly contribution which we will calculate.

In section \ref{sec:higgs}, we relate the QAE contribution to the proton mass to a dynamical Higgs effect. We compare the more standard Higgs mechanism for the fermion mass generation  to the mass generation in the 1+1 non-linear sigma model. We show that in both cases, the scalar part of the Hamiltonian is proportional to the Higgs field, and that mass generation can be measured either through the mass term due to the scalar vacuum condensate or through the response of the Higgs field in the presence of other fields. We then generalize to QCD and relate the QAE contribution to the Higgs-like coupling of scalar resonances to the proton. In the chiral limit and assuming the dominance of the lowest glueball, the glueball-proton coupling is proportional to the proton mass, similar to the Higgs coupling which has been tested at the LHC~\cite{Weinberg:1967tq,Sirunyan:2018koj,Aad:2019mbh}. We also discuss the pion case. The results are consistent with an effective theory~\cite{Ellis:1984jv} for the lowest glueball and its coupling to pions, which is also presented here.

Finally, we draw conclusions and give an outlook. Some technical details are presented in Appendices.

\section{Review on scale generation, RGE and trace anomaly}\label{sec:scale}

In this section we review the mass generation and dimensional transmutation in QCD-like theories. We emphasize that in order to take the continuum limit, the coupling constant of a cutoff theory must depends on the UV cutoff and the physical mass scales of the theory in continuum non-trivially. We show how the trace-anomaly naturally arises as a consequence of this scale-dependency by providing a path-integral based derivation of the trace anomaly $T^{\mu}_{\mu}$ and the renormalization group equation (RGE) for two-piont functions. The advantage of this derivation is that it does not requires the Lorentz invariance in prior and can be applied to lattice-like regularization as well. In section \ref{sec:dec} the same method will be used to investigate the mass-sum rules and to derive the QAE. 
\subsection{Pure $SU(3)$}
To simplify the discussion let us first consider pure SU(3) Yang-Mills (YM) theory. At classical level the theory is massless and has conformal symmetry. As such, it has no mass scale. Any bound state mass must be either zero or infinity, making this theory not very interesting.

However, the conformal symmetry of SU(3) YM theory is broken at the quantum level, where the theory must be defined as limit of a theory with an UV cutoff, such as lattice gauge theory in which the cutoff is given by lattice spacing $a$. In the cutoff theory, the correlation length $\zeta$ (assumed to be  finite) for a gauge-invariant correlation function $\langle O(x)O(0) \rangle_c=\langle O(x)O(0) \rangle-\langle O\rangle^2\sim e^{-x/\zeta}$ is a dimensionless function of the bare coupling constant $g_0$, $\zeta/a=f(g_0)$. The operator $O$ can be chosen to be $F^2$ or some other gauge-invariant operator. One expects $\zeta$ to be the inverse of the glueball mass $M$, $\zeta=1/M$ in the continuum limit $a\rightarrow 0$. Therefore, one has to tune $g_0$ such that $f(g_0)=1/Ma$ goes to infinity as $a\rightarrow 0$. This implicitly introduces a physical scale $M$ into the problem and fixes $g_0=g_0(1/Ma)$ as a function of $1/Ma$. This process has been called ``dimensional transmutation''~\cite{Coleman:1974hr}. In fact, one expects from perturbation theory that as $a\rightarrow 0$, $g_0(1/Ma)$ approaches $0$ logarithmically (asymptotic freedom). The existence of a finite correlation length at small $g_0$ that approaches $\infty$ in unit of $a$ as $g_0\rightarrow 0$ is a genuinely non-perturbtive effect and one of the most important properties of the QCD vacuum.

For pure YM theory, since there is only one free dimensional parameter, the mass scale $M$, which can be extracted from any two-point function, actually determines the full theory. All other physical scales of the theory are proportional to $M$. Among them there are not only masses, but also the string tension $\sigma=cM^2$ that characterizes the linear $\bar q q$ confinement potential. The number $c$ only depends on the $SU(3)$ group.  Despite this relation, we should point out that {\it in general the confinement and spontaneous mass generation is not necessarily related.} The confining phase is characterized by non-vanishing string tension or area law for Wilson-loops. But there is also the standard Higgs phase in which color magnetic charges are confined while color electric charges only got screened, characterised by area law for 't Hooft loops~\cite{tHooft:1977nqb}. And both of the two phases are gapped with finite correlation lengths.  Nevertheless, it is widely believed that the Higgs and confining phase are smoothly connected~\cite{Fradkin:1978dv} and we will explore the similarity between dimensional transmutation and Higgs mechanism in Sec~\ref{Sec:latticesum}. 

\subsection{$SU(3)$ plus fermions}
When one adds fermions with $N_f$ flavors, which we assume to be massless, the classical theory has the $U_A(1)$ symmetry that is known to be broken by the famous $U_A(1)$ anomaly, as well as the $SU(N_f)_L\times SU(N_f)_R$ chiral symmetry for $N_f>1$ that gets spontaneously broken down to $SU(N_f)_V$ by the chiral condensate. In the strictly massless case, the chiral condensate is proportional to the cubic power of the scale $M\sim \Lambda_{\rm QCD}$ introduced before. The same mass scale $M$ is thus related to three different phenomena: mass scale generation, color-confinement and spontaneous chiral symmetry breaking. The contribution from instantons explains the chiral symmetry breaking quite well in the instanton-liquid model~\cite{Diakonov:1995ea,Schafer:1996wv}, and has been supported by lattice results~\cite{Chu:1994vi}. While instantons might account for a large portion of the hadron mass~\cite{Zahed:2021fxk}, it is known~\cite{Greensite:2016pfc} that they can not explain confinement. On the other hand, 
the confinement has been used the main physical mechanism
to generate mass for hadrons in the MIT bag model~\cite{Chodos:1974je}.  

When one includes non-zero fermion masses, say for degenerate $u$ and $d$ quarks, there are two parameters of the theory: the bare coupling constant $g_0$ and the bare quark mass $m_0$. One must fix them by two physical mass scales, namely $M_1$ and $M_2=m_\pi$. One then has the dimensional transmutation relations $m_0=m_\pi f(M_1a,m_\pi a)$ and $g_0=g_0(M_1a,m_\pi a)$.

Due to the presence of mass scales , the naive scaling invariance of the classical theory is broken in the quantum version of the theory. Nevertheless, one still has the extended scale invariance under simultaneous rescaling of space-time and physical mass scales. The Ward identity of this invariance is the RGE or the trace anomaly, as we will discuss next.

\subsection{Trace anomaly from anomalous scale symmetry breaking}\label{Sec:traceanomaly}
After introducing the dimensional transmutation, in this section, using the Euclidean path-integral formalism, we provides a derivation of the trace anomaly and the RGE for two-piont functions. For simplicity we only consider pure-YM like theories. The methods here will be used later to derive QCD Hamiltonian responsible for the mass sum rule and the QAE. Our convention for Euclidean coordinates is $(x_4,\vec{x})$ where $x_4$ is the imaginary time (also in 2D case 
discussed later).  We consider a general interpolating operator $O(x_4,\vec{x})$ for a given hadron, which for simplicity we assume to be renormalization group (RG) invariant. We study the two-point function
\begin{align}
G(T,\vec{p})=\langle O(T,\vec{p})O(0,-\vec{p})\rangle= \frac{\int D\phi e^{-\frac{1}{g_0^2}S[\phi]}O(T,\vec{p})O(0,-\vec{p})}{\int D\phi e^{-\frac{1}{g_0^2}S[\phi]}} 
\end{align}
with some given time $T$. In this expression, $\phi$ is a spin zero field and $\frac{1}{g_0^2}S(\phi)$ is the Euclidean action. It can be  written as
\begin{align}
S[\phi]=\int d^4x {\cal S}(\phi(x))|_{\Lambda_{\rm UV}}\ ,
\end{align}
where we use the symbol $|_{\Lambda_{\rm UV}}$ to denote usage of an UV regulator with value $\Lambda_{\rm UV}$. As we have emphasized, we have $g_0=g_0(\Lambda_{\rm UV}/M)$ where $M$ is the physical scale of the theory, which we choose to be the mass $M$ of the hadron created by $O$ to guarantee the finiteness of the results as we take the limit $\Lambda_{\rm UV}\rightarrow \infty$. At large $T$, the two-point function $G(T,\vec{p})\sim e^{-MT}$ is controlled by the mass of the hadron.  Thus we can extract the hadron mass using:
\begin{align}
M=-\lim_{T\rightarrow \infty}\frac{1}{T}\ln G(T,\vec{p}).
\end{align}
In the following we only consider the rest frame, i.e. $\vec{p}=0$. With these definitions we can derive the RGE and trace anomaly. Let us consider the scale transformation
$x\rightarrow x'=\lambda x$ and $\phi'(x')=\lambda^{-d}\phi(x)$ where $d$ is the naive mass dimension of the theory. In terms of $\phi'$ and $x'$ the two point function becomes:
\begin{align}
G'(T,\vec{0})=\langle O'(T,\vec{0})O'(0,\vec{0})\rangle=\lambda^{-2d_O+3}G(\lambda^{-1}T,0) \ .
\end{align}
where $d_O$ is the naive mass dimension of the operator $O$ and $+3$ comes from the Fourier transformation to momentum space. On the other hand, the action for $\phi'(x')$ becomes
\begin{align}
S[\phi']=\int d^4x' {\cal S}(\phi'(x'))|_{\lambda^{-1}\Lambda_{\rm UV}}\ ,
\end{align}
which leads to the identity
\begin{eqnarray}\label{eq:scaletrans}
\lambda^{-2d_O+3}G(\lambda^{-1}T,\vec{0})
&=& \frac{\int D\phi' e^{-g_0^{-2}\left(\frac{\Lambda_{UV}}{M}\right)\int d^4x' {\cal S}(\phi'(x'))|_{\lambda^{-1}\Lambda_{\rm UV}}}O'(T,\vec{0})O'(0,\vec{0})}{\int D\phi' e^{-g_0^{-2}\left(\frac{\Lambda_{UV}}{M}\right)\int d^4x' {\cal S}(\phi'(x'))|_{\lambda^{-1}\Lambda_{\rm UV}}}} \ .
\end{eqnarray}
If it were not for the $\Lambda$ dependence of the UV regulator, the right hand side would be the same as $G(T,\vec{0})$. The presence of $\Lambda_{\rm UV}$ results in a $\lambda$ dependence from the mismatch between $g_0(\Lambda_{\rm UV}/M)$ and $\lambda^{-1}\Lambda_{\rm UV}$. In fact, in terms of $\Lambda'_{\rm UV}=\lambda^{-1}\Lambda_{\rm UV}$, one simply has $g_0(\Lambda'_{\rm UV}/(\lambda^{-1} M))$. Thus, the equation is equivalent to
\begin{align}\label{eq:rescaledac}
\lambda^{-2d_O+3}G(\lambda^{-1}T,\vec{0})=\frac{\int D\phi' e^{-g_0^{-2}\left(\frac{\Lambda'_{UV}}{\lambda^{-1}M}\right)\int d^4x' {\cal S}(\phi'(x'))|_{\Lambda'_{\rm UV}}}O'(T,\vec{0})O'(0,\vec{0})}{\int D\phi' e^{-g_0^{-2}\left(\frac{\Lambda'_{UV}}{\lambda^{-1}M}\right)\int d^4x' {\cal S}(\phi'(x'))|_{\Lambda'_{\rm UV}}}} \ .
\end{align}
Now, all $\lambda$ dependence has been absorbed into $g_0$. By taking one derivative with respect to $\lambda$ and evaluating it at $\lambda=1$, the equation reads
\begin{eqnarray}\label{eq:twopointward}
\Bigl(-2d_O+3-T\frac{d}{dT}\Bigr)G(T,\vec{0})
&=&\Big\langle\frac{2\beta(g_0)}{g_0^3}\int d^4x {\cal S}(\phi(x)) O(T,\vec{0})O(0,\vec{0}) \Big\rangle_c  \ .
\end{eqnarray}
Here
\begin{align}
\beta(g_0)=\Lambda_{\rm UV}\frac{dg_0(\Lambda_{\rm UV}/M)}{d\Lambda_{\rm UV}}
\end{align}
is the bare beta function and the $\langle \rangle_c$ denotes the connected part. By comparing with the Ward-identity for scale-transformation (Eq.~(42) in Ref.~\cite{Ji:1995sv})
\begin{align}
    \Bigl(-2d_O+3-T\frac{d}{dT}\Bigr)G(T,\vec{0})=\langle \int d^4x T^{\mu}_{\mu}(x) O(T,\vec{0})O(0,\vec{0}) \Big\rangle_c \ ,
\end{align}
one can identify the trace part $T^{\mu}_{\mu}$ of the EMT as 
\begin{align}
T^{\mu}_{\mu}(x)=\frac{2\beta(g_0)}{g_0^3}{\cal S}(\phi(x)) \ .
\end{align}
The derivation here also shows that the operator $\frac{2\beta(g_0)}{g_0^3}{\cal S}(\phi(x))$ must be RG invariant and scheme independent. 

One further notice that at large $T\gg M$, Eq.~(\ref{eq:twopointward}) is dominated by linear terms in $T$
\begin{align}
MT G(T,0)=T\Big\langle\frac{2\beta(g_0)}{g_0^3}\int d^3\vec{x}  O(T,\vec{0}){\cal S}(\phi(0,\vec{x}))O(0,\vec{0}) \Big\rangle_c \ .
\end{align}
Thus one obtains the scale-setting relation 
\begin{eqnarray} \label{eq:tracecon}
M&=&\lim_{T\rightarrow \infty}\frac{\Big\langle \int d^3\vec{x}  O(T,\vec{0})  T^{\mu}_{\mu}(0,\vec{x})O(0,\vec{0}) \Big\rangle_c}{\langle O(T,\vec{0})O(0,\vec{0})\rangle}\nonumber \\
&=&\frac{\langle \vec{P}=0|\int d^3\vec{x} T^{\mu}_{\mu}(\vec{x})| \vec{P}=0\rangle}{\langle \vec{P}=0|\vec{P}=0\rangle} \ .
\end{eqnarray}
Here we use $|\vec{P}=0\rangle$ to denote the lightest hadron state created by the interpolating field $O$.


To summarize, in this section we have provided a review of the far-reaching consequences of dimensional transmutation in QCD-like theories. Starting from the path-integral representation of the theory with a generic UV cutoff, it has been shown that as a result of the scale-dependency of the bare coupling constant, a non-vanishing scalar field naturally emerges in the continuum limit and can be identified as the trace of the energy-momentum tensor. We should mention that although all the results in this section are well-known, our derivation of the trace-anomaly has the advantage that it does not requires Lorentz-invariance in prior and can be applied to lattice-like regularizations as well. In the next section the same method will be adopted to study the Hamiltonian-based mass sum rules.  
\section{Quantum anomalous energy and mass sum rule}\label{sec:dec}
To derive a mass sum rule, we need to examine the traditional derivation of the EMT, in particular, the Hamiltonian density $T^{00}$ more carefully in the presence of a regulator. The standard way to obtain the EMT is to study the change of the Lagrangian under an infinitely small space-time transformation $x^{\mu}\rightarrow x^{\mu}+\delta x^{\mu}(x)$. In the presence of the cutoff, this shift could potentially change the UV cutoff and extra attention must be paid. For example, the scale transformation changes $a\rightarrow \lambda a$ and as we have seen in the previous section, the $\lambda$ dependence of $a$ can be absorbed into $g_0$, resulting in the trace anomaly. In this section we show that individual components of $T^{\mu\nu}$ such as $T^{00}$ in the cutoff scheme can also suffer from such effects, leading to an anomalous contrition to the mass sum rule.

As in the previous section, let's consider the two point function $G(T,\vec{0})$. We would like to derive the mass sum rule as a Ward identity for the following transformation:
\begin{align}\label{eq:tomporesacle}
x_4'=\lambda x_4, \\
\phi'(x_4',\vec{x})=\phi(x_4,\vec{x}) \ .
\end{align}
The two point function in terms of the $\phi'$ field then becomes $G(\lambda^{-1}T,\vec{0})$. 
On the other hand, the same two point function can be obtained from the functional integral over the action for $\phi'$, which can be written in the generic form:
\begin{align}
S[\phi']=\frac{1}{g_0^2}\int d^4x' \left(\lambda {\cal S}_{\tau}(\phi'(x'))+\lambda^{-1}{\cal S}_{s}(\phi'(x'))\right)_{\lambda^{-1}\Lambda_{\rm UV},\Lambda_{\rm UV}} \ .
\end{align}
Here ${\cal S}_{\tau}(\phi'(x'))$, ${\cal S}_{s}(\phi'(x'))$  can be roughly identified with $\frac{1}{2}F^{4i}F_{4i}=-\frac{1}{2}E^2$, $\frac{1}{4}F^{ij}F_{ij}=\frac{1}{2}B^2$, respectively. The symbol $\lambda^{-1}\Lambda_{\rm UV},\Lambda_{\rm UV}$ indicates that the UV cutoff in the $x^4$ direction has been modified to $\lambda^{-1}\Lambda_{\rm UV}$, while in the spatial direction it remained as before. If there were no $\lambda$ dependence of the cutoff, following the same logic as before, evaluating one derivative with respect to $\lambda$ and taking the large $T$ limit, one would end up with the following sum rule:
\begin{align}
M=\frac{\langle \vec{P}=0|\frac{1}{g_0^2}\int d^3\vec{x}\bigg[{\cal S}_{s}(\phi(0,\vec{x}))-{\cal S}_{\tau}(\phi(0,\vec{x}))\bigg]| \vec{P}=0\rangle}{\langle \vec{P}=0|\vec{P}=0\rangle} \ .
\end{align}
This would amount to the ``naive'' Hamiltonian
\begin{align}\label{eq:Hnaive}
H_{c}=\frac{1}{g_0^2}\int d^3\vec{x}\bigg[{\cal S}_{s}(\phi(0,\vec{x}))-{\cal S}_{\tau}(\phi(0,\vec{x}))\bigg] \ .
\end{align}
However, due to the possible $\lambda$ dependence of the cutoff, there can be an anomalous contribution $H_{a}$ to $H$. In the following subsections we investigate this $H_a$ for different regulators.

\subsection{Lattice and cutoff regularization}\label{Sec:latticesum}
In this subsection we investigate the lattice regularization in more detail. We restrict ourselves to the Wilson action for pure YM theory and stay in the infinite volume limit. In terms of the link variables $U_{\mu}(x)$, the standard Wilson action reads:
\begin{align}
S[U]=-\frac{1}{g_0^2}\sum_{x}\sum_{\mu\nu}{\rm Tr}P_{\mu\nu}(U,x) \ ,
\end{align}
in which we sum over the trace of the Wilson-loop $P_{\mu\nu}(U,x)$ for the elementary plaquettes in the $\mu\nu$ plane at position $x$. $P_{\mu\nu}(U,x)$ is defined as the product of the link variables
along the boundary of the plaquette. Neglecting the effects of scale transformations, on which we will focus in this section, in the limit of an infinitesimal $a$ one gets  $U_{\mu}(x)=e^{-iaA^{\mu}(x)}$ and $P_{\mu\nu}(U,x)=e^{-ia^2F_{\mu\nu}(x)}$ and thus $-\int d^4x\frac{1}{4}F^2$ for the action in the continuum.

Given this information, let us construct the rescaled version of the theory, equivalent to
\begin{align}\label{eq:timere}
S[\phi']=\frac{1}{g_0^2}\int d^4x' \left(\lambda {\cal S}_{\tau}(\phi'(x'))+\lambda^{-1}{\cal S}_{s}(\phi'(x'))\right)_{\lambda^{-1}\Lambda_{\rm UV},\Lambda_{\rm UV}} \ ,
\end{align}
where the correlation length in the imaginary time direction is rescaled by a factor $\lambda$. To avoid taking a derivative with respect to the cutoff, one absorbs the $\lambda$ dependence into the parameters of the action which gives
\begin{align}
S[\phi']=\frac{1}{g_0^2(\Lambda_{\rm UV}/M,\lambda)}\int d^4x' \left(\lambda {\cal S}_{\tau}(\phi'(x'))+\lambda^{-1}{\cal S}_{s}(\phi'(x'))\right)_{\Lambda_{\rm UV}}\ .
\end{align}
This expression is equivalent to expression~(\ref{eq:timere}). Our task is thus reduced to determining the $\lambda$ dependence of $g_0$.

In lattice regulation, the above suggests to investigate the action $S_{\lambda}$
\begin{align}\label{eq:reslatt}
S_{\lambda}[U]=-\frac{1}{g_0^2}\sum_{x}\bigg(\lambda P_{\tau}(x)+\frac{1}{\lambda}P_{s}(x) \bigg) \ ,
\end{align}
where $P_{\tau}$ contains the sum over the temporal plaquettes in the $(x_4,x^i)$ plane and $P_{s}$ over all purely spatial plaquettes.  This action has already been investigated in Ref.~\cite{Karsch:1982ve}. Below we provide a brief introduction to its properties. Naively, expanding to leading order in $a$, it seems that Eq.~(\ref{eq:reslatt}) is already sufficient to produce the required rescaling in $x_4$ direction. However, due to the presence of a hard cutoff, the loop integral of the $\lambda$ dependent propagators can not be simply rescaled back by $x_4\rightarrow \lambda^{-1}x_4$, e.g.,
\begin{eqnarray}
\int_{p^2\le \Lambda^2} \frac{d^4p}{(2\pi)^4}\frac{1}{\lambda p_4^2+\lambda^{-1}\vec{p}^2+\lambda^{-1}m^2}
&\ne& \int_{p^2\le \Lambda^2} \frac{d^4p}{(2\pi)^4}\frac{1}{p^2+m^2} \ .
\end{eqnarray}
Therefore, additional $\lambda$ dependencies get introduced by loop integrals that must be compensated by a change in $g_0$. 

Non-perturbatively, this point can be argued in the following way. One notices that since all the field variables and coupling constants are dimensionless, what really characterizes the theory are the correlation lengths $\zeta_{\tau}$ in  temporal and $\zeta_s$ in spatial directions, measured in natural lattice units. Both of them are functions of $g_0$ and $\lambda$. As $g_0\rightarrow 0$, one expects that both approaches $\infty$, but that their ratio approaches $\lambda$:
\begin{align}
\zeta_{\tau}=\lambda f(g_0,\lambda) \ ,\\
\zeta_{s}=f(g_0,\lambda) \ .
\end{align}
In order for the continuum limit to be just a rescaled version of the original theory, one must identify $f(g_0,\lambda)=\frac{1}{Ma}$. As a result, we conclude that $g_0$ in Eq.~(\ref{eq:reslatt}) must be $\lambda$ dependent,
\begin{align}
g_0=g_0(Ma,\lambda)\ ,
\end{align}
for the continuum limit of the theory defined by Eq.~(\ref{eq:reslatt}) to be given simply by rescaling. This is equivalent to stating that the physical scale remains the same, whereas the cutoff in $x_4$ direction is rescaled, which is the viewpoint of Ref.~\cite{Karsch:1982ve}. It has been furthermore proven in that paper that the $\lambda$ derivative of $g_0$ is $\frac{1}{4}$ of the beta function
\begin{align}\label{eq:lambdaderi}
\frac{dg_0(aM,\lambda)}{d\lambda}|_{\lambda=1}=-\frac{1}{4}a\frac{dg_0(Ma)}{da}=\frac{1}{4}\beta(g_0) \ .
\end{align}
We will provide a derivation of Eq.~(\ref{eq:lambdaderi}) in Appendix~\ref{sec:deriproof}.  As one can see from the derivation, the result relies crucially on the lattice symmetry and that the space-time dimension is $4$. In $d$ dimensions the above relation can be generalized with $\frac{1}{4}$ replaced by $\frac{1}{d}$.  

With help of Eq.~(\ref{eq:lambdaderi}), by comparing the $\lambda$ derivatives of two point functions in a way similar to Sec.~\ref{Sec:traceanomaly}, the mass-sum rule can be obtained by combining Eq.~(\ref{eq:time}) and Eq.~(\ref{eq:lambdaderi}):
\begin{align}
M=\frac{\langle \vec{P}=0|H_c+H_a| \vec{P}=0\rangle}{\langle \vec{P}=0|\vec{P}=0\rangle} \ ,
\end{align}
where $H_c$ is given by the lattice QCD version of Eq.~(\ref{eq:Hnaive})
\begin{align}
H_c=\frac{1}{g_0^2}\sum_{\vec{x}}(P_{\tau}(0,\vec{x})-P_{s}(0,\vec{x})) \ ,
\end{align}
and the quantum anomalous energy (QAE) contribution reads
\begin{align}
H_a=\frac{\beta(g_0)}{2g_0^3}\sum_{\vec{x}}\left(P_{\tau}(0,\vec{x})+P_s(0,\vec{x})\right) \ .
\end{align}
Comparing with the tensor decomposition of the Hamiltonian~\cite{Ji:1994av,Ji:1995sv},
\begin{align}
    &H=H_T+H_S=\int d^3 \vec{x} T_{T}^{00}(\vec{x})+\frac{1}{4}\int d^3 \vec{x} T^{\mu}_{\mu}(\vec{x}) \ ,  
\end{align}
where $T^{\mu\nu}_T$ is the traceless (tensor) part of the EMT, one found that in lattice regularization the $H_c$ equals to the tensor part $H_T$ of the Hamiltonian, while $H_a$ equals to the scalar part $H_S$ of the Hamiltonian 
\begin{align} 
    H_T=H_c, \;H_S=H_a \ .
\end{align}
The tensor part of the Hamiltonian $H_T$ contributes $\frac{3}{4}$ of the hadron mass, while the scalar part contributes $\frac{1}{4}$, in consistency with the virial theorem~\cite{Ji:1994av}. It is not difficult to see that the derivation can be adopted to generic cutoff schemes that preserve the lattice symmetry.

We shall also mention that if one mantain a generic $\lambda$ in the derivation above, then the Hamiltonian has the form
\begin{align}\label{eq:genericH}
H=\frac{1}{g_0^2}\sum_{\vec{x}}(-\lambda^2P_{\tau}(0,\vec{x})+P_{s}(0,\vec{x})) + \frac{2\lambda}{g_0^3}\frac{dg_0}{d\lambda}\sum_{\vec{x}}\left(\lambda^2P_{\tau}(0,\vec{x})+P_{s}(0,\vec{x})\right) \ .
\end{align}
Eq.~(\ref{eq:genericH}) will be used later in Sec.~\ref{sec:non}. 

\subsection{Dimensional regularization and 
renormalization of EMT}\label{sec:DR}

In dimensional regularization (DR), the spatial dimension has been changed to $D-1=3-2\epsilon$, while the temporal direction remains one-dimensional. Therefore, the rescaling in temporal direction~(\ref{eq:tomporesacle}) will encounter no conflict with the UV-cutoff and we expect that rescaled action agrees with the naive one
\begin{align}
S=\frac{1}{4g_0^2(\epsilon,g_r(\frac{M}{\mu}))\mu^{2\epsilon}}\int dx^4 d^{3-2\epsilon}\vec{x}\left(2\lambda F^{4i}F^{4i}+\lambda^{-1}F^{ij}F^{ij}\right) \ ,
\end{align}
without any $\lambda$ dependence in $g_0$. As a result, the naive $H_c=\frac{1}{2}\left({\vec E}^2+\vec{B}^2 \right)$ is the full-Hamiltonian and contains both the scalar and tensor parts $H_T$ and $H_S$. A general lesson that we will learn from the non-linear sigma model in Sec.~\ref{sec:non} is that the finner the cutoff in the temporal direction, the larger the proportional in the full Hamiltonian carried by the ``naive'' one.  

The classical-looking Hamiltonian mixes the scalar and  tensor contributions can also be explained by investigating the Lorentz symmetry of the theory.  The dimensional-regularized theory has a $SO(1,3-2\epsilon)$ symmetry group instead of the $SO(1,3)$ in the continuum. In $4$-D,  $\frac{1}{2}\left({\vec E}^2+\vec{B}^2 \right)$ is the $00$ component of a traceless rank two tensor $-F^{\mu\rho}F^{\nu}_{\rho}+\frac{g^{\mu\nu}}{4}F^2$, which remains true in $d$-dimension with a factor of 1/4 in the second term. However, in $4-2\epsilon$ dimensions the traceless tensor is  $-F^{\mu\rho}F^{\nu}_{\rho}+\frac{g^{\mu\nu}}{4-2\epsilon}F^2$ and differs from the EMT in $d$-dimension  by an ``evanescent'' operator proportional to $\epsilon F^2$~\cite{Collins:1976yq,Collins:1984xc}. After taking the $\epsilon \rightarrow 0$ limit, it is $-F^{\mu\rho}F^{\nu}_{\rho}+\frac{g^{\mu\nu}}{4-2\epsilon}F^2$  becomes the tensor part in 4-D, while $\epsilon F^2$ remains finite despite the $\epsilon$ in front due to the presence of $\frac{1}{\epsilon}$ ultraviolet poles and becomes the trace-anomaly. This is a good demonstration of the fact that the difference in $SO(1,3)$ and $SO(1,3-2\epsilon)$ combined with the presence of UV divergence has far reaching consequences for renormalization of tensorial operators and their traces~\cite{Collins:1984xc}, in particular, the EMT for QCD as we will review now. 

Notice that although the discussions up to here are only for pure-YM, one can generalize them to the case of full QCD similarly. In the notation of ~\cite{Hatta:2018sqd,Metz:2020vxd}, the EMT for QCD reads
\begin{align}
T^{\mu\nu}=O_1^{\mu\nu}+\frac{O_2^{\mu\nu}}{4}+O_3^{\mu\nu} \ ,
\end{align}
with the operators:
\begin{eqnarray}
O_1^{\mu\nu}&=&-F^{\mu\rho}F^{\nu}_{\rho}  \ , \\
O_2^{\mu\nu}&=&g^{\mu\nu}F^2 \  ,  \\
O_3^{\mu\nu}&=&\bar \psi i\gamma^{(\mu}D^{\nu)} \psi \ , \\
O_4^{\mu\nu}&=&g^{\mu\nu} m\bar\psi\psi \ .
\end{eqnarray}
Although $T^{\mu\nu}$ is UV finite after summing over all the terms, none of the individual operators above have simple renormalizatin property,  due to the fact that they contain both scalar and tensor representations of the Lorentz group $SO(1,3-2\epsilon)$. To simplify the renormalizatin property and fully utilize the Lorentz structure, a standard way to proceed~\cite{Ji:1994av,Ji:1995sv,Peskin:1995ev} is to decompose them into the tracefull and traceless parts accroding to the Lorentz group $SO(1,3-2\epsilon)$ and renormalize separately 
\begin{align}
T^{\mu\nu}=T^{\mu\nu}_S+T^{\mu\nu}_T \ ,
\end{align}
where
\begin{eqnarray}
T^{\mu\nu}_T&=&\bigg(O^{\mu\nu}_1+\frac{O^{\mu\nu}_2}{4-2\epsilon}\bigg)+\bigg(O^{\mu\nu}_3-\frac{O^{\mu\nu}_4}{4-2\epsilon}\bigg) \ , \\
T^{\mu\nu}_S&=&\frac{g^{\mu\nu}}{4-2\epsilon}\bigg(m\bar\psi\psi-\frac{2\epsilon}{4-2\epsilon}F^2\bigg) \ ,
\end{eqnarray}
are tensor and scalar parts of the EMT in $4-2\epsilon $ dimensions. Under renormalization, operators belonging to tensor and scalar representations do not mix with each other and become the tensor and scalar operators for the renormalized theory in $4$-D after taking $\epsilon \rightarrow 0$ in the end. For more details regarding the standard way of renormalizing the energy-momentum tensor, see Ref.~\cite{Ji:1995sv}. 

Instead of renormalizing the trace and traceless parts of EMT separately, in Ref.~\cite{Hatta:2018sqd,Metz:2020vxd} the renormalization is performed for the operators $O_{1}$, $O_{2}$, $O_{3}$ by directly subtracting the $\frac{1}{\epsilon}$ poles~\cite{Collins:1984xc} without separating the tensor and scalar contributions.  As far as renormalization is concerned, this is perfectly fine. However, this renormalization procedure does not respect the Lorentz symmetry and the resulting finite operators $O_{1,R}$, $O_{2,R}$, $O_{3,R}$ mixes different Lorentz representations. In particular, the renormalized operator $\frac{1}{2}\left(\vec{E}^2+\vec{B}^2 \right)_{R}$ in this renormalization procedure mixes the tensor and scalar representations of the Lorentz group and is physically less useful. In contrary, the notation $\frac{1}{2}\left(\vec{E}^2+\vec{B}^2\right)_R$ has commonly been reserved for the 00-component of the renormalized traceless tensor ~\cite{Luke:1992tm,Ji:1994av,Kharzeev:1995ij} and can be measured directly through deep-inelastic scattering as the momentum fraction carried by
gluons. More importantly, the non-standard renormalization scheme advocated in Refs.~\cite{Hatta:2018sqd, Metz:2020vxd} hides the scheme and scale-independent QAE contribution in the mass.

To summarize, in the subsection we have shown that the ``naive'' Hamiltonian in dimensional regularization mixes the scalar and tensor contribution from two viewpoints: one is based on time-rescaling property and another is based on investigating the Lorentz invariance.  In the next subsection we will argue that a maximally scheme-independent mass sum rule must preserve the Lorentz structure and a clear separation between the tensor and scalar energy contributions. 

\subsection{Scheme-independent mass decomposition}

In the previous two subsections, we argued that the operator form of the mass sum rule is sensitive to the regularization scheme. In a cutoff scheme such as the symmetric lattice scheme for pure YM, we obtain the classical-looking term $H_c=\int d^3x \frac{1}{2g_0^2}\left(\vec{E}^2+\vec{B}^2\right)$ as well as the anomalous term $H_a$, while in DR we only get the ``naive'' one. In other regularization schemes such as the asymmetric lattice scheme, we get another linear combination.

Thus, the contribution of the ``naive'' Hamiltonian $H_c$ to the nucleon mass is scheme dependent and has not the clear physical interpretation one would usually expect. However, what we are interested is a maximally scheme independent decomposition of the nucleon mass. As was already discussed in \cite{Ji:1994av,Ji:1995sv}, the decomposition of the Hamiltonian into a traceless (scalar) and trace (tensor) parts is such, $H=H_T+H_S$. A crucial point is that the trace and trace-less parts of EMT correspond to two different irreducible representations of the Lorentz group. Therefore, this separation is unique and independent of any interpretation issue. In both dimensional and lattice regularization, the trace part is
\begin{align}
H_S=H_a=\frac{\beta(g)}{2g^3}\int d^3\vec{x}{\cal S}(0,\vec{x}) \ ,
\end{align}
while the expressions for the traceless part $H_T$ look different in different schemes but give identical contributions to the hadron mass
\begin{eqnarray}
H_T(\text{lattice})&=&\int\frac{ d^3\vec{x} }{2g_0^2}(\vec{E}^2+\vec{B}^2)  \ ,\nonumber \\
H_T(\text{DR})&=&\int \frac{d^{3-2\epsilon}\vec{x}}{g_0^2\mu^{2\epsilon}}\bigg(\frac{2-2\epsilon}{4-2\epsilon}\vec{E}^2+\frac{2}{4-2\epsilon}\vec{B}^2\bigg)
\nonumber  \ ,\\
\langle M|H_T(\text{lattice})|M\rangle &=& \langle M|H_T(\text{DR})|M\rangle =\frac{3}{4}M \ .
\label{eq:50}
\end{eqnarray}
It is the traceless part (\ref{eq:50}), but not the naive Hamiltonian that naturally corresponds to the gluon kinematic energy. 
The trace part of the energy momentum tensor corresponds to higher twist operators and thus cannot describe the energy contribution of the twist-two gluon parton distribution function, i.e. its second Mellin moment
$\int dx x f_{g}(x)$ which can be obtained by extrapolating experimental data and direct lattice calculations.

Generalizing to the case of full QCD, one has the following decomposition in DR:
\begin{eqnarray}\label{eq:sumDR}
H&=&H_T+H_S\equiv (H_g+H_q)+H_m+H_a \ ,\\
H_S&=&H_a+\frac{1}{4}H_m  \ ,\\
H_T&=&(H_g+H_q)+\frac{3}{4}H_m \ ,\\
\end{eqnarray}
where 
\begin{eqnarray}\label{eq:EB}
  H_a&=&\frac{1}{4} \int d^{3}\vec{x} \bigg(\frac{\beta(g)}{2g}F^2+\gamma_mm\bar\psi \psi\bigg)_{R} \ , \\
 H_m&=&\int d^{3}\vec{x} (m\bar\psi \psi)_{R} \ , \\
 H_g+H_q&=&\int d^{3}\vec{x} \bigg(\frac{2-2\epsilon}{4-2\epsilon}\vec{E}^2+\frac{2}{4-2\epsilon}\vec{B}^2+\bar \psi (-i\vec{\alpha}\cdot D)\psi\bigg)_R \ ,
\end{eqnarray}
in which the lower-script ${\rm R}$ denotes renormalized version. Of these $H_a$, $H_g+H_q$ and $H_m$ are separately scale invariant, while $H_g$ and $H_q$ are not. The decomposition preserves the Lorentz symmetry and can be obtained by looking at the $00$ components of the EMT renormalized using the standard methods in Refs.~\cite{Ji:1994av,Ji:1995sv,Peskin:1995ev} as discussed in Sec.~\ref{sec:DR}. 

Let's now consider the extraction of all the matrix elements in our decomposition. We first denote the nucleon sigma term as
\begin{align}
\sigma_N=\frac{\langle P|m\bar\psi\psi|P\rangle}{2M_N} \ .
\end{align}
We then notice the relation
\begin{align}
\langle P|T_{T}^{\mu\nu}|P \rangle=2\Bigl(P^{\mu}P^{\mu}-\frac{g^{\mu\nu}}{4}\Bigr)(x_q(\mu^2)+x_g(\mu^2))
\end{align}
which relates the matrix element of a twist-2 operator to moments of gluon and quark PDFs. The momentum fractions simply satisfies $x_q(\mu^2)+x_g(\mu^2)=1$.  Given these, we define quantities $M_g$, $M_q$, $M_m$ and $M_a$ by 
\begin{align}
&M_g+M_q=\langle H_g+H_q\rangle=\frac{3M_N}{4}(x_q(\mu^2)+x_g(\mu^2))-\frac{3\sigma_N}{4} \ ,\\
&M_m=\langle H_m\rangle=\sigma_N \ , \\
&M_a=\langle H_a\rangle=\frac{M_N-\sigma_N}{4}\ .
\end{align}
where $M_N$ is the nucleon mass. Therefore, the anomalous energy for the nucleon equals $\frac{1}{4}$ of the nucleon mass minus the nucleon sigma term $\sigma_N$. Essentially, two scale invariant quantities, $M_N$ and $\sigma_N$ are required. It is interesting to see that one has $M_g+M_q=3M_a$, a result
of the virial theorem.

To summarize, in this section we have reviewed the mass sum-rule in various regularization schemes. We have shown that due to the presence of UV cutoff in the temporal direction, the Hamiltonian of QCD-like theories in lattice cutoff requires a term equivalent to the QAE in addition to the ``classical'' one. We have shown that in dimensional regularization, the anomalous energy is hidden in the naive classical-looking Hamiltonian and a scheme independent mass decomposition must preserve the Lorentz structure and treat the scalar and tensor contributions at different footing. We also commented on the renormalization of EMT in QCD. In the next section we use the non-linear sigma model in $1+1$ dimension as an example to further demonstrate these points. 

\section{Mass generation and trace-anomaly in the $1+1$ dimensional non-linear sigma model}\label{sec:non}
As illustration of the results in Sec.~\ref{sec:dec}, we investigate in this section the mass decomposition of the 2-dimensional non-linear sigma model in the large $N$ limit in detail~\cite{Ji:2021pys}. We work exclusively in Euclidean formulation of the theory. The model ~\cite{Novikov:1984ac,Shifman:2012zz} consists of an $N$ component scalar field $\pi=(\pi^1,...\pi^N)$ normalized by  $\sum_{a=1}^N\pi^a\pi^a=1$. The action reads
\begin{align}
S=\frac{1}{2g_0^2}\int d^2x (\partial_{\mu} \pi^a)(\partial_{\mu} \pi^a) \ .
\end{align}
Here, $g_0$ is the dimensionless coupling constant. This model is $O(N)$ rotational invariant $\pi^a\rightarrow O^{ab}\pi^b$. A perturbative analysis of the model can be performed by using the parametrization $\pi=\left(g_0\pi^1,...g_0\pi^{N-1},\sqrt{1-g_0^2\sum_{i=1}^{N-1}\pi_i^2}\right)$ near the north pole which is identified with the perturbative vacuum. In this parametrization, the action for $\pi^1,..\pi^{N-1}$ reads:
\begin{align}
S=\int d^2x \frac{1}{2}\sum_{a=1}^{N-1}(\partial_{\mu} \pi^a)(\partial_{\mu} \pi^a)+\frac{g_0^2\sum_{a=1,b=1}^{N-1}(\pi^a\partial_{\mu} \pi^a)(\pi^b\partial_{\mu} \pi^b)}{2\sqrt{1-g_0^2\sum_{i=1}^{N-1}\pi_i^2}} \ .
\end{align}
The $O(N)$ symmetry is broken to $O(N-1)$ in the perturbative vacuum and the remaining $N-1$ $\pi^a$s are massless Goldstone bosons. One then expands the square root and treats the resulting terms as perturbations.  One can show that the resulting theory is renormalizable to all orders in $g_0$, and that the theory is asymptotically free~\cite{Novikov:1984ac,Shifman:2012zz}. However, due to the infrared divergences in 2d, the perturbative analysis fails to capture the vacuum structure of the theory. Instead of having $N-1$ massless modes, one expects that the theory is gapped and dimensional transmutation occurs in a similar way as for QCD in 4d. The $\pi^a$ fields are all massive and the $SO(N)$ invariance should be unbroken. But unlike QCD, there is no color charges in the theory that got confined. Here, we should notice that while there is quite convincing theoretical and numerical evidence that these statements are true, a formal proof is still missing, as far as we know.   

In the large $N$ limit of the theory, defined by taking $N\rightarrow \infty$ with $\lambda_0=g_0^2N$ being fixed, the theory is exactly savable and one can use it as a tool to investigate the mass structure in asymptotically free theory with mass generation. We will first provide a self-contained review of the model in large $N$ limit in Sec.~\ref{Sec:reviewsigma}, then investigate the mass structure of the model in Sec.~\ref{Sec:4.2}-~\ref{Sec:4.4}. From the discussion in Sec.~\ref{sec:scale}, the trace anomaly of the theory is given by
\begin{align}
    T^{\mu}_{\mu}=\frac{\beta(g_0)}{g_0}\sum_{a=1}^{N} \left(\partial^{\mu}\pi^a \partial_{\mu} \pi^a \right) \ , 
\end{align}
where the $\beta(g_0)$ is the beta function of the theory to be given later, which implies that the anomalous energy term reads:
\begin{align}
H_a=\frac{\beta(g_0)}{2g_0}\int dx_1 \sum_{a=1}^N\bigg((\partial_1 \pi^a)^2+(\partial_4 \pi^a)^2\bigg) \ .
\end{align}
We will show that regardless of the regularization scheme, the operator form of $H_a$ remains the same and contributes to half of the $\pi^a$ mass. 
On the other hand, we will show that the naive kinematic plus potential energy contribution $T+V$ or $H_c$, similar to the $\frac{1}{2}\left(\vec{E}^2+\vec{B}^2 \right)$ in gauge theory is sensitive to the regularization scheme and mixes the traceless and trace contributions in the case of DR. According to the discussions in Sec.~\ref{Sec:latticesum} and~\ref{sec:DR}, the explicit operator forms that reduce to the Hamiltonian $H$ in the continuum limit is regulator dependent. However, the total contribution of $H$ remains the same in all the regularization schemes.

\subsection{Review of the large $N$ limit of the model}\label{Sec:reviewsigma}

For the convenience for our discussion here we present here a self-contained introduction to the solution in the large $N$ limit defined similar to that of QCD as $\lambda=g_0^2N$ fixed while $N\rightarrow \infty$, for more details see Ref.~\cite{Novikov:1984ac,Shifman:2012zz}.  

One first introduces an auxiliary field $\hat \sigma$ and rewrites the action as:
\begin{align}\label{eq:sigmaxauxi}
S=\frac{1}{2g_0^2}\int d^2x (\partial_{\mu} \pi^a)(\partial_{\mu} \pi^a)+i\int d^2x \frac{\hat \sigma}{2g_0^2}\Bigl(\sum_a\pi^a\pi^a-1\Bigr) \ .
\end{align}
(By integrating out $\hat \sigma$, one recover the constraint $\sum_a\pi^a\pi^a=1$.) One then integrates out $\pi^a$ instead and ends up with the following effective action for $\hat \sigma$
\begin{align}
S[\sigma]&=\frac{N}{2}{\rm Tr}\ln(-\partial^2+i\hat \sigma)-\frac{i}{2g_0^2}\int d^2x \hat \sigma \nonumber \\
         &= N\bigg(\frac{1}{2}{\rm Tr}\ln(-\partial^2+i\hat\sigma)-\frac{i}{2\lambda_0}\int d^2x \hat\sigma\bigg) \ .
\end{align}
At large $N$, one expects the action to be dominated by the saddle point at $\hat\sigma=-im^2$ where $m$ is the fundamental mass scale of the model that will become the mass of the $\pi^a$ fields as we will see. The saddle point satisfies the gap equation:
\begin{align}\label{eq:gapsigma}
\int \frac{d^2k}{(2\pi)^2}\frac{1}{k^2+m^2}=\frac{1}{\lambda_0} \ ,
\end{align}
which determines the bare $\lambda_0$ as a function of the mass $m$ and the UV cutoff. To check the dominance of the saddle-point, one can expand the effective action around it $\hat\sigma=-im^2+\sigma$. The linear term vanishes due to the gap equation, and the quadratic term for $\sigma$ reads
\begin{align}
S_{2}[\sigma]=\frac{N}{4}{\rm Tr}\bigg(\frac{1}{-\partial^2+m^2}\sigma\frac{1}{-\partial^2+m^2}\sigma\bigg) =\frac{N}{2}\int\frac{d^2p}{(2\pi)^2}\sigma^{\dagger}(p)\Sigma^{-1}(p)\sigma(p)
\end{align}
where the ${\rm Tr}$ denotes the trace in coordinate or momentum space and the inverse propagator is:
\begin{align}
\Sigma^{-1}(p)=\int \frac{d^2k}{2(2\pi)^2}\frac{1}{\left((p-k)^2+m^2\right)\left(k^2+m^2\right)}
\end{align}
which is convergent and positive definite. At zero momentum, we have $\Sigma^{-1}(0)=\frac{1}{8\pi m^2}$. This guarantees the stability of the saddle point. One further notices that all the higher order terms for $\sigma$ are proportional to $N$, therefore, after rescaling $\sigma \rightarrow \frac{1}{\sqrt{N}}\sigma$, the action reads:
\begin{align}
S=NS_0+\frac{1}{4}{\rm Tr}\bigg(\frac{1}{-\partial^2+m^2}\sigma\frac{1}{-\partial^2+m^2}\sigma\bigg)+\sum_{i\ge 3}\frac{1}{N^{\frac{i}{2}-1}}S_{i} \ ,
\end{align}
and we obtain a systematic expansion in $1/\sqrt{N}$. This shows the dominance of the large $N$ saddle-point. . 

\begin{figure}[t]
\centering
\includegraphics[width=0.4\columnwidth]{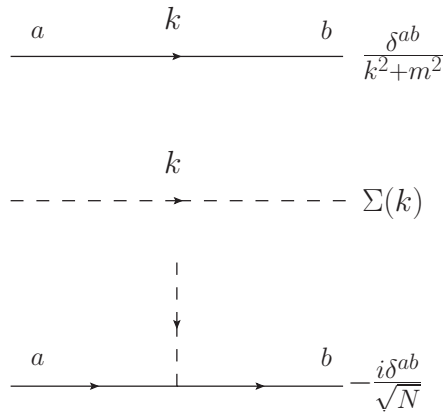}
\caption{The Feynman rules for the sigma model. The solid line and dashed line represents the $\pi$ and $\sigma$ propagators, respectively.}
\label{fig:rules}
\end{figure}

Given the large $N$ solution based on the auxiliary field, we move to the original fields $\pi^a$ in Eq.~(\ref{eq:sigmaxauxi}). To leading order in $N$, ~~$m^2\pi^a\pi^a$ is the mass term for $\pi^a$. Therefore, the large $N$ saddle point indicates that the $\pi^a$ form a massive $SO(N)$ vector multiplet in contrast to the $SO(N-1)$ multiplet obtained in the perturbative approach. All the higher order contributions can be generated from the following Feynman rules.
\begin{itemize}
  \item The massive field $\pi^a\rightarrow g_0\pi^a$ represented by a solid line has the propagator $\frac{\delta^{ab}}{k^2+m^2}$.
  \item The $\sigma$ field represented by a dashed line has the propagator $\Sigma(p)$. At zero momentum it reads $\Sigma(0)=8\pi m^2$.
  \item The interaction between two $\pi^a$ and one $\sigma$ is represented by the vertex $-\frac{i}{\sqrt{N}}$.
  \item The one-loop self-energy diagram for the $\sigma$ propagator, as well as the one-loop tadpole diagrams for $\sigma $ have to be discarded.
\end{itemize}
On can show that the resulting theory is renormalizable to all orders in $\frac{1}{N}$. There are three types of divergent diagrams. The two point function for $\pi$ is quadratically divergent, the tadpole-diagram for $\sigma$ is logarithmically divergent and the $\sigma-\pi$ vertex is also logarithmically divergent. These divergencies can be removed by corresponding charge and field renormalization, and the resulting theory is equivalent to the original nonlinear sigma model with a running $\frac{1}{g_0^2}$ beyond the leading order result. The Feynman rules for the theory are shown in Fig.~\ref{fig:rules}.

After introducing the large $N$ solution, we should mention that beyond the large $N$ limit, the $O(N)$ non-linear sigma model is widely believed to be integrable and has been investigated in Refs.~\cite{Shankar:1977cm,Wiegmann:1985jt}. Moreover, in the $O(3)$ case, there are instanton solutions which has been suggested to be responsible for the mass generation~\cite{Iwasaki:1980hd}. It is  interesting to combine the various viewpoints to produce a deeper understanding of the mass structure for the model. These are left for a future work.
 
\subsection{Mass decomposition in hard cutoff $k^2\le \Lambda_{UV}^2$ and lattice regularization}\label{Sec:4.2}

In this subsection we study the mass-decomposition in the hard cutoff $k^2\le \Lambda_{UV}^2$ and lattice regularization. We first consider the hard-cutoff. With this cutoff, $\lambda_0=Ng^2_0$ is fixed by the gap equation Eq.~(\ref{eq:gapsigma})
\begin{align}
\frac{1}{g_0^2N}=\int_{k^2\le\Lambda_{\rm UV}^2}~\frac{d^2k}{(2\pi)^2}~\frac{1}{k^2+m^2}=\frac{1}{4\pi}\ln \frac{\Lambda^2_{UV}}{m^2} \ .
\end{align}
As a result, the $\beta$ function reads:
\begin{align}
\beta(g_0)=\Lambda_{\rm UV}\frac{dg_0}{d\Lambda_{\rm UV}}=-\frac{Ng_0^3}{4\pi} \ .
\end{align}
Therefore, the anomalous term $H_a$ is:
\begin{align}
H_a=-\frac{Ng_0^2}{8\pi}\int dx_1 \sum_{a=1}^N\bigg((\partial_1 \pi^a)^2+(\partial_4 \pi^a)^2\bigg) \ .
\end{align}
According to the discussion in Sec.~\ref{Sec:latticesum}, in the symmetric cutoff the Hamiltonian has the form 
\begin{align}\label{eq:Hsymmetric}
    &H=H_c+H_a  \ , \\ 
    &H_c=\frac{1}{2}\int dx_1 \sum_{a=1}^N\bigg((\partial_1 \pi^a)^2-(\partial_4 \pi^a)^2\bigg) \ .
\end{align}
We now evaluate the matrix element of $H_c$ and $H_a$ in the a massive $\pi^a$ state $|\pi^a,\vec{P}=0\rangle$ at rest.

\begin{figure}[t]
\centering
\includegraphics[width=0.4\columnwidth]{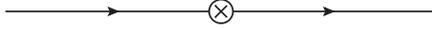}
\caption{The tree level diagram for the $H_c$ and $H_a$ operator insertions. The crossed circle represents a generic $\pi$-bilinear operator. }
\label{fig:tree}
\end{figure}

\begin{figure}[t]
\centering
\includegraphics[width=0.4\columnwidth]{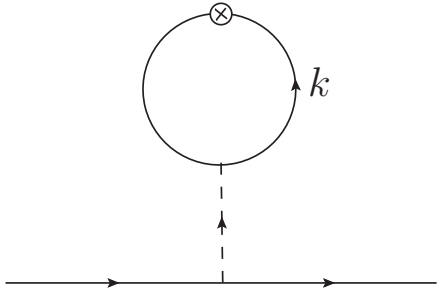}
\caption{The one-loop tadpole diagram for the $H_c$ and $H_a$ operator insertions. This diagram is leading in $N$ because the two factors $\frac{1}{\sqrt{N}}$ from the $\pi$-$\sigma$ coupling cancels the factor $N$ from the loop. }
\label{fig:tadpole}
\end{figure}

We first study $H_a$, more precisely, we calculate the forward matrix-element of $H_a$ in the $\pi^a$ state at rest. In the leading order of $N$ there are two diagrams. The first one, shown in Fig.~\ref{fig:tree}, is of tree level and the second one shown in Fig.~\ref{fig:tadpole} is of one-loop level. It is leading because the factor $N$ from the loop cancels the factor $(\frac{1}{\sqrt{N}})^2$ from the vertices. The diagram gives (note that $k^2=(k_4)^2+(k_1)^2$):
\begin{align}
E_a=\frac{1}{2m}\bigg(\frac{Ng_0^2}{4\pi}m^2+\frac{Ng_0^2}{8\pi}\Sigma(0)\int_{k^2\le \Lambda^2_{\rm UV}}\frac{d^2k}{(2\pi)^2}\frac{k^2}{(k^2+m^2)^2}\bigg) \ .
\end{align}
Using $\Sigma(0)=8\pi m^2$ one obtains
\begin{align}
E_a&=\frac{\langle \pi^a,\vec{P}=0|H_a|\pi^a,\vec{P}=0\rangle}{\langle \pi^a,\vec{P}=0|\pi^a,\vec{P}=0\rangle}=\frac{1}{2m}\frac{Ng_0^2}{4\pi}m^2\bigg(1-\frac{\Sigma(0)}{2}\int_{k^2\le \Lambda^2_{\rm UV}}\frac{d^2k}{(2\pi)^2}\frac{1}{(k^2+m^2)^2}\bigg)\nonumber \\
&+\frac{1}{2m} Ng_0^2m^2\int_{k^2\le \Lambda^2_{\rm UV}}\frac{d^2k}{(2\pi)^2}\frac{1}{(k^2+m^2)} \ .
\end{align}
The first line vanishes due to the identity $\frac{1}{2}\Sigma(0)\int \frac{d^2k}{(2\pi)^2}\frac{1}{(k^2+m^2)^2}=1$, even in the presence of the regulator, while the second line equals  $\frac{m^2}{2m}=\frac{m}{2}$ after using the gap equation Eq.~(\ref{eq:gapsigma}).
Therefore, we conclude that
\begin{align}
E_a=\frac{m}{2} \ ,
\end{align}
in consistency with the virial theorem. 

We then study the contribution $E_c$ of $H_c$ in the $\pi^a$ state. At leading order in $N$ there are again two contributions given by the tree-level diagram in Fig.~\ref{fig:tree} and the 
one-loop diagram in  Fig.~\ref{fig:tadpole}. The tree level diagram can be easily calculated and contributes to $\frac{m}{2}$, while the tadpole diagram contribution reads
\begin{align}
E_c|_{\text{Fig.~\ref{fig:tadpole}}}=\frac{1}{2m}N(-\frac{i}{\sqrt{N}})^2\Sigma(0)\frac{1}{2}\int_{k^2\le \Lambda^2_{\rm UV}}\frac{d^2k}{(2\pi)^2}\frac{(k_4)^2-(k_1)^2}{(k^2+m^2)^2}=0 \ .
\end{align}
Therefore, we conclude that in the cutoff scheme $k^2\le \Lambda^2_{\rm UV}$ the kinematic energy, i.e. the expectation value of $H_c$, contributes half of the mass. Therefore the average of $H$ equals to $m$, consistent with the fact that the $\pi^a$ field has mass $m$.  

Similarly, the lattice cutoff can be treated using the same methods. The only difference is that $\partial^2$ becomes the lattice version of the Laplacian. The gap equation reads in lattice regularization
\begin{align}
\frac{1}{g_0^2N}=\int_{-\frac{\pi}{a}}^{\frac{\pi}{a}}\frac{dk_4dk_1}{(2\pi)^2}\frac{1}{m^2+\frac{2}{a^2}(2-\cos ak_4-\cos ak_1)} \ .
\end{align}
With the notation $A(k,a)=\frac{2}{a^2}(2-\cos ak_4-\cos ak_1)$,~ $\Sigma(p)$ is given by:
\begin{align}
\Sigma^{-1}(p,a)=\frac{1}{2}\int_{-\frac{\pi}{a}}^{\frac{\pi}{a}}\frac{dk_4dk_1}{(2\pi)^2}\frac{1}{m^2+A(p-k,a)}\frac{1}{m^2+A(k,a)} \ .
\end{align}
It is straightforward to show that  $\Sigma(p,a)\rightarrow \Sigma(p)$ as $a\rightarrow 0$, therefore the propagator for the sigma field remains the same in lattice regularization.

These equations allow us to check whether all sum rules for the lattice cutoff are the same as for the hard cutoff $k^2\le \Lambda_{\rm UV}$ in the continuum limit, which is the case.
For example, the contribution $H_a$ reads:
\begin{align}
E_a&=\frac{1}{2m}\frac{Ng_0^2}{4\pi}m^2\bigg(1-\frac{\Sigma(0)}{2}\int_{-\frac{\pi}{a}}^{\frac{\pi}{a}}\frac{d^2k}{(2\pi)^2}\frac{1}{(A(k,a)+m^2)^2}\bigg)\nonumber \\
&+\frac{1}{2m} Ng_0^2m^2\int_{-\frac{\pi}{a}}^{\frac{\pi}{a}}\frac{d^2k}{(2\pi)^2}\frac{1}{A(k,a)+m^2} 
\end{align}
which is $\frac{m}{2}$ thanks to the gap equation.
\subsection{Mass decomposition in hard cutoff $|k_1|\le \Lambda_{UV}$}\label{Sec:4.3}
We then investigate the scheme in which we only impose a hard cutoff $|k_1|\le \Lambda_{UV}$ in the spatial direction while the $k_4$ is unbounded. This regulator breaks the rotational invariance explicitly. Since the $k_4$ integral can be rescaled freely, we expect that in this case the naive Hamiltonian $H_c$ term generates the complete hadron mass and 
\begin{align}
    H=H_c \ .
\end{align}
Most of above calculations remain valid. The gap equation reads
\begin{align}
\frac{1}{g_0^2N}=\int_{|k_1|\le\Lambda_{\rm UV}}\frac{d^2k}{(2\pi)^2}\frac{1}{k^2+m^2}=\frac{1}{2\pi}\ln \frac{\Lambda_{\rm UV}}{m}+\frac{\ln 2}{2\pi} \ .
\end{align}
Also the beta function remains the same. We first study the contribution of $H_a$. The diagrams for $H_a$ are again given by Fig.~\ref{fig:tree} and Fig.~\ref{fig:tadpole}. The total result reads
\begin{align}
E_a&=\frac{1}{2m}\frac{Ng_0^2}{4\pi}m^2\bigg(1-\frac{\Sigma(0)}{2}\int_{|k_1|\le \Lambda_{\rm UV}}\frac{d^2k}{(2\pi)^2}\frac{1}{(k^2+m^2)^2}\bigg)\nonumber \\
&+\frac{1}{2m} Ng_0^2m^2\int_{|k_1|\le \Lambda_{\rm UV}}\frac{d^2k}{(2\pi)^2}\frac{1}{(k^2+m^2)} \ .
\end{align}
which can be shown to become $\frac{m}{2}$ in the limit $\Lambda_{\rm UV}\rightarrow \infty$ by using the gap equation.

We then study the contribution for $H=H_c$, the tree diagram remains the same and equals $\frac{m}{2}$, while the tadpole diagram becomes
\begin{align}
E_c|_{\text{Fig.~\ref{fig:tadpole}}}=\frac{1}{2m}N\Bigl(-\frac{i}{\sqrt{N}}\Bigr)^2\Sigma(0)\frac{1}{2}\int_{|k_1|\le \Lambda_{\rm UV}}\frac{d^2k}{(2\pi)^2}\frac{k_4^2-k_1^2}{(k^2+m^2)^2} \ .
\end{align}
Unlike in the case of a symmetric cutoff, the integral does not vanish
\begin{align}
\int_{|k_1|\le \Lambda_{UV}}\frac{d^2k}{(2\pi)^2}\frac{k_4^2-k_1^2}{(k^2+m^2)^2}=\frac{1}{4\pi} 
\end{align}
in the $\Lambda_{\rm UV}\rightarrow \infty$ limit. The fact that a naively vanishing integral is non-vanishing due to the presence of cutoff signals the anomalous nature of this one-loop contribution. Using $\Sigma(0)=8\pi m^2$ we obtain again $\frac{m}{2}$. Thus in the $|p_1|\le \Lambda_{UV}$ scheme one found that the average of $H$ equals to $m$.

Again, the contribution of the total Hamiltonian equals to the mass of the $\pi^a$ field. We should emphasize that although the contribution of $H_c$ in this cutoff equals to the total mass of $\pi^a$, the one-loop part is in fact of anomalous nature and equals to that of $H_a$. The part that can be identified as ``classical'' $T+V$ contribution remains to be $\frac{m}{2}$.

More generally, we can show that in the regularization scheme $\frac{k_4^2}{\lambda^2}+k_1^2\le \Lambda^2_{UV}$ where $\lambda>0$ is positive, corresponding to the time-rescaled theory $x_4'=\lambda x_4$ discussed in Sec.~\ref{Sec:latticesum}. In this case, the running of $g_0$ is $\lambda$ dependent:
\begin{align}
\frac{1}{g_0^2N}=\frac{1}{2\pi}\ln \frac{\Lambda_{\rm UV}}{m}-\frac{1}{2\pi}\ln\left(\frac{\lambda+1}{2\lambda}\right) \ .
\end{align}
From which one has 
\begin{align}\label{eq:dg0dlambda}
    \lambda\frac{dg_0}{d\lambda}= \frac{\beta(g_0)}{1+\lambda} \ .
\end{align}
At $\lambda=1$, corresponding to the symmetric cutoff case, one has $\lambda\frac{dg_0}{d\lambda}= \frac{\beta(g_0)}{2}$. At generic $\lambda$,  from Eq.~(\ref{eq:genericH}), the total Hamiltonian reads 
\begin{align}\label{eq:geneticHsigma}
    H=H_c+\frac{\lambda}{g_0}\frac{dg_0}{d\lambda}\int dx_1 \sum_{a=1}^N\bigg((\partial_1 \pi^a)^2+(\partial_4 \pi^a)^2\bigg)  \ ,
\end{align}
where we have used Eq.~(\ref{eq:dg0dlambda}) in the last equality. The contribution of $H_a$ , which is contained in the one-loop diagram of $H_c$ as well as the last term in Eq.~(\ref{eq:geneticHsigma}) can be calculated as
\begin{align}
    E_a&=\frac{1}{2m}\frac{Ng_0^2}{4\pi}m^2\bigg(1-\frac{\Sigma(0)}{2}\int_{\frac{k_4^2}{\lambda^2}+k_1^2\le \Lambda^2_{UV}}\frac{d^2k}{(2\pi)^2}\frac{1}{(k^2+m^2)^2}\bigg)\nonumber \\
&+\frac{1}{2m} Ng_0^2m^2\int_{\frac{k_4^2}{\lambda^2}+k_1^2\le \Lambda^2_{UV}}\frac{d^2k}{(2\pi)^2}\frac{1}{(k^2+m^2)}=\frac{m}{2} \ ,
\end{align}
and remains to the $\frac{m}{2}$, while the average of the $H_c$ part in this case can be calculated as $\frac{m\lambda}{1+\lambda}$, thus using Eq.~(\ref{eq:geneticHsigma}) one has
\begin{align}
\frac{\langle \pi^a,\vec{P}=0|H|\pi^a,\vec{P}=0\rangle}{\langle \pi^a,\vec{P}=0|\pi^a,\vec{P}=0\rangle}= \frac{m\lambda}{1+\lambda}+\frac{m}{1+\lambda}=m \ ,
\end{align}
which equals to the mass of the $\pi^a$.
\subsection{Mass decomposition in dimensional regularization}\label{Sec:4.4}
\begin{table*}
\caption{\label{tab:mass} Mass decomposition in various schemes: While the naive Hamiltonian $H_c$ gives scheme-dependent results, the anomaly contribution is always present and scheme-independent. The total mass is of course fixed.}
\vskip 0.3 cm
\begin{center}
\begin{tabular}{|c|c|c|c|}
\hline
scheme & $H_c$  &  $H_a\equiv H_S$ & $H$  \\
\hline\hline
$k^2\le \Lambda^2_{\rm UV}$&	$\frac{m}{2}$ &	$\frac{m}{2}$& $m$\\
\hline
lattice &	$\frac{m}{2}$&	$\frac{m}{2}$&	$m$\\
\hline
$|k_1|\le \Lambda_{\rm UV}$&	$m$ &	$\frac{m}{2}$&	$m$\\
\hline
DR &	$m$&	$\frac{m}{2}$&	$m$ \\
\hline
$\frac{k_4^2}{\lambda^2}+k_1^2\le \Lambda^2_{\rm UV}$ &	$\frac{\lambda m}{1+\lambda}$&	$\frac{m}{2}$&	$m$ \\
\hline
\end{tabular}
\end{center}
\end{table*}
Finally, we study DR. For DR with space-time dimension $d=2-2\epsilon$ one should change $g_0^2\rightarrow g_0^2\mu^{2\epsilon}$. The gap equation reads:
\begin{align}
\frac{1}{g_0^2N}=\mu^{2\epsilon}\int\frac{d^{2-2\epsilon}k}{(2\pi)^{2-2\epsilon}}\frac{1}{k^2+m^2}=\Bigl(\frac{\mu}{m}\Bigr)^{2\epsilon}\frac{\Gamma(\epsilon)}{(4\pi)^{1-\epsilon}}\nonumber \\ =\frac{1}{4 \pi  \epsilon }+\frac{2 \ln (\frac{\mu}{m})-\gamma_{\rm E} +\ln (4 \pi )}{4 \pi } \ .
\end{align}
The beta function is again given by $-\frac{g_0^3}{4\pi}$. In DR, the total Hamiltonian reads
\begin{align}
    H=H_c  \ ,
\end{align}
similar to the $|k_1|\le \Lambda_{UV}$ case. Lets first study the contribution of $H_a$. The diagrams are again given in Fig.~\ref{fig:tree} and Fig.~\ref{fig:tadpole}. The result reads
\begin{align}
E_a&=\frac{1}{2m}\frac{Ng_0^2}{4\pi}m^2\bigg(1-\frac{1}{2}\Sigma(0)\mu^{2\epsilon}\int\frac{d^{2-2\epsilon}k}{(2\pi)^{2-2\epsilon}}\frac{1}{(k^2+m^2)^2}\bigg)\nonumber \\ &+\frac{1}{2m} Ng_0^2\mu^{2\epsilon}m^2\int\frac{d^{2-2\epsilon}k}{(2\pi)^{2-2\epsilon}}\frac{1}{(k^2+m^2)} \ ,
\end{align}
which again equals $\frac{m}{2}$ as $\epsilon\rightarrow 0$ thanks to the gap equation.

We then discuss the $H_c$. The tree level diagram remains the same as before and equals $\frac{m}{2}$. The tadpole diagram now reads:
\begin{align}
E_c|_{\text{Fig.~\ref{fig:tadpole}}}=\frac{1}{2m}N\Bigl(-\frac{i}{\sqrt{N}}\Bigr)^2\Sigma(0)\frac{1}{2}\mu^{2\epsilon}\int\frac{dk_4d^{1-2\epsilon}k_1}{(2\pi)^{2-2\epsilon}}\frac{k_4^2-k_1^2}{(k^2+m^2)^2} \ .
\end{align}
Since we have split the space-time of dimension $2-2\epsilon$ into a 1-dimensional time and a $1-2\epsilon$ dimensional space, the integral reads:
\begin{align}
\int\frac{dk_4d^{1-2\epsilon}k_1}{(2\pi)^{2-2\epsilon}}\frac{k_4^2-k_1^2}{(k^2+m^2)^2}=\frac{\epsilon}{1-\epsilon}\int \frac{d^{2-2\epsilon}k}{(2\pi)^{2-2\epsilon}}\frac{k^2}{(k^2+m^2)^2} \rightarrow \frac{1}{4\pi}\ ,
\end{align}
where the factor $\frac{\epsilon}{1-\epsilon}$ comes from $\frac{1}{2-2\epsilon}-\frac{1-2\epsilon}{(2-2\epsilon)}$. As $\epsilon \rightarrow 0$, there is a $\frac{1}{4\pi \epsilon}$ pole from the integral which cancels the $\epsilon$ in front, leading to a finite result. This is identical to the mechanism of trace-anomaly in DR and we again see that the {\it non-vanishing one-loop diagram for $H_c$ is of anomalous nature}. By using $\Sigma(0)=8\pi m^2$ one obtains $\frac{m}{2}$. Thus we found that the contribution of $H$ again equals to the mass of the $\pi^a$ field. 
 
In dimensional regularization, one can check that the traceless part of the EMT produces indeed $\frac{m}{2}$. $H_T$ in $2-2\epsilon$ dimensions reads:
\begin{align}
H_T=\frac{1}{\mu^{2\epsilon}}\int d^{1-2\epsilon}x \sum_a\Bigl(-\frac{1-2\epsilon}{2-2\epsilon}(\partial_4\pi^a)^2+\frac{1}{2-2\epsilon}(\partial_1 \pi^a)^2\Bigr)
\end{align}
with which the Hamiltonian can be equivalently written as 
\begin{align}
    H=H_T+H_a \ .
\end{align}
We now verify that the contribution of $H_T$ is indeed $\frac{m}{2}$. The tree level diagram remains $\frac{m}{2}$. The tadpole diagram is now proportional to
\begin{align}
E_T|_{\text{Fig.~\ref{fig:tadpole}}}=\int\frac{dk_4d^{1-2\epsilon}k_1}{(2\pi)^{2-2\epsilon}}\frac{1}{(k^2+m^2)^2}\Bigl(-\frac{1-2\epsilon}{2-2\epsilon}k_4^2+\frac{1}{2-2\epsilon}k_1^2\Bigr) \ ,
\end{align}
which in turn is proportional to
\begin{align}
\bigg(-\frac{1-2\epsilon}{2-2\epsilon}\frac{1}{2-2\epsilon}+\frac{1}{2-2\epsilon}\frac{1-2\epsilon}{2-2\epsilon}\bigg) \equiv 0 \ .
\end{align}
The vanishing of the one-loop contribution for $H_T$ is identical to that of $H_c$ case in symmetric cutoff. Again, only the tree-level contribution to $H_c$ can be identified as the ``classical'' $T+V$ contribution. 
  
To summarize, in this section we have investigated the mass structure of $1+1$ non-linear sigma model in detail.  In all the schemes, the anomalous term $H_a$ has the same operator form and contributes half of the $\pi^a$ mass. The total Hamiltonian, although differs in operator forms in different regularization schemes, contribute to the total $\pi^a$ mass. On the other hand, although the naive Hamiltonian $H_c$ has the same form as the classical one, the contribution is actually regulator dependent and has no universal physical meaning. In regularization schemes that treat both directions equally, $H_c$ can be identified as $H_T$ while in schemes in which the $k_4$ integral can be re-scaled back and forth, such as the $|k_1|\le \Lambda_{\rm UV}$ scheme and the dimensional regularization scheme, $H_c$ can be identified as the full Hamiltonian. However, even in regulators where $H_c$ equals to $H$ formally, we still found that the one-loop contribution of $H_c$ is of anomalous nature and can be identified as $H_a$.  
We summarize the various contributions in table.~\ref{tab:mass}. The example of the 1+1 dimensional sigma model thus illustrates that the QAE is part of the total mass of $\pi^a$ regardless regularization schemes. This suggests that its contribution to the total mass might be related to some very general property of the theory under study.

\section{Virial theorem and perturbative anomalous energy contribution in QED}\label{sec:qed}
In the previous sections, it has been shown that due to the cutoff dependency of the coupling constants, a trace anomaly is generated and contributes to the energy. In this section we study the case of QED. In QED, although there is no dynamical scale generation, there are still UV divergences that lead to the beta function $\beta(e)=\frac{e^3}{12\pi^2}$ and mass anomalous dimension $\gamma_m=\frac{3e^2}{8\pi^2}$.  As a consequence, there is an anomalous contribution to the energy
\begin{align}\label{eq:Haqed}
H_a=\frac{1}{4} \int d^{3}\vec{x} \bigg(\frac{\beta(e)}{2e}F^2+\gamma_mm\bar\psi \psi\bigg) \ ,
\end{align}
in addition to the mass term
\begin{align}\label{eq:Hmqed}
H_m=\int d^3\vec{x} m\bar\psi\psi \ ,
\end{align}
where $F^{\mu\nu}$ is the electromagnetic field strength and $\psi$ is the electron field.  In this section we study $H_a$ for free electrons and for bond-states in a background field.

\subsection{Virial theorem and non-relativistic reduction}

Before coming to the QAE, let's consider the virial theorem for QED in the non-relativistic limit and show that it reduces to the corresponding virial theorem in quantum mechanics. One first notices that the scalar part of the Hamiltonian reads
\begin{align}\label{eq:Hsqed}
    H_S=\frac{1}{4}H_m+H_a \ ,
\end{align}
where $H_a$ and $H_m$ are given in Eqs.~(\ref{eq:Haqed},\ref{eq:Hmqed}). 
In the mean time, the full Hamiltonian reads in symmetric regularization scheme:
\begin{align}
H=\int d^3 \vec{x} \bigg(\frac{1}{2}(\vec{E}^2+\vec{B}^2)+\psi^{\dagger}(-i\vec{\alpha}\cdot\vec{D})\psi\bigg)+H_m+H_a \ .
  \label{eq:5-4}
\end{align}
Therefore, the virial theorem,
$E_S=E/4$, indicates that
\begin{align} \label{eq:virialqed}
    \Big \langle \vec{P}=0 \left| \int d^3 \vec{x} \bigg(\frac{1}{2}(\vec{E}^2+\vec{B}^2)+\psi^{\dagger}(-i\vec{\alpha}\cdot\vec{D})\psi\bigg) \right| \vec{P}=0 \Big\rangle=3\langle  \vec{P}=0| H_a | \vec{P}=0\rangle \ ,
\end{align}
where $|\vec{P}=0\rangle$ is a state in the rest frame.
One then notices that in the non-relativistic limit, the anomalous contributions is unimportant. Neglecting $H_a$ in Eq.~(\ref{eq:virialqed}), one obtains the relation:
\begin{align}\label{eq:Virialcou}
\langle \vec{P}=0|(\vec{E}^2+\vec{B}^2)/2+(-i\vec{\alpha}\cdot\vec{D})|\vec{P}=0\rangle=0  \ .
\end{align}
We now show that Eq.~(\ref{eq:Virialcou}) reduces to the virial theorem for the hydrogen-like systems in non-relativistic quantum mechanics.

We chose the Coulomb gauge $\nabla \cdot \vec{A}=0$ where $\vec{A}$ contains only transverse part $\vec{A}=\vec{A}_T$ . The quantization of QED in this gauge is explained in many textbooks~\cite{Weinberg:1995mt}. The temporal component $A^0$ decouples from the transverse part of the gauge field and is expressed as,
\begin{align}\label{eq:A0solution}
A^0=-\frac{e}{\nabla^2}\bar \psi\gamma^0\psi \ .
\end{align}
By using $\vec{E}=-\partial_t \vec{A}_T-\nabla A_0$ and the explicit solution of $A_0$ in Eq.~(\ref{eq:A0solution}), the transverse and longitudinal parts of the electric field decouple from each other and one has the relation 
\begin{align}\label{eq:decomposeE}
\int d^3\vec{x}\frac{1}{2}\left(\vec{E}^2+\vec{B}^2\right)=\int d^3\vec{x}\frac{1}{2}\left(\vec{E}_T^2+\vec{B}_T^2\right)+\frac{e^2}{2}\int d^3\vec{x}d^3\vec{y}\frac{\psi^{\dagger}\psi(\vec{x})\psi^{\dagger}\psi(\vec{y})}{4\pi |\vec{x}-\vec{y}|} \ ,
\end{align}
where $\vec{E}_T=-\partial_t \vec{A}_T$ and $\vec{B}_T=\nabla \times \vec{A}_T$ are the radiactive part of the photon field. 
For a positronium state, to leading order in a non-relativistic expansion the contribution of the transverse radiation fields $\vec{A}_T$ is negligible, thus the virial theorem Eq.~(\ref{eq:Virialcou}) reduces to
\begin{align}\label{eq:virialnon}
\langle  \vec{P}=0|\psi^{\dagger}(-i\vec{\alpha}\cdot \vec{\nabla})\psi| \vec{P}=0\rangle +\Big\langle \vec{P}=0\left| \frac{e^2}{2}\int d^3\vec{x}d^3\vec{y}\frac{\psi^{\dagger}\psi(\vec{x})\psi^{\dagger}\psi(\vec{y})}{4\pi |\vec{x}-\vec{y}|}  \right| \vec{P}=0\Big\rangle=0 \ ,
\end{align}
by using Eq.~(\ref{eq:decomposeE}). 

We now investigate the consequence of Eq.~(\ref{eq:virialnon}) on the leading component of the positronium state 
\begin{align}
|\vec{P}=0\rangle=\int\frac{d^3\vec{k}}{2(2\pi)^{3}E_k}\psi_{ss'}(\vec{k})a_s^{\dagger}(\vec{k})b_{s'}^{\dagger}(-\vec{k})|0\rangle \,
\end{align}
where the creation and anihilation operators are normalized as $[a_s(\vec{k}),a_{s'}^{\dagger}(\vec{k}')]_+=(2\pi)^32E_k\delta_{s,s'}\delta^3(\vec{k}-\vec{k'})$ , and similarly for $b$ and $b^{\dagger}$. The non-relativistic wave function is normalized as
\begin{align}
\int \frac{d^3\vec{k}}{(2\pi)^3}\psi^{\dagger}_{ss'}(\vec{k})\psi_{ss'}(\vec{k})=1 \ .
\end{align}
Using the free field
\begin{align}
\psi(x)=\int \frac{d^3\vec{k}}{(2\pi)^32E_k}\sum_s\bigg(a_s(\vec{k})u_s(k)e^{-ik\cdot x}+b_s^{\dagger}(\vec{k})v_s(k)e^{ik\cdot x}\bigg)\ ,
\end{align}
in the non-relativistic limit the matrix elements can be calculated as 
\begin{align}
  \langle \vec{P}=0|\psi^{\dagger}(-i\vec{\alpha}\cdot \vec{\nabla})\psi|\vec{P}=0\rangle=2\int \frac{d^3\vec{k}}{(2\pi)^3}\psi^{\dagger}_{ss'}(\vec{k})\psi_{ss'}(\vec{k})\frac{k^2}{m} \ ,
\end{align}
 and 
 \begin{align}
     \left\langle \vec{P}=0\left|\frac{e^2}{2}\int d^3\vec{x}d^3\vec{y}\frac{\psi^{\dagger}\psi(\vec{x})\psi^{\dagger}\psi(\vec{y})}{4\pi|\vec{x}-\vec{y}|}\right|\vec{P}=0\right \rangle=-e^2\int\frac{d^3\vec{k}d^3\vec{q}}{(2\pi)^3}\frac{\psi^{\dagger}_{ss'}(\vec{k})\psi_{ss'}(\vec{q})}{|\vec{k}-\vec{q}|^2} \ .
 \end{align}
 Therefore, the relation Eq.~(\ref{eq:virialnon}) simply reduces to the non-relativistic virial theorem 
\begin{align}
 \langle V\rangle =-2\langle T\rangle  \ ,
\end{align}
where
\begin{align}
 \langle V\rangle =-e^2\int\frac{d^3\vec{k}d^3\vec{q}}{(2\pi)^3}\frac{\psi^{\dagger}_{ss'}(\vec{k})\psi_{ss'}(\vec{q})}{|\vec{k}-\vec{q}|^2} \ ,\\
\langle T\rangle =\int \frac{d^3\vec{k}}{(2\pi)^3}\psi^{\dagger}_{ss'}(\vec{k})\psi_{ss'}(\vec{k})~\frac{k^2}{m} \ ,
\end{align}
are the kinematic and Coulomb energy. In conclusion, for non-relativistic hydrogen-like systems the virial theorem simply reduces to the classical relation $\langle V \rangle=-2\langle T\rangle$.
\subsection{The anomalous contribution to the electron pole mass}
After discussing the virial theorem, we now return to the perturbative QAE in QED. The simplest quantity is the electron pole mass $m_e$, defined as the pole of the inverse electron propagator~\cite{Kronfeld:1998di}. Here we show that although small, the perturbative QAE does contributes to the electron pole mass. 

We first consider the $\overline{MS}$ scheme at renormalization scale $\mu$, where the minimally renormalized electron mass $m=m(\mu)$ that appears in the Lagrangian is not the electron pole mass $m_e$, instead, at one-loop order one has the relation~\cite{Broadhurst:1991fy} 
\begin{align} \label{eq:oneloopole}
    m_e=m(\mu)\left(1+\frac{\alpha(\mu)}{\pi}\right) \ .
\end{align}
The one-loop beta function reads $\beta(e)=\frac{e^3}{12\pi^2}$ and $\gamma_m=\frac{3e^2}{8\pi^2}$.  At one-loop order, the $F^2$ term does not contribute and only the $\gamma_m m \bar \psi \psi$ term contributes to the anomalous energy $H_a$
\begin{align}
E_a=\frac{3\alpha}{8\pi}m(\mu) \ .
\end{align}
On the other hand, one can show that at one loop order the mass term contributes as 
\begin{align}
    E_m=\frac{\langle\vec{P}=0| H_m|\vec{P}=0 \rangle}{\langle\vec{P}=0 |\vec{P}=0\rangle}=\left(1-\frac{\alpha}{2\pi}\right) m(\mu) \ ,
\end{align}
which gives the correct result $ E_a+\frac{1}{4}E_m=\frac{m_e}{4}$.

We then consider the on-shell renormalization scheme~\cite{Peskin:1995ev,Weinberg:1995mt}, where the mass parameter of the renormalized Lagrangian equals to the pole mass . At one-loop order the mass anomalous dimension equals to $\gamma_m=\frac{3e^2}{8\pi^2}$ which is the same as in $\overline{MS}$. Therefore the contribution of the QAE is
\begin{align}
    E_a=\frac{3\alpha}{8\pi}m_e \ .
\end{align}
On the other hand, the contribution of the mass terms now reads
\begin{align}
    E_m=\left(1-\frac{3\alpha}{2\pi}\right)m_e \ ,
\end{align}
which is again consistent with the relation $E_a+\frac{1}{4}E_m=\frac{m_e}{4}$. 

\subsection{The anomalous contribution in a background field (Lamb shift)}

In this subsection we investigate the contribution of the anomalous term to the QED radiative correction of hydrogen atom binding energies. The naive Lagrangian density of the system with a background Coulomb field reads:
\begin{align}
{\cal L}=\bar \psi (i\gamma \cdot\partial -m)\psi -e\bar \psi\gamma^{\mu}\psi (A_{\mu}+{\cal A}_{\mu})-\frac{1}{4}F^{\mu\nu}F_{\mu\nu} \ ,
\end{align}
where ${\cal A}^{\mu}=\frac{Ze}{4\pi|\vec{x}|}\delta^{\mu}_0$ is the Coulomb field of the heavy nucleus and $A^{\mu}$ is the QED photon field. We first study the trace anomaly of such a system, which is related to its renormalization properties. It is easy to see that all the UV divergences of the system can be taken care of using the standard wave function renomalization  $Z_1-1=Z_2-1$ for electron and  $Z_3-1$ for photon, as well as the mass counter-term $m\delta_m$ order by order in perturbation theory. The counter-terms in terms of the renormalized fields $A_{R},\bar \psi_R$ are
\begin{align}
\delta{\cal L}=&(Z_1-1)\bar \psi_R (i\gamma \cdot \partial-m) \psi_R-e(Z_1-1)\bar\psi_R \gamma^{\mu}\psi( A_{\mu,R}+{\cal A}_{\mu})\nonumber \\
-& \delta_m m Z_1\bar\psi_R \psi_R-\frac{Z_3-1}{4}F^{\mu\nu}_RF_{\mu\nu,R} -\frac{Z_3-1}{2}F_R^{\mu\nu}{\cal F}_{\mu\nu} \ ,
\end{align}
where the last term is needed to cancel the UV divergence of vacuum polarization diagrams with one $A_{\mu}$ and one ${\cal A}_{\mu}$ insertion. Again,  $Z_2-1=Z_1-1$ follows from the Ward identity and the bare charge is related to the renormalized charge by $e_{R}A_{R}=e_{0}A_{0}$. Thus $e_0=Z_3^{-\frac{1}{2}}e_R$ and the beta function is purely determined by $Z_3$. By re-scaling with $e_0A_0\rightarrow A_0$ and $e{\cal A}\rightarrow {\cal A}$, in terms of the bare fields the Lagrangian reads:
\begin{align}
&{\cal L}=\psi_0 i\gamma \cdot(\partial+A_0+{\cal A})\psi_0-\frac{Z_3}{4e^2}F_0^2-\frac{Z_3-1}{2e^2}F_0^{\mu\nu}{\cal F}_{\mu\nu}-(1+\delta_m) m\bar \psi_0\psi_0 \ ,
\end{align}
where all the cutoff dependencies are absorbed into the renormalization constants $Z_3$ and $\delta_m$, which are precisely those which determine the QED beta function and the electron mass anomalous dimension. From now on until the end of this section, without mention all the field operators are renormalized and we will omit the lower-script $R$ for all the renormalized quantities. Thus, one has for the trace anomaly in the background field:
\begin{align}
T^{\mu}_{\mu}=\frac{\beta(e)}{2e}F^{\mu\nu}F_{\mu\nu}+\frac{\beta(e)}{e}F^{\mu\nu}{\cal F}_{\mu\nu}+m(1+\gamma_m)\bar\psi\psi \ ,
\end{align}
which implies that
\begin{align}
H_a=\frac{1}{4}\int d^3\vec{x} \bigg(\frac{\beta(e)}{2e}F^{\mu\nu}F_{\mu\nu}+\frac{\beta(e)}{e}F^{\mu\nu}{\cal F}_{\mu\nu}+m\gamma_m\bar\psi\psi\bigg)\ .
\end{align}
Let's investigate the contribution of $H_a$, $\langle N|H_a|N\rangle$ to the energy shift for a given bound state $|N\rangle$ in the background field  ${\cal A}$ when radiative corrections are included .

\begin{figure}[t]
\includegraphics[width=1\columnwidth]{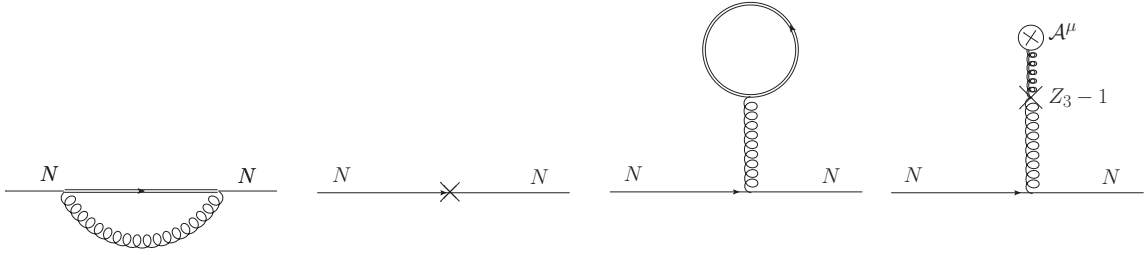}
\caption{The Feynman diagram for the one-loop electron self energy. The double lines are dressed electron propagators and the crosses are counter terms. They contribute to the energy shift $\delta E_N$. Notice that the forth diagram corresponds to the counter term $\frac{Z_3-1}{2}F^{\mu\nu}{\cal F}_{\mu\nu} $ that couples the background and radiative photon fields.  }
\label{fig:oneloopselfenergy}
\end{figure}

We first review the field theoretical calculation of the energy-shift in the presence of a background, following the notation and approach in Weinberg's book~\cite{Weinberg:1995mt}. One first quantize the theory in the background field ${\cal A}$ without QED corrections. The energy levels are labelled as $|N\rangle$, while the positive and negative energy solutions to the Dirac equation are denoted by $u_N(\vec{x})$ and $v_N(\vec{x})$. This theory is regarded as the ``free'' theory. We then add QED interactions and treat them as perturbations. The resulting perturbation theory in background field differs from the standard QED by that all the electron propagators are ``dressed''~\cite{Weinberg:1995mt} in the external field, but otherwise remains very similar.  One can show~\cite{Weinberg:1995mt} that the one-loop correction $\delta E_N$ to the energy level $N$ can be calculated 
efficiently using covariant perturbation theory as
\begin{align}
\delta E_N=\int \frac{d^3\vec{p}d^3\vec{p}\,'}{(2\pi)^6} \bar u_N(\vec{p}\,')\Sigma(E_N,\vec{p}\,'; E_N,\vec{p}\,)u_N(\vec{p}\,) \ ,
\end{align}
where $\Sigma(E_N,\vec{p}\,'; E_N, \vec{p}\,)$ is the one-loop electron self energy diagram  with dressed propagator in the background field. See Fig.~\ref{fig:oneloopselfenergy} for the one-loop self energy diagrams that contributes to $\delta E_N$. 

Similarly, when interactions are added, the state $|N\rangle$ in the background field will also receive radiative corrections.  We need to calculate the QAE contribution 
\begin{align}\label{eq:deltaENa}
    \delta E_N^a=\langle N|\int d^3\vec{x} \bigg(\frac{\beta(e)}{8e}F^{\mu\nu}F_{\mu\nu}+\frac{\beta(e)}{4e}F^{\mu\nu}{\cal F}_{\mu\nu}+\frac{1}{4}m\gamma_m\bar\psi\psi\bigg) |N\rangle \ ,
\end{align}
in the perturbed wave function $|N\rangle$.  It can be calculated in covariant pertubation theory using standard Feynman rules for operator insertions, paying attention to insertion of quark-bilinear operators at external legs.  More conveniently, using the fact that the trace anomaly can be obtained as mass-derivatives of counter-terms in the Hamiltonian as explained in Appendix \ref{Sec:FeynmanHellman}, $\delta E_N^a$ can also be obtained from the Feynman diagrams for $\delta E_N$ by taking mass derivatives in the counter-terms $(Z_3-1)$ and $\delta_m$. To lowest order in radiative correction, the $\frac{\beta(e)}{8e} F^{\mu\nu} F_{\mu\nu}$ term in Eq.~(\ref{eq:deltaENa}) does not contribute, therefore one only needs to consider the $\frac{\beta(e)}{4e}F^{\mu\nu}{\cal F}_{\mu\nu}$ and $\gamma m\bar \psi \psi$ terms.   

We first calculate  $\langle N|\int d^3\vec{x}\frac{\beta(e)}{4e}F^{\mu\nu}{\cal F}_{\mu\nu}|N\rangle$. To lowest order in radiative corrections, the Feynman diagram is identical to the last diagram in Fig.~\ref{fig:oneloopselfenergy} with $(Z_3-1)$ replaced by the mass derivative or the beta function
\begin{align}
   \frac{\beta(e)}{2e}=\frac{m}{4}\frac{d}{dm}(Z_3-1)= \frac{\alpha}{6\pi} \ ,
\end{align}
and can be directly calculated as 
\begin{align}
\langle N|\int d^3\vec{x}\frac{\beta(e)}{4e}F^{\mu\nu}{\cal F}_{\mu\nu}|N\rangle=\frac{\alpha}{6\pi} ie \int \frac{ d^3\vec{p} ~d^3\vec{p}'}{(2\pi)^6} \bar u_{N}(\vec{p}\,')\Gamma^{\mu}(E_N,\vec{p}\,';E_N,\vec{p}\,)u_N(\vec{p}\,){\cal A}_{\mu}(\vec{p}\,'-\vec{p}\,) \ ,
\end{align}
where
\begin{align}
&\Gamma^{\mu}(p',p)=\frac{i}{(p'-p)^2+i0}\bigg(-(p-p')^2g^{\mu\nu}+(p-p')^{\mu}(p-p')^{\nu}\bigg)\gamma_{\nu}  \ .
\end{align}
For ${\cal A}^{\mu}$ that only has temporal component,  the last term is proportional to $E_N-E_N$ and vanishes. Therefore, one can simplify the result to
\begin{align}\label{eq:lambFF}
\langle N| \int d^3\vec{x}\frac{\beta(e)}{4e}F^{\mu\nu}{\cal F}_{\mu\nu}|N\rangle
= \frac{\alpha^2}{6\pi}\int d^3\vec{x}\frac{u_{N}^{\dagger}(\vec{x})u_{N}(\vec{x})}{|\vec{x}|} \ ,
\end{align}
which is of order $m\alpha^3$.  Our results Eq.~(\ref{eq:lambFF}) for the photonic contribution to QAE differs has an additional $-2$ factor compared with Eq.~(7) in Ref.~\cite{Sun:2020ksc}. However, the total results in Ref.~\cite{Sun:2020ksc} for the energy shift is still correct possibly due to another discrepancy when evaluating the contribution of $m\bar\psi \psi$.

We then calculate the mass anomalous dimension term. The contribution is identical to the second diagram in Fig.~\ref{fig:oneloopselfenergy} with $\delta_m$ replaced by $m\gamma_m$. It can be evaluated simply as
\begin{align}
\langle N| \int d^3\vec{x} \frac{1}{4}m\gamma_m\bar \psi \psi| N\rangle=\frac{1}{4}\gamma_m m\int d^3\vec{x}\bar u_N(\vec{x})u_N(\vec{x})=\frac{1}{4}\gamma_m E_N \ .
\end{align}
Therefore, by summing the above contributions one gets for the total anomalous part:
\begin{align}\label{eq:tracelamb}
\delta E^a_{n,j}=\frac{\alpha^2}{6\pi}\int d^3\vec{x}\frac{u_{n,j}^{\dagger}(\vec{x})u_{n,j}(\vec{x})}{|\vec{x}|}+\frac{3\alpha}{8\pi}E_{n,j} \ ,
\end{align}
for the energy level $N\equiv n,j$, where $n$ is the radial quantum number and $j$ is the total spin. The first term is the photonic contribution while the second terms is the fermionic contribution. Here $u_{n,j}(\vec{x})$ is the quantum-mechanical wave function that solves the Dirac equation in a static Coulumb field, and $E_{n,j}$ is the bound state energy. In the non-relativistic limit, Eq.~(\ref{eq:tracelamb}) can be further expanded in $\alpha$ and contains contributions at ${\cal O}(\alpha)$, ${\cal O}(\alpha^3)$ and ${\cal O}(\alpha^5)$. The ${\cal O}(\alpha)$ and ${\cal O}(\alpha^3)$ contributions will be cancelled by other terms,
while the contribution at ${\cal O}(\alpha^5)$ reads
\begin{align}
\delta E^{a,(5)}_{n,j}=-\frac{7m_e\alpha^5}{24\pi n^4}\left(\frac{3}{8}-\frac{1}{2j+1}\right) \ .
\end{align}
This contributes to the famous Lamb shift at ${\cal O}(\alpha^5)$. 


We shall also mention that the electron mass $H_m$ gives a non-trivial contribution to the bound-state energy as well. In appendix \ref{Sec:FeynmanHellman}, we show that after adding the electron mass contribution,  the total scalar energy contribution is $\frac{1}{4}$ of the bond-state energy, consistent with the virial theorem.

\section{Anomalous energy contribution as Higgs mechanism }\label{sec:higgs}

In both QCD and the non-linear sigma-model, the QAE generates a non-perturbative contribution characterized by a new mass scale (dimensional transmutation~\cite{Coleman:1974hr}). It is then natural to consider the QAE itself responsible for the scale generation.  
In this section we take this view seriously and show that the anomalous scalar field can be considered as a dynamical Higgs field~\cite{Ji:2021pys}, and the QAE contribution to the
mass comes from its dynamical response to the matter, 
in analogy to the Higgs mechanism~\cite{Peskin:1995ev} for the fermion masses in the standard model. 

To see this analogy, let's first review the Higgs mechanism for fermion mass generation
in a simplified context without gauge symmetry. Introduce a complex scalar $\phi=\frac{1}{\sqrt{2}}(\sigma+i\pi)$ with action
\begin{align}
{\cal L}=\phi^{\dagger}(-\partial_{\mu}\partial^{\mu}+\mu^2)\phi-\frac{\lambda}{4!}(\phi^\dagger\phi)^2 \ .
\end{align}
where $\mu^2>0$. The saddle point of the potential $V(\phi)=-\mu^2 |\phi|^2+\frac{\lambda}{4!}|\phi|^4$ satisfies 
\begin{align}
2\mu^2|\phi_c|=\frac{\lambda}{3!}|\phi_c|^3 \ .
\end{align}
Lets expand the field around the saddle point $\phi=(|\phi_c|+\frac{1}{\sqrt{2}}h+i\frac{1}{\sqrt{2}}\pi)$.
To leading order, the field $h$ is massive with $M_h^2=2\mu^2$, while $\pi$ is massless. In the classical theory, the canonical EMT reads:
\begin{align}
T^{\mu\nu}=2\phi^{\dagger}\partial^{\mu}\partial^{\nu}\phi-g^{\mu\nu}{\cal L} \ ,
\end{align}
In quantum theory, Collins and others have shown that one needs to add the term $-\frac{1}{6}(\partial^{\mu}\partial^{\nu}-g^{\mu\nu}\partial^2)\phi^2$ to make all the matrix elements finite. 
To our desired order in $\lambda$, up to total derivative terms, the trace of the EMT reads
\begin{align}
T^{\mu}_{\mu}=-2\mu^2|\phi^2| \ .
\end{align}
which can be easily derived using the equation of motion. Therefore, the scalar part of the Hamiltonian reads:
\begin{align}
H_S=-\int d^3\vec{x}\left(\frac{1}{2}|\phi_c|h+\frac{1}{4}h^2\right) \ .
\end{align}

\begin{figure}[t]
\centering
\includegraphics[width=0.3\columnwidth]{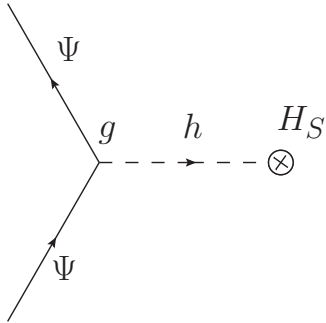}
\caption{Dynamical response of the scalar Hamiltonian $H_S$
in the presence of the fermion $\Psi$, generating a contribution to the fermion mass  The dotted line
represents the dynamical Higgs particles $h$ and the crossed circle denotes the scalar Hamiltonian linear in $h$. The coupling $g$ between the Higgs field and the fermion is proportional to fermion mass. }
\label{fig:higgs}
\end{figure}

In the presence of the a massless fermion $\Psi$ with Yukawa type coupling
\begin{align}
{\cal L}_{\rm int}=-\bar \Psi \Psi g\frac{\phi+\phi^{\dagger}}{\sqrt{2}} \ ,
\end{align}
which generated a fermion mass term with $m_\Psi = \sqrt{2}g\phi_c$. The above 
also yield a dynamical coupling between the fermion and the Higgs particle, $(-g)h\bar \Psi \Psi$, which is proportional to the fermion mass $m_\Psi$. This dynamical 
coupling generates a response of the Higgs field in the presence
of the fermion, which contributes to the fermion mass,
\begin{align}
\langle \Psi|H_S|\Psi\rangle = \frac{(-g)f_s}{m_h^2} = \frac{1}{4}g\phi_c = \frac{1}{4}m_\Psi \ ,
\end{align}
where $f_s=-\frac{1}{2}\mu^2\phi_c$ is a scalar decay constant. The $\frac{1}{m_h^2}$ is due to the zero-momentum propagator of the Higgs field. Therefore, the scalar part of the Hamiltonian contributes 1/4 of the fermion mass through the dynamical Higgs. See Fig.~\ref{fig:higgs} for a depiction of the mechanism. This simple example demonstrates that the mass of the fermions can also be measured by the response of the fluctuating part of the scalar field in the presence of the matter~\cite{Ji:2021pys}.

Similarly, for the non-linear sigma model, the QAE contribution to the meson mass
can be explained in term of a dynamical Higgs mechanism as follows. One first notices that using the equation of motion, the anomalous Hamiltonian can also be re-written in terms of the auxiliary scalar
\begin{align}
H_a=-\frac{iNm^2}{8\pi}\int dx_1\sigma \ ,
\end{align}
where the dimensionless scalar $\sigma= (\Sigma-\langle \Sigma \rangle)/m^2$ contains the quantum fluctuation part. This is similar to the Higgs example above, in that the scalar part of the Hamiltonian is linear in the sigma field.
Its contribution to the pion mass is determined by $\langle \pi^i|\sigma|\pi^i\rangle$.
By using the $\pi\pi\sigma$ vertices in Eq.~(10), and the dominance of the
zero-momentum $\sigma$ propagator $\langle \sigma(0)\sigma(0) \rangle=8\pi /(Nm^2)$
in the intermediate state, the response of the scalar $\sigma$ to
$\pi^i$ state exactly makes $H_a$ contributing $\frac{1}{2}$ of the $\pi^i$ mass. We shall mention that the propagator of $\sigma$~\cite{Shifman:2012zz} contains only a cut starting at the two-$\pi$ threshold $p^2=4m^2$ but no poles, unlike the Higgs field $h$ in the previous example. Nevertheless, the zero-momentum propagator of $\sigma$ contributes to the average of the anomalous Hamiltonian exactly the same way as the zero-momentum propagator of the Higgs field $h$.

\subsection{Dynamical scalar and QAE contribution to the nucleon mass and pressure}

The idea that the mass is generated from 
the response of the scalar field in the presence
of the external source can be generalized to QCD. 
For simplicity, we consider the limiting case of massless up and down quarks. The anomalous
Hamiltonian comes entirely from the gluon composite scalar, $H_a = \int d^3\vec{x} \Phi(x)$,
where $\Phi(x) = \beta(g)/(8g)F^{\mu\nu}F_{\mu\nu}(x)$. As in the non-linear sigma model,
its contribution to the nucleon mass can be seen as a form of dynamical Higgs-mechanism,
which is consistent with that the Higgs and confining phases of matter-coupled gauge theory are smoothly connected~\cite{Fradkin:1978dv,Banks:1979fi}.

It is interesting to recall that for the infinite-heavy $\bar Q Q$ state separated by $r$ in pure gauge theory, it has been shown~\cite{Rothe:1995hu,Rothe:1995av} that the non-perturbative
contribution of $H_a$ to the static potential is $\frac{1}{4}(V(r)+rV'(r))$. At large $r$ where the confinement potential
dominate $V(r)\sim \sigma r$, the anomalous contribution is exactly one half of the confinement potential.

The scalar field $\Phi(x)$ has a vacuum condensate $\Phi_0=\langle 0|\Phi|0\rangle$~\cite{Shifman:1978bx,Shifman:1978by}.
However, in the presence of the nucleon, the quantum response is measured by
\begin{equation}
      \phi (x) = \Phi(x) - \Phi_0 \ ,
\end{equation}
which is a dynamical version of the MIT bag-model constant $B$~\cite{Chodos:1974je}.
Its contribution to the nucleon mass can be seen as the response
of the scalar field to the nucleon source,
\begin{equation}
    E_a = \langle \phi\rangle_N = \langle N|\phi(x)|N\rangle \ ,
\end{equation}
where the nucleon state is normalized as $\langle N|N\rangle=(2\pi)^3\delta^3(0)$.
If $\phi(x)$ is a static constant $B$ inside the nucleon, $E_a$ will be of order $
BV$, where $V$ is the effective volume in which the valence quarks
are present. In the MIT bag model, the nucleon mass is entirely determined by the bag constant, in line
with the view that the QAE determines the mass scale.

We should point out that it has been proposed  that static response of the composite gluon scalar $\phi$ in the nucleon state
can be measured in the electro-production of heavy quarkonium on the proton~\cite{Kharzeev:1995ij,Kharzeev:1998bz,Brodsky:2000zc,Hatta:2019lxo,Wang:2019mza,Du:2020bqj,Zeng:2020coc} or leptoproduction of heavy quarkonium at large photon virtuality~\cite{Boussarie:2020vmu}. The color dipole from the quarkonium will be an effective probe of the $F^2$. This also provides a direct determination of the QAE contribution to the mass. Nevertheless, it has also been found recently that the near threshold production of heavy meson is dominated by the twist-two tensor contribution instead of the $0^{++}$ scalar contribution based on holographic QCD~\cite{Mamo:2019mka,Mamo:2021krl} or perturbative analysis~\cite{Guo:2021ibg,Sun:2021gmi}. To further clarify the role of trace anomaly in the heavy-meson production requires more elaborate QCD analysis and is left for future work. 

\begin{figure}[t]
\centering
\includegraphics[width=0.3\columnwidth]{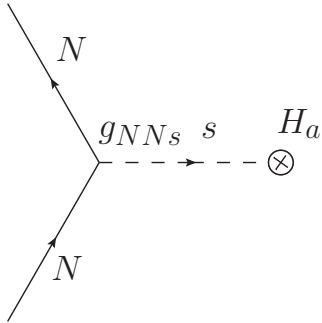}
\caption{Quantum anomalous energy contribution to the nucleon mass seen as
a dynamical response of the anomalous scalar field in the presence of the nucleon. The dotted line
represents the intermediate scalar particles with couplings $g_{NNs}$ proportional to the nucleon mass, which is
dominated by a single Higgs particle in the Higgs mechanism.}
\label{fig:dispersion}
\end{figure}

Similar to fermion masses in elementary particle physics, we can also consider a dynamical
response of the $\phi$ in the presence of the nucleon through a tower of scalar $0^{++}$
spectral states, as in the Higgs model.  Assume an effective
coupling between the nucleon and scalar $g_{NN\phi}\bar NN \phi$,
the QAE contribution to the mass can be related to the scalar field response function,
\begin{equation}\label{eq:disper1}
    \langle N|\phi| N\rangle =ig_{NN\phi}\langle \phi(0) \phi(0) \rangle
\end{equation}
where $\langle \phi(0) \phi(0) \rangle$ is the zero-momentum propagator of the scalar field $\phi$.
If the propagator is dominated by a series of scalar resonances,
or $\langle \phi(0) \phi(0) \rangle=\sum_s \frac{if_s^2}{-m_s^2}$, one has~\cite{Ji:2021pys}
\begin{align}\label{eq:disper2}
 \langle N|\phi| N\rangle =\sum_s \frac{g_{NNs}f_s}{m_s^2} \ .
\end{align}
Here $m_s$ is the mass of the scalar resonances, $f_s=\langle s|\phi|0\rangle$ is the decay
constant and $g_{NNs}\equiv g_{NN\phi}f_s$ is the coupling of the nucleon to the scalars.
See Fig.~\ref{fig:dispersion} for a depiction.

One might assume the dominance of the lowest mass scalar glueball-like state, generically called $\sigma$, for the above equation. If the coupling constant $g_{NNs}$ can be extracted through experiment, one can perform a consistency check on the $\sigma$ dominance picture by combining the glueball masses and the decay constants extracted from lattice QCD calculations~\cite{Morningstar:1999rf,Chen:2005mg}. In fact, for the lowest glueball state $\sigma$, one can say more. In~\cite{Ellis:1984jv}, an effective action for $\sigma$ that is consistent at tree level with the Ward-identity for $T^{\mu}_{\mu}$ has been constructed. Introduce a dimension-1 scalar field $\Sigma$, which replaces the scalar field $\Phi$ through the relation
\begin{align}
\Phi\equiv -\frac{m_\sigma^4}{64|\Phi_0|}\Sigma^4 \ ,
\end{align}
where $m_\sigma$ is a mass parameter the meaning of which will be explained later. The effective action for $\Sigma$ is
\begin{align}
{\cal L}=\frac{1}{2}\partial_{\mu}\Sigma \partial^{\mu}\Sigma-V(\Sigma) \ ,
\end{align}
where the effective potential $V(\Sigma)$ is
\begin{align}
V(\Sigma)=\frac{m_\sigma^4}{256|\Phi_0|}\Sigma^4 \ln \frac{\Sigma}{C}  \ .
\end{align}
The potential has a minimum at $\Sigma=\sigma_0$ constrained by the relation $4\ln \frac{\sigma_0}{C}=-1$.  In terms of $\sigma_0$ and $m_\sigma$, the vacuum condensate can be expressed as
\begin{align}\label{eq:phi0}
|\Phi_0|^2=\frac{m_{\sigma}^4}{256}\sigma_0^4 \ .
\end{align}
which determines $C$ as a function of $m_\sigma$ and $|\Phi_0|$, the two independent parameters of the theory. The dilatation symmetry that was broken at quantum level in the original theory is broken in the effective theory at classical level by the dynamically generated potential $V(\Sigma)$.  This can be viewed as the realization of dimensional transmutation in an effective Higgs phase.  

By expanding $\Sigma=\sigma_0+\sigma$, one can show that the $\sigma$ is a massive scalar with mass equals to $m_\sigma$. Furthermore, one expand $\Phi$ to leading order in $\sigma$
\begin{align}
\phi=\Phi-\Phi_0=\frac{m_\sigma^4\sigma_0^3}{64|\Phi_0|}\sigma+{\cal O}(\sigma^2) \ ,
\end{align}
from which the decay constant $f_{\sigma}$ can be extracted as~\cite{Ellis:1984jv}
\begin{align}\label{eq:fsigma}
f_{\sigma}=m_\sigma\sqrt{|\Phi_0|} \ .
\end{align}
Assume that the coupling between the nucleon $N$ and the scalar $\Sigma$ is given by the Yukawa coupling
\begin{align}
{\cal L}_{\rm Yukawa}=-g_{NN\sigma}\bar N N \Sigma \ .
\end{align}
The nucleon mass can be measured exactly through two ways. One way is through the mass term generated by the vacuum expectation value $\sigma_0$ or $m_N=g_{NN\sigma }\sigma_0$, from which one can extract~\cite{Ji:2021pys}
\begin{align}\label{eq:gnns}
g_{NN\sigma}=\frac{m_Nm_\sigma}{4\sqrt{|\Phi_0 |}}
\end{align}
in the chiral limit. 

The second way is through the response of $\Sigma$ in the presence of the nucleon. By assuming the $\sigma$ dominance in the intermediate state, one has
\begin{align}
\langle N|\phi|N\rangle =\frac{f_\sigma g_{NN\sigma}}{m_\sigma^2} = \frac{m_N}{4} \ ,
\end{align}
which we have used Eq.~(\ref{eq:gnns}) 
and the formula for $f_{\sigma}$ in Eq.~(\ref{eq:fsigma}). Thus we confirm 
that the scalar response is proportional to the nucleon mass, exactly as 1/4 predicted by
virial theorem. This also exactly corresponds to the Higgs model mentioned earlier.  

Of course, the above simple picture is
modified strongly by the presence of light quarks. The scalar spectrum in QCD 
is more complicated, and is in between the simple Higgs and the 1+1 sigma models. However, the coupling
between the nucleon (or any other hadrons) and with the scalars must be proportional to the
to its mass, same as in the Higgs case which has been tested recently at LHC~\cite{Weinberg:1967tq,Sirunyan:2018koj,Aad:2019mbh}.

\subsection{Anomalous energy contribution to pion mass}
Finally, we show that the anomalous contribution also plays an important role to the pion mass. For simplicity we only consider the two-flavor case with equal quark masses $m$ for up and down quarks $u$ and $d$. The theory has two mass scales, the quark mass $m$ and ${\Lambda}_{QCD}$. The scalar part of the Hamiltonian reads
\begin{align}
H_{S}=\int d^3 \vec{x}\frac{1}{4}m (\bar u u +\bar dd)+H_a \ ,
\end{align}
where $H_a$ is the anomalous part of the Hamiltonian. In the $m \ll \Lambda_{QCD}$ limit, it is well known~\cite{Peskin:1995ev} that the mass $M_\pi$ for the Golstone boson or the pion relates to the quark mass and the chiral-condensate through the Gell-Mann-Oakes-Renner (GMOR)~\cite{GellMann:1968rz} relation:
\begin{align}\label{eq:pionm}
M_{\pi}^2=-\frac{1}{f_{\pi}^2} \langle 0|m(\bar u u +\bar dd)|0 \rangle \ .
\end{align}
Therefore, the contribution of the mass-term to the pion mass can be calculated as
\begin{align}
\langle \pi |\int d^3 \vec{x}\frac{1}{4}m (\bar u u +\bar dd)|\pi \rangle =\frac{1}{4}m\frac{\partial M_\pi}{\partial m}=\frac{1}{8}M_\pi \ ,
\end{align}
where the pion state is normalized as $\langle \pi|\pi \rangle=(2\pi)^3\delta^3(0)$ and we have applied the Feynman-Hellman theorem~\cite{Feynman:1939zza} in the first equality and Eq.~(\ref{eq:pionm}) in the last equality. One can also derive this relation by inserting pion intermediate state in the correlation function $\langle 0|\partial_{\mu}j^{\mu5}(x) m(\bar u u+ \bar d d)(y) \partial_{\nu}j^{\nu 5}(z)|0\rangle$ and using the Ward-identity for the axial current. This results, in the context of QCD mass decomposition, is noticed originally in Ref.~\cite{Ji:1995sv}. Since the scalar part of the Hamiltonian contributes to $\frac{1}{4}$ of the pion mass in total due to the virial-theorem, we found that the anomalous contribution to the pion mass equals to
\begin{align}\label{eq:piona}
\langle \pi |H_a|\pi \rangle=\frac{1}{8}M_\pi \ .
\end{align}
Although the pion being the lowest-energy state, the anomalous part contributes to a significant amount of its mass, equally to the quark mass term. The quark mass term fails to dominate in the scalar part. This is however understandable since even in the chiral limit the theory still has dimensional transmutation which gives rise to the $\Lambda_{\rm QCD}$. The pion mass in Eq.~(\ref{eq:pionm}) involves a mixing between the quark mass $m$ and the chiral-condensate measured by $\Lambda_{\rm QCD}$. Therefore, it is natural that both the quark mass term and the anomalous term contributes equally to the pion mass.

Assuming the sigma dominance, an effective action for the coupling between the pion and the lightest glueball field has been constructed in Ref.~\cite{Lanik:1984fc} and reads
\begin{align}
{\cal L}=\frac{f_\pi^2}{4}\left(\frac{\Sigma}{\sigma_0}\right)^2 {\rm Tr} \partial^{\mu} U^{\dagger} \partial_{\mu} U-\left(\frac{\Sigma}{\sigma_0}\right)^3m \langle 0| \bar u u|0 \rangle {\rm Tr}(U^{\dagger}+U)-\frac{1}{3}m \langle 0| \bar u u|0 \rangle \left(\frac{\Sigma}{\sigma_0}\right)^4\ ,
\end{align}
where $U=\exp\left(\frac{i}{f_\pi}\sum_{a=1}^3\pi^a \tau^a\right)$ is the standard SU(2) matrix for the pion and $\Sigma$, $\sigma_0$ are given in previous subsection. The last term is required to maintain $\sigma_0$ as the minimal of the potential for $\Sigma$. By expanding the Lagrangian above, the coupling between the pion and the $\sigma$ is proportional to the mass square $M_\pi^2$ of the pion.

More specifically, the $\pi \pi \sigma$ coupling term ${\cal L}_{\pi\pi \sigma}$ reads
\begin{align}
{\cal L}_{\pi \pi \sigma}=\partial_{\mu} \pi^a \partial^{\mu} \pi^a \frac{\sigma}{\sigma_0} -\frac{3}{2}M_\pi^2 \pi^a \pi^a \frac{\sigma}{\sigma_0}=-\frac{1}{2}M_\pi^2 \pi^a \pi^a \frac{\sigma}{\sigma_0}\ ,
\end{align}
where we have used the equation of motion in the first term. Therefore, in the massive pion state the expectation value of $\phi$ can be calculated as
\begin{align}
    \langle \pi|\phi| \pi\rangle=\frac{1}{2M_\pi}\times \frac{M_\pi^2}{\sigma_0}\frac{f_\sigma}{m_\sigma^2}=\frac{M_\pi}{8} \ ,
\end{align}
where we have used the relation $\frac{f_\sigma}{m_\sigma^2\sigma_0}=\frac{1}{4}$ which follows from Eq.~(\ref{eq:phi0}) and Eq.~(\ref{eq:fsigma}). The factor $\frac{1}{2M_\pi}$ is due to overall normalization of the state. This is consistent with Eq.~(\ref{eq:piona}) that the anomalous contribution is responsible for $\frac{1}{8}$ of the pion mass. Furthermore, it indicates
that the scalar field response inside the pion vanishes
in the chiral limit, consistent with the expectation that the Goldstone boson arises from the chiral rotation of the QCD vacua.

\section{Conclusion}

In this paper we expand our previous study on the implications of anomalous scale symmetry breaking effect on the nucleon mass structure in QCD and other relativistic quantum field theories~\cite{Ji:2021pys}. The scale symmetry breaking generates a non-perturbative anomalous
contribution to the QCD energy called QAE and therefore to all hadron masses. The QAE also sets the scale for the contributions of more familiar quark and gluon kinetic energies. 

We start by explaining the role of UV divergences in generating the mass scale of QCD through  the so-called dimensional transmutation. 
We demonstrate through a path integral formulation of a two-point function that the trace anomaly naturally arises as consequences of the UV cut-off dependence in QCD-like theories. Furthermore, the QAE contribution to the hadron mass can be derived as resulting from UV cut-off dependence of couplings and quark masses by investigating the time-rescaling property of the theory. Contrary to some mis-understandings in the literature, the naive expression for the Hamiltonian is scheme dependent, but the QAE contrition is scheme independent and is a key physical part of the nucleon mass. We emphasize the importance of Lorentz invariance when renormalizing tensor operators in DR and show that the maximally-scheme-independent decomposition of the hadron mass is facilitated by separating the trace and traceless contributions. 

We then study the scale symmetry breaking effect and the mass structure in the large $N$ non-linear sigma model in 1+1 dimensions. We  demonstrate explicitly in different UV regularization schemes that the QAE is indeed scheme independent and is a crucial part of the mass of the $\pi^a$ particles. On the other hand, the naive Hamiltonian is regulator sensitive and lacks universal physical meaning. Furthermore, the fundamental mass scale in the model is generated through a scalar field that develops a non-vanishing vacuum condensate,  resembling the Higgs mechanism of generating quark and electron masses. 
We also show that in QED, although there is no scale generation, the effect of QAE is perturbative and non-zero, and contributes to the electron pole mass and the famous Lamb-shift.

Finally, inspired by the non-linear sigma model, we explore the similarity between the QAE contribution to proton mass generation and the Higgs effect. Similar to the sigma model and standard Higgs mechanism, the nucleon mass can be measured through either static or dynamical response of the Higgs field in the presence of the nucleon. By interpreting the Higgs particles as scalar glueballs, one can determine that the nucleon-glueballs coupling is proportional to nucleon mass, similar to the case of standard Higgs mechanism. We show how these ideas works in an effective theory in which the dimensional transmutation is realized in an effective Higgs phase and the anomalous scalar field acquires a dynamical generated potential. In the chiral limit the QAE contributes to $\frac{1}{8}$ of the pion mass and the pion-glueball coupling is proportional to pion mass square. 

However, the connection between the anomalous scalar field $F^2$ and the laws of fundamental QCD is not clear to us at present time. In particular, $F^2$ may play crucial role in color confinement as well as spontaneous chiral symmetry breaking. In the MIT bag model, the response of the $F^2$ is related to the bag constant $B$ which plays a role of negative pressure to confine quarks. A more microscopic model for the $F^2$-assisted confinement is provided by 't Hooft~\cite{tHooft:2002pmx} in which the Lagrangian of $F^2$ is modified by scalar coupling and generates the flux tubes between colored sources. 
In the instanton liquid model~\cite{Zahed:2021fxk}, the $F^2$ contribution is related to the average density of instantons that sets up the fundamental mass scale of the theory. However,  a full picture for the role of $F^2$ in the mass generation of the proton and confinement awaits further studies and
understanding of lattice QCD simulations~\cite{Biddle:2019gke}. 

\acknowledgments
We thank K.-F. Liu, Z.-E. Meziani, F. Yuan, and I. Zahed for discussions related to the proton mass. This material is supported by the U.S. Department of Energy, Office of Science, Office of Nuclear Physics, under contract number DE-SC0020682. A.S. acknowledges support from the German Research Foundation (DFG) through the Trans Regional Collaborative Research Center number 55 (SFB/TRR-55).

\appendix
\section{Derivation of Eq.~(\ref{eq:lambdaderi})}\label{sec:deriproof}

In this appendix we provide a derivation of Eq.~(\ref{eq:lambdaderi}). The general idea, similar to that in Ref.~\cite{Karsch:1982ve}, is to explore the lattice symmetry . Lets consider two two-point functions in the ensemble given by the action $S_{\lambda}$ in Eq.~(\ref{eq:reslatt}). One of them extends in temporal direction as before, while the other extends in the spacial direction $e_i$ where $i=1,2,3$
\begin{align}
G_{\tau}(T,0)=  \frac{\int DU O(Te_4,\vec{p}=0)O(0,\vec{p}=0) e^{-\frac{1}{g_0^2(\lambda)}S_{\lambda}[U]}}{\int DU e^{-\frac{1}{g_0^2(\lambda)}S_{\lambda}[U]}}   \ , \\ 
G_{i}(T,0)=     \frac{\int DU O(Te_i,\vec{p}=0)O(0,\vec{p}=0) e^{-\frac{1}{g_0^2(\lambda)}S_{\lambda}[U]}}{\int DU e^{-\frac{1}{g_0^2(\lambda)}S_{\lambda}[U]}} \ . 
\end{align}
Here, $e_4$ and $e_i$ are unit vectors along the (imaginary) time and the $i$-th spacial directions. The scalar operator $O$ is assumed to be local and scale invariant.  The notation $(Te_i,\vec{x})$ is meant to indicate that the $i$-th coordinate is $T$, and the remaining three directions are labeled by $\vec{x}$ with Fourier conjugating variable $\vec{p}$.  

One notice that, in the continuum limit, $G_{\tau}(T,0)$ becomes the time rescaled version of the two-point function and behaves as $e^{-MT/\lambda}$ at large $T$, while $G_i(T,0)$ is not rescaled and behaves as $e^{-MT}$.  By comparing the $\lambda$ derivatives as before, one obtains the relations:
\begin{eqnarray}
-\lambda^2\Big \langle  \frac{1}{g_0^2}\sum_{\vec{x}} P_\tau(\vec{x})\Big\rangle_{\tau}+\lambda \Big \langle \frac{1}{g_0^2}\sum_{\vec{x}} P_s(\vec{x})\Big \rangle_{\tau}+\frac{2\lambda^2}{g_0^3}\frac{dg_0}{d\lambda}\Big\langle \sum_{\vec{x}} {\cal S}(\vec{x}) \Big\rangle_{\tau}&=&M \ , \label{eq:time}\\
-\lambda^2\Big \langle  \frac{1}{g_0^2}\sum_{\vec{x}} P_\tau(\vec{x})\Big\rangle_{i}+\lambda \Big \langle \frac{1}{g_0^2}\sum_{\vec{x}} P_s(\vec{x})\Big \rangle_{i}+\frac{2\lambda^2}{g_0^3}\frac{dg_0}{d\lambda}\Big\langle \sum_{\vec{x}} {\cal S}(\vec{x}) \Big\rangle_{i}&=&0 \label{eq:spatial}\ .
\end{eqnarray}
Here the averages $\langle \rangle _i$ and $\langle \rangle_{\tau}$ are defiend as
\begin{eqnarray}
\langle A(\vec{x})\rangle_{i}&=&\lim_{T\rightarrow \infty}\frac{\langle O(Te^i,\vec{p}=0)A (0_4,\vec{x})O(0,\vec{p}=0)\rangle_c}{\langle O(Te^i,\vec{p}=0)O(0,\vec{p}=0)\rangle} \ ,\\
\langle A(\vec{x})\rangle_{\tau}&=&\lim_{T\rightarrow \infty}\frac{\langle O(Te^4,\vec{p}=0)A(0_i,\vec{x}) O(0,\vec{p}=0)\rangle_c}{\langle O(Te^4,\vec{p}=0)O(0,\vec{p}=0)\rangle} \ ,
\end{eqnarray}
 where $0_i\equiv 0e_i$ and $0_4\equiv 0e_4$.  At $\lambda=1$ the symmetry of the hypercubic lattice implies the following relations between the averages 
 \begin{eqnarray}
  \Big \langle  \sum_{\vec{x}} P_{4 i}(\vec{x})\Big\rangle_{\tau}&=& \Big \langle  \sum_{\vec{x}} P_{kj}(\vec{x})\Big\rangle_{k}|_{k \ne j} \ , \label{eq:Prelations1}\\
     \Big \langle \sum_{\vec{x}} P_{ ij}(\vec{x})\Big\rangle_{\tau}&=& \Big \langle  \sum_{\vec{x}} P_{lm}(\vec{x})\Big\rangle_{k}|_{k \ne l,m} \ , \label{eq:Prelations2} \\ 
    \Big \langle \sum_{\vec{x}} {\cal S}(\vec{x}) \rangle_{\tau}& =&\langle\sum_{\vec{x}} {\cal S}(\vec{x}) \Big \rangle_{i} \ ,
 \end{eqnarray} 
where $P_{4i}$ and $P_{ij}$ denotes plaquettes in $4i$ and $ij$ planes. One now adds Eq.~(\ref{eq:time}) to the sum of Eqs.~(\ref{eq:spatial}) over $i=1,2,3$ and take $\lambda=1$. All the averages over $P_{s}$ and $P_{\tau}$ in the first two terms of Eq.~(\ref{eq:time}) and Eq.~(\ref{eq:spatial}) cancel out thanks to Eqs.~(\ref{eq:Prelations1}) and~(\ref{eq:Prelations2}), left only with 
\begin{align}
M=\left.4\times \frac{2}{g_0^3}\frac{dg_0}{d\lambda}\right|_{\lambda=1}~\Big\langle\int d^3\vec{x} {\cal S}(0,\vec{x}) \Big\rangle_{\tau} \ .
\end{align}
Comparing with equation Eq.~(\ref{eq:tracecon}) one finds
\begin{align}
\left.\frac{dg_0}{d\lambda}\right|_{\lambda=1}=\frac{\beta(g_0)}{4} \ ,
\end{align}
which finishes the derivation of Eq.~(\ref{eq:lambdaderi}).

\section{QAE from mass derivative of Hamiltonian}\label{Sec:FeynmanHellman}

In this appendix we show that the trace anomaly in QED can be obtained from mass derivative of the Hamiltonian in on-shell renormalization scheme where UV cut-off can be traded with the electron mass as the infrared cut-off. This idea can be generalized to other renormalization scheme which we will not consider here.

More speficially, the mass-derivative of the QED Hamiltonian gives the trace anomaly
\begin{align}\label{eq:massderi}
 m\frac{\partial H}{\partial m}=\int d^3 \vec{x}T^{\mu}_{\mu}(\vec{x}) \equiv 4H_S\ ,
\end{align}
where $m$ is on-shell electron mass. In terms of bare fields and the conjugate momenta, the Hamiltonian in Coulomb gauge reads
\begin{align}\label{eq:halqed}
H=\int d^3\vec{x} \left (\frac{e^2_0}{2}(\vec{\Pi})^2+\frac{1}{2e_0^2}\vec{B}^2\right)+\int d^3\vec{x} \bar \psi (-i\vec{\gamma} \cdot \vec{D}+mZ_m)\psi +e_0^2\int d^3 \vec{x}d^3\vec{y}\frac{\psi^{\dagger}\psi(\vec{x}) \psi^{\dagger}\psi(\vec{y})}{4\pi |\vec{x}-\vec{y}|} \ ,
\end{align}
where $\vec{A}$ and $\vec{\Pi}$ are gauge vector potential and the 
conjugating fields, satisfying the transverse commutation relation $[\Pi_i(\vec{k}),A_j(\vec{x})]=-i(\delta_{ij}-\frac{k_ik_i}{k^2})e^{-i\vec{k}\cdot \vec{x}}$. Notice that the non-standard appearances of $e_0$ after re-scaling the gauge potential by this factor.
The electron pole mass $m$ relates to the bare mass through $m_0=mZ_m$. Treating $e_0$ and $Z_m$ as dimensionless parameters, the naive mass derivative of the Hamiltonian looks like $ \int d^3\vec{x}mZ_m \bar \psi  \psi $. However, to make the theory UV finite, the bare coupling constants $e_0$ and the mass renormalization constant $Z_m$ must be function of the $m$ and the UV cutoff $\Lambda$:
\begin{align}
e_0^2=e^2(1+\frac{e^2}{12\pi^2}\ln\frac{\Lambda^2}{m^2}+..) \ , \\
Z_m=1-\frac{3e^2}{16\pi^2}\ln\frac{\Lambda^2}{m^2}+.. \ .
\end{align}
Therefore, the mass derivative of the Hamiltonian also depends on the beta function $\beta(e_0)=-m\frac{\partial e_0}{\partial m}$ and the mass anomalous dimension $\gamma_m=m\frac{\partial \ln Z_m}{\partial m}$. Using the relation $F^2=2(\vec{B}^2-\vec{E}^2)$ and $\vec{E}=g_0^2 \vec{\Pi}$, one has the mass derivative of the Hamiltonian
\begin{align}
m\frac{\partial H}{\partial m}=\int d^3 \vec{x} \bigg(mZ_m(1+\gamma_m)\bar \psi \psi+\frac{\beta(e_0)}{2e_0^3}F^2\bigg) \ ,
\end{align}
where we have combined the $\vec{E}^2$ from the radiation field and the Coulomb field into the total $F^2$, which is just Eq. (\ref{eq:massderi}).

Beside the operator proof, here we also provides a diagrammatic argument of the above derivation, using the QED in background field in Sec.~\ref{sec:qed} as an example. We show that: taking mass derivatives in one-loop Feynman diagrams Fig.~\ref{fig:oneloopselfenergy} for $\delta E_N$ will exactly produce the one-loop Feynman diagrams for insertion of $4H_S$. The mass derivative has four origins: the explicit mass dependency of the electron propagator, the implicit mass dependency in the energy level $E_N$,  the mass dependencies in renormalization constants $\delta_m $ and $Z_3-1$, and the implicit mass dependency in the wave function $u_N$.  

\begin{figure}[h]
\centering
\includegraphics[width=0.3\columnwidth]{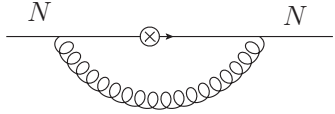}
\caption{The $m\bar\psi\psi$ insertion at internal line of the first diagram in Fig.~\ref{fig:oneloopselfenergy}. It corresponds to the mass derivative of the internal electron propagator. Notice that all the electron lines are dressed.}
\label{fig:inside}
\end{figure}

\begin{figure}[h]
\includegraphics[width=0.3\columnwidth]{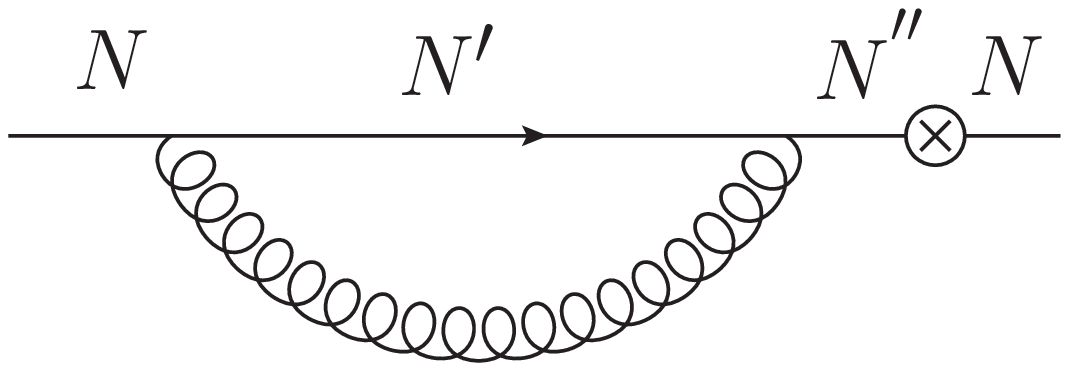}
\includegraphics[width=0.3\columnwidth]{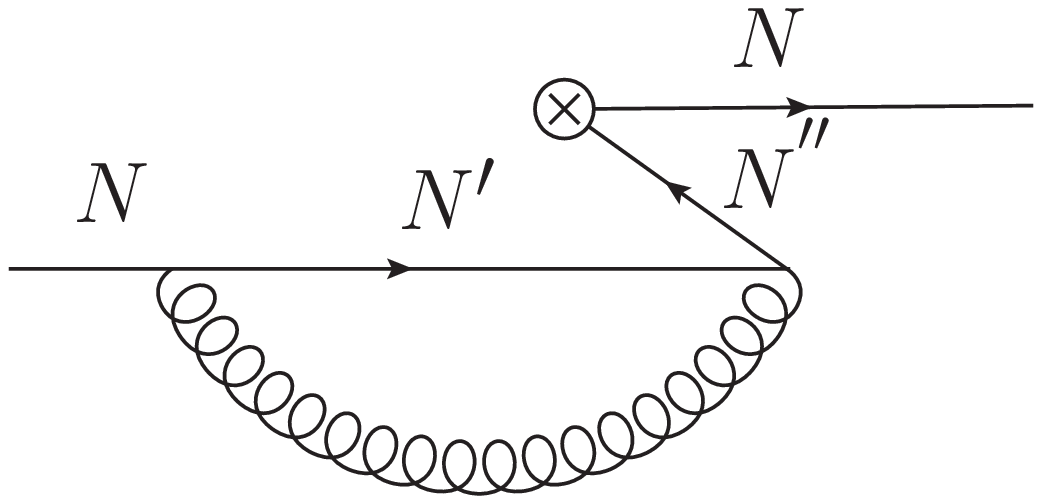}
\includegraphics[width=0.3\columnwidth]{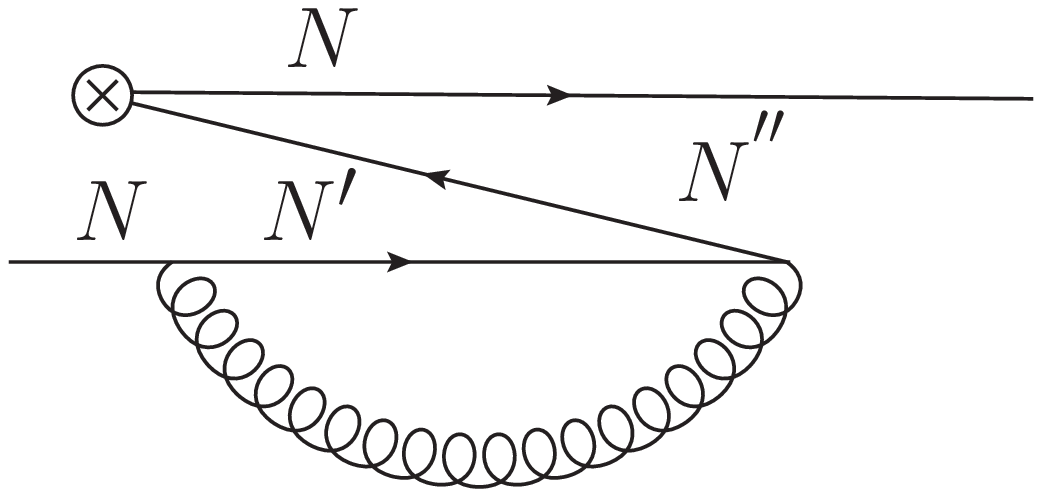}
\caption{The $m\bar\psi\psi$ insertion at external line of the first diagram in Fig.~\ref{fig:oneloopselfenergy}.  It corresponds to the mass derivative of the external Dirac wave function. For simplicity we only draw the case in which the internal electron is forward moving and the derivative is taken at the out-going wave function. The other cases are similar. The first diagram corresponds to the first term at right hand side of equality in Eq.~(\ref{eq:derie}), while the other two diagrams with back-moving lines combine to produce the second term in Eq.~(\ref{eq:derie}). }
\label{fig:backward}
\end{figure}

The mass derivative of the fermion propagator $\frac{1}{i\gamma \cdot D-m}$ simply reduces to $m\bar \psi \psi$ operator insertion in the internal electron line as shown in Fig.~\ref{fig:inside}. The mass dependency in $E_N$ will leads to the wave function renormalization in external legs. The mass dependencies in renormalization constants $\delta_m$ and $Z_3-1$ will exactly leads to the anomalous energy contribution. Finally, the mass derivative of the external wave function $u_N$ is more complicated, which is shown the remaining diagrams where the $m\bar\psi \psi$ are inserted at external legs.  One can see this by using the relation
\begin{align}\label{eq:derie}
m\frac{\partial}{\partial m} u_N(\vec{x})=\sum_{N'\ne N}\frac{u_{N'}(\vec{x})\int d^3\vec{y} m\bar u_{N'}(\vec{y})u_N(\vec{y})}{E_N-E_N'}+\sum_{N'\ne N}\frac{v_{N'}(\vec{x})\int d^3\vec{y} m\bar v_{N'}(\vec{y})u_N(\vec{y})}{E_N+E_N'} \ .
\end{align}
Let's apply this results to one of the two time-ordered self-energy diagrams, for example the forward moving diagram. Then the first term will directly be identified as a forward moving diagram for external $m\bar \psi \psi$ insertion, while the second term corresponds to the combination of two backward moving diagrams using the relation in energy denominators
\begin{align}
\left(\frac{1}{E_N-E_{N'}}+\frac{1}{E_N-(2E_{N}+E_{N''})}\right)\frac{1}{E_{N'}+E_{N''}}=\frac{1}{E_N-E_{N'}}\frac{1}{E_N+E_{N''}}  \ .
\end{align}
They are exactly the diagrams where the $m\bar \psi \psi$ are inserted in the external legs. See Fig.~\ref{fig:backward} for a depiction of the mass derivatives for the first diagram in Fig.~\ref{fig:oneloopselfenergy}. For other diagrams the analysis are similar.

\bibliographystyle{apsrev4-1}
\bibliography{bibliography}

\begin{thebibliography}{70}%
\makeatletter
\providecommand \@ifxundefined [1]{%
 \@ifx{#1\undefined}
}%
\providecommand \@ifnum [1]{%
 \ifnum #1\expandafter \@firstoftwo
 \else \expandafter \@secondoftwo
 \fi
}%
\providecommand \@ifx [1]{%
 \ifx #1\expandafter \@firstoftwo
 \else \expandafter \@secondoftwo
 \fi
}%
\providecommand \natexlab [1]{#1}%
\providecommand \enquote  [1]{``#1''}%
\providecommand \bibnamefont  [1]{#1}%
\providecommand \bibfnamefont [1]{#1}%
\providecommand \citenamefont [1]{#1}%
\providecommand \href@noop [0]{\@secondoftwo}%
\providecommand \href [0]{\begingroup \@sanitize@url \@href}%
\providecommand \@href[1]{\@@startlink{#1}\@@href}%
\providecommand \@@href[1]{\endgroup#1\@@endlink}%
\providecommand \@sanitize@url [0]{\catcode `\\12\catcode `\$12\catcode
  `\&12\catcode `\#12\catcode `\^12\catcode `\_12\catcode `\%12\relax}%
\providecommand \@@startlink[1]{}%
\providecommand \@@endlink[0]{}%
\providecommand \url  [0]{\begingroup\@sanitize@url \@url }%
\providecommand \@url [1]{\endgroup\@href {#1}{\urlprefix }}%
\providecommand \urlprefix  [0]{URL }%
\providecommand \Eprint [0]{\href }%
\providecommand \doibase [0]{http://dx.doi.org/}%
\providecommand \selectlanguage [0]{\@gobble}%
\providecommand \bibinfo  [0]{\@secondoftwo}%
\providecommand \bibfield  [0]{\@secondoftwo}%
\providecommand \translation [1]{[#1]}%
\providecommand \BibitemOpen [0]{}%
\providecommand \bibitemStop [0]{}%
\providecommand \bibitemNoStop [0]{.\EOS\space}%
\providecommand \EOS [0]{\spacefactor3000\relax}%
\providecommand \BibitemShut  [1]{\csname bibitem#1\endcsname}%
\let\auto@bib@innerbib\@empty
\bibitem [{\citenamefont {Peskin}\ and\ \citenamefont
  {Schroeder}(1995)}]{Peskin:1995ev}%
  \BibitemOpen
  \bibfield  {author} {\bibinfo {author} {\bibfnamefont {M.~E.}\ \bibnamefont
  {Peskin}}\ and\ \bibinfo {author} {\bibfnamefont {D.~V.}\ \bibnamefont
  {Schroeder}},\ }\href@noop {} {\emph {\bibinfo {title} {{An Introduction to
  quantum field theory}}}}\ (\bibinfo  {publisher} {Addison-Wesley},\ \bibinfo
  {address} {Reading, USA},\ \bibinfo {year} {1995})\BibitemShut {NoStop}%
\bibitem [{\citenamefont {Ji}(1995{\natexlab{a}})}]{Ji:1994av}%
  \BibitemOpen
  \bibfield  {author} {\bibinfo {author} {\bibfnamefont {X.-D.}\ \bibnamefont
  {Ji}},\ }\href {\doibase 10.1103/PhysRevLett.74.1071} {\bibfield  {journal}
  {\bibinfo  {journal} {Phys. Rev. Lett.}\ }\textbf {\bibinfo {volume} {74}},\
  \bibinfo {pages} {1071} (\bibinfo {year} {1995}{\natexlab{a}})},\ \Eprint
  {http://arxiv.org/abs/hep-ph/9410274} {arXiv:hep-ph/9410274} \BibitemShut
  {NoStop}%
\bibitem [{\citenamefont {Ji}(1995{\natexlab{b}})}]{Ji:1995sv}%
  \BibitemOpen
  \bibfield  {author} {\bibinfo {author} {\bibfnamefont {X.-D.}\ \bibnamefont
  {Ji}},\ }\href {\doibase 10.1103/PhysRevD.52.271} {\bibfield  {journal}
  {\bibinfo  {journal} {Phys. Rev. D}\ }\textbf {\bibinfo {volume} {52}},\
  \bibinfo {pages} {271} (\bibinfo {year} {1995}{\natexlab{b}})},\ \Eprint
  {http://arxiv.org/abs/hep-ph/9502213} {arXiv:hep-ph/9502213} \BibitemShut
  {NoStop}%
\bibitem [{\citenamefont {Rothe}(1995{\natexlab{a}})}]{Rothe:1995hu}%
  \BibitemOpen
  \bibfield  {author} {\bibinfo {author} {\bibfnamefont {H.~J.}\ \bibnamefont
  {Rothe}},\ }\href {\doibase 10.1016/0370-2693(95)00663-6} {\bibfield
  {journal} {\bibinfo  {journal} {Phys. Lett. B}\ }\textbf {\bibinfo {volume}
  {355}},\ \bibinfo {pages} {260} (\bibinfo {year} {1995}{\natexlab{a}})},\
  \Eprint {http://arxiv.org/abs/hep-lat/9504012} {arXiv:hep-lat/9504012}
  \BibitemShut {NoStop}%
\bibitem [{\citenamefont {Lorce}(2018)}]{Lorce:2017xzd}%
  \BibitemOpen
  \bibfield  {author} {\bibinfo {author} {\bibfnamefont {C.}~\bibnamefont
  {Lorce}},\ }\href {\doibase 10.1140/epjc/s10052-018-5561-2} {\bibfield
  {journal} {\bibinfo  {journal} {Eur. Phys. J. C}\ }\textbf {\bibinfo {volume}
  {78}},\ \bibinfo {pages} {120} (\bibinfo {year} {2018})},\ \Eprint
  {http://arxiv.org/abs/1706.05853} {arXiv:1706.05853 [hep-ph]} \BibitemShut
  {NoStop}%
\bibitem [{\citenamefont {Hatta}\ \emph {et~al.}(2018)\citenamefont {Hatta},
  \citenamefont {Rajan},\ and\ \citenamefont {Tanaka}}]{Hatta:2018sqd}%
  \BibitemOpen
  \bibfield  {author} {\bibinfo {author} {\bibfnamefont {Y.}~\bibnamefont
  {Hatta}}, \bibinfo {author} {\bibfnamefont {A.}~\bibnamefont {Rajan}}, \ and\
  \bibinfo {author} {\bibfnamefont {K.}~\bibnamefont {Tanaka}},\ }\href
  {\doibase 10.1007/JHEP12(2018)008} {\bibfield  {journal} {\bibinfo  {journal}
  {JHEP}\ }\textbf {\bibinfo {volume} {12}},\ \bibinfo {pages} {008} (\bibinfo
  {year} {2018})},\ \Eprint {http://arxiv.org/abs/1810.05116} {arXiv:1810.05116
  [hep-ph]} \BibitemShut {NoStop}%
\bibitem [{\citenamefont {Metz}\ \emph {et~al.}(2020)\citenamefont {Metz},
  \citenamefont {Pasquini},\ and\ \citenamefont {Rodini}}]{Metz:2020vxd}%
  \BibitemOpen
  \bibfield  {author} {\bibinfo {author} {\bibfnamefont {A.}~\bibnamefont
  {Metz}}, \bibinfo {author} {\bibfnamefont {B.}~\bibnamefont {Pasquini}}, \
  and\ \bibinfo {author} {\bibfnamefont {S.}~\bibnamefont {Rodini}},\
  }\href@noop {} {\  (\bibinfo {year} {2020})},\ \Eprint
  {http://arxiv.org/abs/2006.11171} {arXiv:2006.11171 [hep-ph]} \BibitemShut
  {NoStop}%
\bibitem [{\citenamefont {Ji}(2021)}]{Ji:2021mtz}%
  \BibitemOpen
  \bibfield  {author} {\bibinfo {author} {\bibfnamefont {X.}~\bibnamefont
  {Ji}},\ }\href {\doibase 10.1007/s11467-021-1065-x} {\bibfield  {journal}
  {\bibinfo  {journal} {Front. Phys. (Beijing)}\ }\textbf {\bibinfo {volume}
  {16}},\ \bibinfo {pages} {64601} (\bibinfo {year} {2021})},\ \Eprint
  {http://arxiv.org/abs/2102.07830} {arXiv:2102.07830 [hep-ph]} \BibitemShut
  {NoStop}%
\bibitem [{\citenamefont {Liu}(2021)}]{Liu:2021gco}%
  \BibitemOpen
  \bibfield  {author} {\bibinfo {author} {\bibfnamefont {K.-F.}\ \bibnamefont
  {Liu}},\ }\href@noop {} {\  (\bibinfo {year} {2021})},\ \Eprint
  {http://arxiv.org/abs/2103.15768} {arXiv:2103.15768 [hep-ph]} \BibitemShut
  {NoStop}%
\bibitem [{\citenamefont {Zahed}(2021)}]{Zahed:2021fxk}%
  \BibitemOpen
  \bibfield  {author} {\bibinfo {author} {\bibfnamefont {I.}~\bibnamefont
  {Zahed}},\ }\href@noop {} {\  (\bibinfo {year} {2021})},\ \Eprint
  {http://arxiv.org/abs/2102.08191} {arXiv:2102.08191 [hep-ph]} \BibitemShut
  {NoStop}%
\bibitem [{\citenamefont {Kharzeev}\ \emph {et~al.}(1999)\citenamefont
  {Kharzeev}, \citenamefont {Satz}, \citenamefont {Syamtomov},\ and\
  \citenamefont {Zinovjev}}]{Kharzeev:1998bz}%
  \BibitemOpen
  \bibfield  {author} {\bibinfo {author} {\bibfnamefont {D.}~\bibnamefont
  {Kharzeev}}, \bibinfo {author} {\bibfnamefont {H.}~\bibnamefont {Satz}},
  \bibinfo {author} {\bibfnamefont {A.}~\bibnamefont {Syamtomov}}, \ and\
  \bibinfo {author} {\bibfnamefont {G.}~\bibnamefont {Zinovjev}},\ }\href
  {\doibase 10.1007/s100529900047} {\bibfield  {journal} {\bibinfo  {journal}
  {Eur. Phys. J. C}\ }\textbf {\bibinfo {volume} {9}},\ \bibinfo {pages} {459}
  (\bibinfo {year} {1999})},\ \Eprint {http://arxiv.org/abs/hep-ph/9901375}
  {arXiv:hep-ph/9901375} \BibitemShut {NoStop}%
\bibitem [{\citenamefont {Hatta}\ and\ \citenamefont
  {Yang}(2018)}]{Hatta:2018ina}%
  \BibitemOpen
  \bibfield  {author} {\bibinfo {author} {\bibfnamefont {Y.}~\bibnamefont
  {Hatta}}\ and\ \bibinfo {author} {\bibfnamefont {D.-L.}\ \bibnamefont
  {Yang}},\ }\href {\doibase 10.1103/PhysRevD.98.074003} {\bibfield  {journal}
  {\bibinfo  {journal} {Phys. Rev. D}\ }\textbf {\bibinfo {volume} {98}},\
  \bibinfo {pages} {074003} (\bibinfo {year} {2018})},\ \Eprint
  {http://arxiv.org/abs/1808.02163} {arXiv:1808.02163 [hep-ph]} \BibitemShut
  {NoStop}%
\bibitem [{\citenamefont {Meziani}\ and\ \citenamefont
  {Joosten}(2020)}]{Meziani:2020oks}%
  \BibitemOpen
  \bibfield  {author} {\bibinfo {author} {\bibfnamefont {Z.-E.}\ \bibnamefont
  {Meziani}}\ and\ \bibinfo {author} {\bibfnamefont {S.}~\bibnamefont
  {Joosten}},\ }in\ \href@noop {} {\emph {\bibinfo {booktitle} {{Probing
  Nucleons and Nuclei in High Energy Collisions}: {Dedicated to the Physics of
  the Electron Ion Collider}}}}\ (\bibinfo {year} {2020})\ pp.\ \bibinfo
  {pages} {234--237},\ \bibinfo {note}
  {{doi:\url{10.1142/9789811214950_0048}}}\BibitemShut {NoStop}%
\bibitem [{\citenamefont {Durr}\ \emph {et~al.}(2008)\citenamefont {Durr} \emph
  {et~al.}}]{Durr:2008zz}%
  \BibitemOpen
  \bibfield  {author} {\bibinfo {author} {\bibfnamefont {S.}~\bibnamefont
  {Durr}} \emph {et~al.},\ }\href {\doibase 10.1126/science.1163233} {\bibfield
   {journal} {\bibinfo  {journal} {Science}\ }\textbf {\bibinfo {volume}
  {322}},\ \bibinfo {pages} {1224} (\bibinfo {year} {2008})},\ \Eprint
  {http://arxiv.org/abs/0906.3599} {arXiv:0906.3599 [hep-lat]} \BibitemShut
  {NoStop}%
\bibitem [{\citenamefont {Fodor}\ and\ \citenamefont
  {Hoelbling}(2012)}]{Fodor:2012gf}%
  \BibitemOpen
  \bibfield  {author} {\bibinfo {author} {\bibfnamefont {Z.}~\bibnamefont
  {Fodor}}\ and\ \bibinfo {author} {\bibfnamefont {C.}~\bibnamefont
  {Hoelbling}},\ }\href {\doibase 10.1103/RevModPhys.84.449} {\bibfield
  {journal} {\bibinfo  {journal} {Rev. Mod. Phys.}\ }\textbf {\bibinfo {volume}
  {84}},\ \bibinfo {pages} {449} (\bibinfo {year} {2012})},\ \Eprint
  {http://arxiv.org/abs/1203.4789} {arXiv:1203.4789 [hep-lat]} \BibitemShut
  {NoStop}%
\bibitem [{\citenamefont {Borsanyi}\ \emph {et~al.}(2015)\citenamefont
  {Borsanyi} \emph {et~al.}}]{Borsanyi:2014jba}%
  \BibitemOpen
  \bibfield  {author} {\bibinfo {author} {\bibfnamefont {S.}~\bibnamefont
  {Borsanyi}} \emph {et~al.},\ }\href {\doibase 10.1126/science.1257050}
  {\bibfield  {journal} {\bibinfo  {journal} {Science}\ }\textbf {\bibinfo
  {volume} {347}},\ \bibinfo {pages} {1452} (\bibinfo {year} {2015})},\ \Eprint
  {http://arxiv.org/abs/1406.4088} {arXiv:1406.4088 [hep-lat]} \BibitemShut
  {NoStop}%
\bibitem [{\citenamefont {Walker-Loud}(2018)}]{Walker-Loud:2018gvh}%
  \BibitemOpen
  \bibfield  {author} {\bibinfo {author} {\bibfnamefont {A.}~\bibnamefont
  {Walker-Loud}},\ }\href {\doibase 10.1103/Physics.11.118} {\bibfield
  {journal} {\bibinfo  {journal} {APS Physics}\ }\textbf {\bibinfo {volume}
  {11}},\ \bibinfo {pages} {118} (\bibinfo {year} {2018})}\BibitemShut
  {NoStop}%
\bibitem [{\citenamefont {Abdel-Rehim}\ \emph {et~al.}(2016)\citenamefont
  {Abdel-Rehim}, \citenamefont {Alexandrou}, \citenamefont {Constantinou},
  \citenamefont {Hadjiyiannakou}, \citenamefont {Jansen}, \citenamefont
  {Kallidonis}, \citenamefont {Koutsou},\ and\ \citenamefont {Vaquero
  Aviles-Casco}}]{Abdel-Rehim:2016won}%
  \BibitemOpen
  \bibfield  {author} {\bibinfo {author} {\bibfnamefont {A.}~\bibnamefont
  {Abdel-Rehim}}, \bibinfo {author} {\bibfnamefont {C.}~\bibnamefont
  {Alexandrou}}, \bibinfo {author} {\bibfnamefont {M.}~\bibnamefont
  {Constantinou}}, \bibinfo {author} {\bibfnamefont {K.}~\bibnamefont
  {Hadjiyiannakou}}, \bibinfo {author} {\bibfnamefont {K.}~\bibnamefont
  {Jansen}}, \bibinfo {author} {\bibfnamefont {C.}~\bibnamefont {Kallidonis}},
  \bibinfo {author} {\bibfnamefont {G.}~\bibnamefont {Koutsou}}, \ and\
  \bibinfo {author} {\bibfnamefont {A.}~\bibnamefont {Vaquero Aviles-Casco}}
  (\bibinfo {collaboration} {ETM}),\ }\href {\doibase
  10.1103/PhysRevLett.116.252001} {\bibfield  {journal} {\bibinfo  {journal}
  {Phys. Rev. Lett.}\ }\textbf {\bibinfo {volume} {116}},\ \bibinfo {pages}
  {252001} (\bibinfo {year} {2016})},\ \Eprint
  {http://arxiv.org/abs/1601.01624} {arXiv:1601.01624 [hep-lat]} \BibitemShut
  {NoStop}%
\bibitem [{\citenamefont {Alexandrou}\ \emph {et~al.}(2017)\citenamefont
  {Alexandrou}, \citenamefont {Constantinou}, \citenamefont {Hadjiyiannakou},
  \citenamefont {Jansen}, \citenamefont {Kallidonis}, \citenamefont {Koutsou},
  \citenamefont {Vaquero Avil\'es-Casco},\ and\ \citenamefont
  {Wiese}}]{Alexandrou:2017oeh}%
  \BibitemOpen
  \bibfield  {author} {\bibinfo {author} {\bibfnamefont {C.}~\bibnamefont
  {Alexandrou}}, \bibinfo {author} {\bibfnamefont {M.}~\bibnamefont
  {Constantinou}}, \bibinfo {author} {\bibfnamefont {K.}~\bibnamefont
  {Hadjiyiannakou}}, \bibinfo {author} {\bibfnamefont {K.}~\bibnamefont
  {Jansen}}, \bibinfo {author} {\bibfnamefont {C.}~\bibnamefont {Kallidonis}},
  \bibinfo {author} {\bibfnamefont {G.}~\bibnamefont {Koutsou}}, \bibinfo
  {author} {\bibfnamefont {A.}~\bibnamefont {Vaquero Avil\'es-Casco}}, \ and\
  \bibinfo {author} {\bibfnamefont {C.}~\bibnamefont {Wiese}},\ }\href
  {\doibase 10.1103/PhysRevLett.119.142002} {\bibfield  {journal} {\bibinfo
  {journal} {Phys. Rev. Lett.}\ }\textbf {\bibinfo {volume} {119}},\ \bibinfo
  {pages} {142002} (\bibinfo {year} {2017})},\ \Eprint
  {http://arxiv.org/abs/1706.02973} {arXiv:1706.02973 [hep-lat]} \BibitemShut
  {NoStop}%
\bibitem [{\citenamefont {Yang}\ \emph {et~al.}(2018)\citenamefont {Yang},
  \citenamefont {Liang}, \citenamefont {Bi}, \citenamefont {Chen},
  \citenamefont {Draper}, \citenamefont {Liu},\ and\ \citenamefont
  {Liu}}]{Yang:2018nqn}%
  \BibitemOpen
  \bibfield  {author} {\bibinfo {author} {\bibfnamefont {Y.-B.}\ \bibnamefont
  {Yang}}, \bibinfo {author} {\bibfnamefont {J.}~\bibnamefont {Liang}},
  \bibinfo {author} {\bibfnamefont {Y.-J.}\ \bibnamefont {Bi}}, \bibinfo
  {author} {\bibfnamefont {Y.}~\bibnamefont {Chen}}, \bibinfo {author}
  {\bibfnamefont {T.}~\bibnamefont {Draper}}, \bibinfo {author} {\bibfnamefont
  {K.-F.}\ \bibnamefont {Liu}}, \ and\ \bibinfo {author} {\bibfnamefont
  {Z.}~\bibnamefont {Liu}},\ }\href {\doibase 10.1103/PhysRevLett.121.212001}
  {\bibfield  {journal} {\bibinfo  {journal} {Phys. Rev. Lett.}\ }\textbf
  {\bibinfo {volume} {121}},\ \bibinfo {pages} {212001} (\bibinfo {year}
  {2018})},\ \Eprint {http://arxiv.org/abs/1808.08677} {arXiv:1808.08677
  [hep-lat]} \BibitemShut {NoStop}%
\bibitem [{\citenamefont {He}\ \emph {et~al.}(2021)\citenamefont {He},
  \citenamefont {Sun},\ and\ \citenamefont {Yang}}]{He:2021bof}%
  \BibitemOpen
  \bibfield  {author} {\bibinfo {author} {\bibfnamefont {F.}~\bibnamefont
  {He}}, \bibinfo {author} {\bibfnamefont {P.}~\bibnamefont {Sun}}, \ and\
  \bibinfo {author} {\bibfnamefont {Y.-B.}\ \bibnamefont {Yang}},\ }\href@noop
  {} {\  (\bibinfo {year} {2021})},\ \Eprint {http://arxiv.org/abs/2101.04942}
  {arXiv:2101.04942 [hep-lat]} \BibitemShut {NoStop}%
\bibitem [{\citenamefont {Collins}\ \emph {et~al.}(1977)\citenamefont
  {Collins}, \citenamefont {Duncan},\ and\ \citenamefont
  {Joglekar}}]{Collins:1976yq}%
  \BibitemOpen
  \bibfield  {author} {\bibinfo {author} {\bibfnamefont {J.~C.}\ \bibnamefont
  {Collins}}, \bibinfo {author} {\bibfnamefont {A.}~\bibnamefont {Duncan}}, \
  and\ \bibinfo {author} {\bibfnamefont {S.~D.}\ \bibnamefont {Joglekar}},\
  }\href {\doibase 10.1103/PhysRevD.16.438} {\bibfield  {journal} {\bibinfo
  {journal} {Phys. Rev. D}\ }\textbf {\bibinfo {volume} {16}},\ \bibinfo
  {pages} {438} (\bibinfo {year} {1977})}\BibitemShut {NoStop}%
\bibitem [{\citenamefont {Shifman}\ \emph {et~al.}(1978)\citenamefont
  {Shifman}, \citenamefont {Vainshtein},\ and\ \citenamefont
  {Zakharov}}]{Shifman:1978zn}%
  \BibitemOpen
  \bibfield  {author} {\bibinfo {author} {\bibfnamefont {M.~A.}\ \bibnamefont
  {Shifman}}, \bibinfo {author} {\bibfnamefont {A.}~\bibnamefont {Vainshtein}},
  \ and\ \bibinfo {author} {\bibfnamefont {V.~I.}\ \bibnamefont {Zakharov}},\
  }\href {\doibase 10.1016/0370-2693(78)90481-1} {\bibfield  {journal}
  {\bibinfo  {journal} {Phys. Lett. B}\ }\textbf {\bibinfo {volume} {78}},\
  \bibinfo {pages} {443} (\bibinfo {year} {1978})}\BibitemShut {NoStop}%
\bibitem [{\citenamefont {Ji}\ and\ \citenamefont {Liu}(2021)}]{Ji:2021pys}%
  \BibitemOpen
  \bibfield  {author} {\bibinfo {author} {\bibfnamefont {X.}~\bibnamefont
  {Ji}}\ and\ \bibinfo {author} {\bibfnamefont {Y.}~\bibnamefont {Liu}},\
  }\href@noop {} {\  (\bibinfo {year} {2021})},\ \Eprint
  {http://arxiv.org/abs/2101.04483} {arXiv:2101.04483 [hep-ph]} \BibitemShut
  {NoStop}%
\bibitem [{\citenamefont {Rothe}(1995{\natexlab{b}})}]{Rothe:1995av}%
  \BibitemOpen
  \bibfield  {author} {\bibinfo {author} {\bibfnamefont {H.~J.}\ \bibnamefont
  {Rothe}},\ }\href {\doibase 10.1016/0370-2693(95)01253-2} {\bibfield
  {journal} {\bibinfo  {journal} {Phys. Lett. B}\ }\textbf {\bibinfo {volume}
  {364}},\ \bibinfo {pages} {227} (\bibinfo {year} {1995}{\natexlab{b}})},\
  \Eprint {http://arxiv.org/abs/hep-lat/9508005} {arXiv:hep-lat/9508005}
  \BibitemShut {NoStop}%
\bibitem [{\citenamefont {Weinberg}(1967)}]{Weinberg:1967tq}%
  \BibitemOpen
  \bibfield  {author} {\bibinfo {author} {\bibfnamefont {S.}~\bibnamefont
  {Weinberg}},\ }\href {\doibase 10.1103/PhysRevLett.19.1264} {\bibfield
  {journal} {\bibinfo  {journal} {Phys. Rev. Lett.}\ }\textbf {\bibinfo
  {volume} {19}},\ \bibinfo {pages} {1264} (\bibinfo {year}
  {1967})}\BibitemShut {NoStop}%
\bibitem [{\citenamefont {Sirunyan}\ \emph {et~al.}(2019)\citenamefont
  {Sirunyan} \emph {et~al.}}]{Sirunyan:2018koj}%
  \BibitemOpen
  \bibfield  {author} {\bibinfo {author} {\bibfnamefont {A.~M.}\ \bibnamefont
  {Sirunyan}} \emph {et~al.} (\bibinfo {collaboration} {CMS}),\ }\href
  {\doibase 10.1140/epjc/s10052-019-6909-y} {\bibfield  {journal} {\bibinfo
  {journal} {Eur. Phys. J. C}\ }\textbf {\bibinfo {volume} {79}},\ \bibinfo
  {pages} {421} (\bibinfo {year} {2019})},\ \Eprint
  {http://arxiv.org/abs/1809.10733} {arXiv:1809.10733 [hep-ex]} \BibitemShut
  {NoStop}%
\bibitem [{\citenamefont {Aad}\ \emph {et~al.}(2020)\citenamefont {Aad} \emph
  {et~al.}}]{Aad:2019mbh}%
  \BibitemOpen
  \bibfield  {author} {\bibinfo {author} {\bibfnamefont {G.}~\bibnamefont
  {Aad}} \emph {et~al.} (\bibinfo {collaboration} {ATLAS}),\ }\href {\doibase
  10.1103/PhysRevD.101.012002} {\bibfield  {journal} {\bibinfo  {journal}
  {Phys. Rev. D}\ }\textbf {\bibinfo {volume} {101}},\ \bibinfo {pages}
  {012002} (\bibinfo {year} {2020})},\ \Eprint
  {http://arxiv.org/abs/1909.02845} {arXiv:1909.02845 [hep-ex]} \BibitemShut
  {NoStop}%
\bibitem [{\citenamefont {Ellis}\ and\ \citenamefont
  {Lanik}(1985)}]{Ellis:1984jv}%
  \BibitemOpen
  \bibfield  {author} {\bibinfo {author} {\bibfnamefont {J.~R.}\ \bibnamefont
  {Ellis}}\ and\ \bibinfo {author} {\bibfnamefont {J.}~\bibnamefont {Lanik}},\
  }\href {\doibase 10.1016/0370-2693(85)91013-5} {\bibfield  {journal}
  {\bibinfo  {journal} {Phys. Lett. B}\ }\textbf {\bibinfo {volume} {150}},\
  \bibinfo {pages} {289} (\bibinfo {year} {1985})}\BibitemShut {NoStop}%
\bibitem [{\citenamefont {Coleman}(1975)}]{Coleman:1974hr}%
  \BibitemOpen
  \bibfield  {author} {\bibinfo {author} {\bibfnamefont {S.~R.}\ \bibnamefont
  {Coleman}},\ }\href@noop {} {\bibfield  {journal} {\bibinfo  {journal}
  {Subnucl. Ser.}\ }\textbf {\bibinfo {volume} {11}},\ \bibinfo {pages} {139}
  (\bibinfo {year} {1975})}\BibitemShut {NoStop}%
\bibitem [{\citenamefont {'t~Hooft}(1978)}]{tHooft:1977nqb}%
  \BibitemOpen
  \bibfield  {author} {\bibinfo {author} {\bibfnamefont {G.}~\bibnamefont
  {'t~Hooft}},\ }\href {\doibase 10.1016/0550-3213(78)90153-0} {\bibfield
  {journal} {\bibinfo  {journal} {Nucl. Phys. B}\ }\textbf {\bibinfo {volume}
  {138}},\ \bibinfo {pages} {1} (\bibinfo {year} {1978})}\BibitemShut {NoStop}%
\bibitem [{\citenamefont {Fradkin}\ and\ \citenamefont
  {Shenker}(1979)}]{Fradkin:1978dv}%
  \BibitemOpen
  \bibfield  {author} {\bibinfo {author} {\bibfnamefont {E.~H.}\ \bibnamefont
  {Fradkin}}\ and\ \bibinfo {author} {\bibfnamefont {S.~H.}\ \bibnamefont
  {Shenker}},\ }\href {\doibase 10.1103/PhysRevD.19.3682} {\bibfield  {journal}
  {\bibinfo  {journal} {Phys. Rev. D}\ }\textbf {\bibinfo {volume} {19}},\
  \bibinfo {pages} {3682} (\bibinfo {year} {1979})}\BibitemShut {NoStop}%
\bibitem [{\citenamefont {Diakonov}(1996)}]{Diakonov:1995ea}%
  \BibitemOpen
  \bibfield  {author} {\bibinfo {author} {\bibfnamefont {D.}~\bibnamefont
  {Diakonov}},\ }\href {\doibase 10.3254/978-1-61499-215-8-397} {\bibfield
  {journal} {\bibinfo  {journal} {Proc. Int. Sch. Phys. Fermi}\ }\textbf
  {\bibinfo {volume} {130}},\ \bibinfo {pages} {397} (\bibinfo {year}
  {1996})},\ \Eprint {http://arxiv.org/abs/hep-ph/9602375}
  {arXiv:hep-ph/9602375} \BibitemShut {NoStop}%
\bibitem [{\citenamefont {Sch\"afer}\ and\ \citenamefont
  {Shuryak}(1998)}]{Schafer:1996wv}%
  \BibitemOpen
  \bibfield  {author} {\bibinfo {author} {\bibfnamefont {T.}~\bibnamefont
  {Sch\"afer}}\ and\ \bibinfo {author} {\bibfnamefont {E.~V.}\ \bibnamefont
  {Shuryak}},\ }\href {\doibase 10.1103/RevModPhys.70.323} {\bibfield
  {journal} {\bibinfo  {journal} {Rev. Mod. Phys.}\ }\textbf {\bibinfo {volume}
  {70}},\ \bibinfo {pages} {323} (\bibinfo {year} {1998})},\ \Eprint
  {http://arxiv.org/abs/hep-ph/9610451} {arXiv:hep-ph/9610451} \BibitemShut
  {NoStop}%
\bibitem [{\citenamefont {Chu}\ \emph {et~al.}(1994)\citenamefont {Chu},
  \citenamefont {Grandy}, \citenamefont {Huang},\ and\ \citenamefont
  {Negele}}]{Chu:1994vi}%
  \BibitemOpen
  \bibfield  {author} {\bibinfo {author} {\bibfnamefont {M.~C.}\ \bibnamefont
  {Chu}}, \bibinfo {author} {\bibfnamefont {J.~M.}\ \bibnamefont {Grandy}},
  \bibinfo {author} {\bibfnamefont {S.}~\bibnamefont {Huang}}, \ and\ \bibinfo
  {author} {\bibfnamefont {J.~W.}\ \bibnamefont {Negele}},\ }\href {\doibase
  10.1103/PhysRevD.49.6039} {\bibfield  {journal} {\bibinfo  {journal} {Phys.
  Rev. D}\ }\textbf {\bibinfo {volume} {49}},\ \bibinfo {pages} {6039}
  (\bibinfo {year} {1994})},\ \Eprint {http://arxiv.org/abs/hep-lat/9312071}
  {arXiv:hep-lat/9312071} \BibitemShut {NoStop}%
\bibitem [{\citenamefont {Greensite}(2017)}]{Greensite:2016pfc}%
  \BibitemOpen
  \bibfield  {author} {\bibinfo {author} {\bibfnamefont {J.}~\bibnamefont
  {Greensite}},\ }\href {\doibase 10.1051/epjconf/201713701009} {\bibfield
  {journal} {\bibinfo  {journal} {EPJ Web Conf.}\ }\textbf {\bibinfo {volume}
  {137}},\ \bibinfo {pages} {01009} (\bibinfo {year} {2017})},\ \Eprint
  {http://arxiv.org/abs/1610.06221} {arXiv:1610.06221 [hep-lat]} \BibitemShut
  {NoStop}%
\bibitem [{\citenamefont {Chodos}\ \emph {et~al.}(1974)\citenamefont {Chodos},
  \citenamefont {Jaffe}, \citenamefont {Johnson}, \citenamefont {Thorn},\ and\
  \citenamefont {Weisskopf}}]{Chodos:1974je}%
  \BibitemOpen
  \bibfield  {author} {\bibinfo {author} {\bibfnamefont {A.}~\bibnamefont
  {Chodos}}, \bibinfo {author} {\bibfnamefont {R.}~\bibnamefont {Jaffe}},
  \bibinfo {author} {\bibfnamefont {K.}~\bibnamefont {Johnson}}, \bibinfo
  {author} {\bibfnamefont {C.~B.}\ \bibnamefont {Thorn}}, \ and\ \bibinfo
  {author} {\bibfnamefont {V.}~\bibnamefont {Weisskopf}},\ }\href {\doibase
  10.1103/PhysRevD.9.3471} {\bibfield  {journal} {\bibinfo  {journal} {Phys.
  Rev. D}\ }\textbf {\bibinfo {volume} {9}},\ \bibinfo {pages} {3471} (\bibinfo
  {year} {1974})}\BibitemShut {NoStop}%
\bibitem [{\citenamefont {Karsch}(1982)}]{Karsch:1982ve}%
  \BibitemOpen
  \bibfield  {author} {\bibinfo {author} {\bibfnamefont {F.}~\bibnamefont
  {Karsch}},\ }\href {\doibase 10.1016/0550-3213(82)90390-X} {\bibfield
  {journal} {\bibinfo  {journal} {Nucl. Phys. B}\ }\textbf {\bibinfo {volume}
  {205}},\ \bibinfo {pages} {285} (\bibinfo {year} {1982})}\BibitemShut
  {NoStop}%
\bibitem [{\citenamefont {Collins}(1986)}]{Collins:1984xc}%
  \BibitemOpen
  \bibfield  {author} {\bibinfo {author} {\bibfnamefont {J.~C.}\ \bibnamefont
  {Collins}},\ }\href {\doibase 10.1017/CBO9780511622656} {\emph {\bibinfo
  {title} {{Renormalization}: {An Introduction to Renormalization, The
  Renormalization Group, and the Operator Product Expansion}}}},\ \bibinfo
  {series} {Cambridge Monographs on Mathematical Physics}, Vol.~\bibinfo
  {volume} {26}\ (\bibinfo  {publisher} {Cambridge University Press},\ \bibinfo
  {address} {Cambridge},\ \bibinfo {year} {1986})\BibitemShut {NoStop}%
\bibitem [{\citenamefont {Luke}\ \emph {et~al.}(1992)\citenamefont {Luke},
  \citenamefont {Manohar},\ and\ \citenamefont {Savage}}]{Luke:1992tm}%
  \BibitemOpen
  \bibfield  {author} {\bibinfo {author} {\bibfnamefont {M.~E.}\ \bibnamefont
  {Luke}}, \bibinfo {author} {\bibfnamefont {A.~V.}\ \bibnamefont {Manohar}}, \
  and\ \bibinfo {author} {\bibfnamefont {M.~J.}\ \bibnamefont {Savage}},\
  }\href {\doibase 10.1016/0370-2693(92)91114-O} {\bibfield  {journal}
  {\bibinfo  {journal} {Phys. Lett. B}\ }\textbf {\bibinfo {volume} {288}},\
  \bibinfo {pages} {355} (\bibinfo {year} {1992})},\ \Eprint
  {http://arxiv.org/abs/hep-ph/9204219} {arXiv:hep-ph/9204219} \BibitemShut
  {NoStop}%
\bibitem [{\citenamefont {Kharzeev}(1996)}]{Kharzeev:1995ij}%
  \BibitemOpen
  \bibfield  {author} {\bibinfo {author} {\bibfnamefont {D.}~\bibnamefont
  {Kharzeev}},\ }\href {\doibase 10.3254/978-1-61499-215-8-105} {\bibfield
  {journal} {\bibinfo  {journal} {Proc. Int. Sch. Phys. Fermi}\ }\textbf
  {\bibinfo {volume} {130}},\ \bibinfo {pages} {105} (\bibinfo {year}
  {1996})},\ \Eprint {http://arxiv.org/abs/nucl-th/9601029}
  {arXiv:nucl-th/9601029} \BibitemShut {NoStop}%
\bibitem [{\citenamefont {Novikov}\ \emph {et~al.}(1984)\citenamefont
  {Novikov}, \citenamefont {Shifman}, \citenamefont {Vainshtein},\ and\
  \citenamefont {Zakharov}}]{Novikov:1984ac}%
  \BibitemOpen
  \bibfield  {author} {\bibinfo {author} {\bibfnamefont {V.}~\bibnamefont
  {Novikov}}, \bibinfo {author} {\bibfnamefont {M.~A.}\ \bibnamefont
  {Shifman}}, \bibinfo {author} {\bibfnamefont {A.}~\bibnamefont {Vainshtein}},
  \ and\ \bibinfo {author} {\bibfnamefont {V.~I.}\ \bibnamefont {Zakharov}},\
  }\href {\doibase 10.1016/0370-1573(84)90021-8} {\bibfield  {journal}
  {\bibinfo  {journal} {Phys. Rept.}\ }\textbf {\bibinfo {volume} {116}},\
  \bibinfo {pages} {103} (\bibinfo {year} {1984})}\BibitemShut {NoStop}%
\bibitem [{\citenamefont {Shifman}(2012)}]{Shifman:2012zz}%
  \BibitemOpen
  \bibfield  {author} {\bibinfo {author} {\bibfnamefont {M.}~\bibnamefont
  {Shifman}},\ }\href@noop {} {\emph {\bibinfo {title} {{Advanced topics in
  quantum field theory.}: {A lecture course}}}}\ (\bibinfo  {publisher}
  {Cambridge Univ. Press},\ \bibinfo {address} {Cambridge, UK},\ \bibinfo
  {year} {2012})\BibitemShut {NoStop}%
\bibitem [{\citenamefont {Shankar}\ and\ \citenamefont
  {Witten}(1978)}]{Shankar:1977cm}%
  \BibitemOpen
  \bibfield  {author} {\bibinfo {author} {\bibfnamefont {R.}~\bibnamefont
  {Shankar}}\ and\ \bibinfo {author} {\bibfnamefont {E.}~\bibnamefont
  {Witten}},\ }\href {\doibase 10.1103/PhysRevD.17.2134} {\bibfield  {journal}
  {\bibinfo  {journal} {Phys. Rev. D}\ }\textbf {\bibinfo {volume} {17}},\
  \bibinfo {pages} {2134} (\bibinfo {year} {1978})}\BibitemShut {NoStop}%
\bibitem [{\citenamefont {Wiegmann}(1985)}]{Wiegmann:1985jt}%
  \BibitemOpen
  \bibfield  {author} {\bibinfo {author} {\bibfnamefont {P.~B.}\ \bibnamefont
  {Wiegmann}},\ }\href {\doibase 10.1016/0370-2693(85)91171-2} {\bibfield
  {journal} {\bibinfo  {journal} {Phys. Lett. B}\ }\textbf {\bibinfo {volume}
  {152}},\ \bibinfo {pages} {209} (\bibinfo {year} {1985})}\BibitemShut
  {NoStop}%
\bibitem [{\citenamefont {Iwasaki}(1981)}]{Iwasaki:1980hd}%
  \BibitemOpen
  \bibfield  {author} {\bibinfo {author} {\bibfnamefont {Y.}~\bibnamefont
  {Iwasaki}},\ }\href {\doibase 10.1103/PhysRevLett.47.754} {\bibfield
  {journal} {\bibinfo  {journal} {Phys. Rev. Lett.}\ }\textbf {\bibinfo
  {volume} {47}},\ \bibinfo {pages} {754} (\bibinfo {year} {1981})}\BibitemShut
  {NoStop}%
\bibitem [{\citenamefont {Weinberg}(2005)}]{Weinberg:1995mt}%
  \BibitemOpen
  \bibfield  {author} {\bibinfo {author} {\bibfnamefont {S.}~\bibnamefont
  {Weinberg}},\ }\href@noop {} {\emph {\bibinfo {title} {{The Quantum theory of
  fields. Vol. 1: Foundations}}}}\ (\bibinfo  {publisher} {Cambridge University
  Press},\ \bibinfo {year} {2005})\BibitemShut {NoStop}%
\bibitem [{\citenamefont {Kronfeld}(1998)}]{Kronfeld:1998di}%
  \BibitemOpen
  \bibfield  {author} {\bibinfo {author} {\bibfnamefont {A.~S.}\ \bibnamefont
  {Kronfeld}},\ }\href {\doibase 10.1103/PhysRevD.58.051501} {\bibfield
  {journal} {\bibinfo  {journal} {Phys. Rev. D}\ }\textbf {\bibinfo {volume}
  {58}},\ \bibinfo {pages} {051501} (\bibinfo {year} {1998})},\ \Eprint
  {http://arxiv.org/abs/hep-ph/9805215} {arXiv:hep-ph/9805215} \BibitemShut
  {NoStop}%
\bibitem [{\citenamefont {Broadhurst}\ \emph {et~al.}(1991)\citenamefont
  {Broadhurst}, \citenamefont {Gray},\ and\ \citenamefont
  {Schilcher}}]{Broadhurst:1991fy}%
  \BibitemOpen
  \bibfield  {author} {\bibinfo {author} {\bibfnamefont {D.~J.}\ \bibnamefont
  {Broadhurst}}, \bibinfo {author} {\bibfnamefont {N.}~\bibnamefont {Gray}}, \
  and\ \bibinfo {author} {\bibfnamefont {K.}~\bibnamefont {Schilcher}},\ }\href
  {\doibase 10.1007/BF01412333} {\bibfield  {journal} {\bibinfo  {journal} {Z.
  Phys. C}\ }\textbf {\bibinfo {volume} {52}},\ \bibinfo {pages} {111}
  (\bibinfo {year} {1991})}\BibitemShut {NoStop}%
\bibitem [{\citenamefont {Sun}\ \emph {et~al.}(2020)\citenamefont {Sun},
  \citenamefont {Sun},\ and\ \citenamefont {Zhou}}]{Sun:2020ksc}%
  \BibitemOpen
  \bibfield  {author} {\bibinfo {author} {\bibfnamefont {B.-d.}\ \bibnamefont
  {Sun}}, \bibinfo {author} {\bibfnamefont {Z.-h.}\ \bibnamefont {Sun}}, \ and\
  \bibinfo {author} {\bibfnamefont {J.}~\bibnamefont {Zhou}},\ }\href@noop {}
  {\  (\bibinfo {year} {2020})},\ \Eprint {http://arxiv.org/abs/2012.09443}
  {arXiv:2012.09443 [hep-ph]} \BibitemShut {NoStop}%
\bibitem [{\citenamefont {Banks}\ and\ \citenamefont
  {Rabinovici}(1979)}]{Banks:1979fi}%
  \BibitemOpen
  \bibfield  {author} {\bibinfo {author} {\bibfnamefont {T.}~\bibnamefont
  {Banks}}\ and\ \bibinfo {author} {\bibfnamefont {E.}~\bibnamefont
  {Rabinovici}},\ }\href {\doibase 10.1016/0550-3213(79)90064-6} {\bibfield
  {journal} {\bibinfo  {journal} {Nucl. Phys. B}\ }\textbf {\bibinfo {volume}
  {160}},\ \bibinfo {pages} {349} (\bibinfo {year} {1979})}\BibitemShut
  {NoStop}%
\bibitem [{\citenamefont {Shifman}\ \emph
  {et~al.}(1979{\natexlab{a}})\citenamefont {Shifman}, \citenamefont
  {Vainshtein},\ and\ \citenamefont {Zakharov}}]{Shifman:1978bx}%
  \BibitemOpen
  \bibfield  {author} {\bibinfo {author} {\bibfnamefont {M.~A.}\ \bibnamefont
  {Shifman}}, \bibinfo {author} {\bibfnamefont {A.}~\bibnamefont {Vainshtein}},
  \ and\ \bibinfo {author} {\bibfnamefont {V.~I.}\ \bibnamefont {Zakharov}},\
  }\href {\doibase 10.1016/0550-3213(79)90022-1} {\bibfield  {journal}
  {\bibinfo  {journal} {Nucl. Phys. B}\ }\textbf {\bibinfo {volume} {147}},\
  \bibinfo {pages} {385} (\bibinfo {year} {1979}{\natexlab{a}})}\BibitemShut
  {NoStop}%
\bibitem [{\citenamefont {Shifman}\ \emph
  {et~al.}(1979{\natexlab{b}})\citenamefont {Shifman}, \citenamefont
  {Vainshtein},\ and\ \citenamefont {Zakharov}}]{Shifman:1978by}%
  \BibitemOpen
  \bibfield  {author} {\bibinfo {author} {\bibfnamefont {M.~A.}\ \bibnamefont
  {Shifman}}, \bibinfo {author} {\bibfnamefont {A.}~\bibnamefont {Vainshtein}},
  \ and\ \bibinfo {author} {\bibfnamefont {V.~I.}\ \bibnamefont {Zakharov}},\
  }\href {\doibase 10.1016/0550-3213(79)90023-3} {\bibfield  {journal}
  {\bibinfo  {journal} {Nucl. Phys. B}\ }\textbf {\bibinfo {volume} {147}},\
  \bibinfo {pages} {448} (\bibinfo {year} {1979}{\natexlab{b}})}\BibitemShut
  {NoStop}%
\bibitem [{\citenamefont {Brodsky}\ \emph {et~al.}(2001)\citenamefont
  {Brodsky}, \citenamefont {Chudakov}, \citenamefont {Hoyer},\ and\
  \citenamefont {Laget}}]{Brodsky:2000zc}%
  \BibitemOpen
  \bibfield  {author} {\bibinfo {author} {\bibfnamefont {S.}~\bibnamefont
  {Brodsky}}, \bibinfo {author} {\bibfnamefont {E.}~\bibnamefont {Chudakov}},
  \bibinfo {author} {\bibfnamefont {P.}~\bibnamefont {Hoyer}}, \ and\ \bibinfo
  {author} {\bibfnamefont {J.}~\bibnamefont {Laget}},\ }\href {\doibase
  10.1016/S0370-2693(00)01373-3} {\bibfield  {journal} {\bibinfo  {journal}
  {Phys. Lett. B}\ }\textbf {\bibinfo {volume} {498}},\ \bibinfo {pages} {23}
  (\bibinfo {year} {2001})},\ \Eprint {http://arxiv.org/abs/hep-ph/0010343}
  {arXiv:hep-ph/0010343} \BibitemShut {NoStop}%
\bibitem [{\citenamefont {Hatta}\ \emph {et~al.}(2019)\citenamefont {Hatta},
  \citenamefont {Rajan},\ and\ \citenamefont {Yang}}]{Hatta:2019lxo}%
  \BibitemOpen
  \bibfield  {author} {\bibinfo {author} {\bibfnamefont {Y.}~\bibnamefont
  {Hatta}}, \bibinfo {author} {\bibfnamefont {A.}~\bibnamefont {Rajan}}, \ and\
  \bibinfo {author} {\bibfnamefont {D.-L.}\ \bibnamefont {Yang}},\ }\href
  {\doibase 10.1103/PhysRevD.100.014032} {\bibfield  {journal} {\bibinfo
  {journal} {Phys. Rev. D}\ }\textbf {\bibinfo {volume} {100}},\ \bibinfo
  {pages} {014032} (\bibinfo {year} {2019})},\ \Eprint
  {http://arxiv.org/abs/1906.00894} {arXiv:1906.00894 [hep-ph]} \BibitemShut
  {NoStop}%
\bibitem [{\citenamefont {Wang}\ \emph {et~al.}(2020)\citenamefont {Wang},
  \citenamefont {Evslin},\ and\ \citenamefont {Chen}}]{Wang:2019mza}%
  \BibitemOpen
  \bibfield  {author} {\bibinfo {author} {\bibfnamefont {R.}~\bibnamefont
  {Wang}}, \bibinfo {author} {\bibfnamefont {J.}~\bibnamefont {Evslin}}, \ and\
  \bibinfo {author} {\bibfnamefont {X.}~\bibnamefont {Chen}},\ }\href {\doibase
  10.1140/epjc/s10052-020-8057-9} {\bibfield  {journal} {\bibinfo  {journal}
  {Eur. Phys. J. C}\ }\textbf {\bibinfo {volume} {80}},\ \bibinfo {pages} {507}
  (\bibinfo {year} {2020})},\ \Eprint {http://arxiv.org/abs/1912.12040}
  {arXiv:1912.12040 [hep-ph]} \BibitemShut {NoStop}%
\bibitem [{\citenamefont {Du}\ \emph {et~al.}(2020)\citenamefont {Du},
  \citenamefont {Baru}, \citenamefont {Guo}, \citenamefont {Hanhart},
  \citenamefont {Mei\ss{}ner}, \citenamefont {Nefediev},\ and\ \citenamefont
  {Strakovsky}}]{Du:2020bqj}%
  \BibitemOpen
  \bibfield  {author} {\bibinfo {author} {\bibfnamefont {M.-L.}\ \bibnamefont
  {Du}}, \bibinfo {author} {\bibfnamefont {V.}~\bibnamefont {Baru}}, \bibinfo
  {author} {\bibfnamefont {F.-K.}\ \bibnamefont {Guo}}, \bibinfo {author}
  {\bibfnamefont {C.}~\bibnamefont {Hanhart}}, \bibinfo {author} {\bibfnamefont
  {U.-G.}\ \bibnamefont {Mei\ss{}ner}}, \bibinfo {author} {\bibfnamefont
  {A.}~\bibnamefont {Nefediev}}, \ and\ \bibinfo {author} {\bibfnamefont
  {I.}~\bibnamefont {Strakovsky}},\ }\href {\doibase
  10.1140/epjc/s10052-020-08620-5} {\bibfield  {journal} {\bibinfo  {journal}
  {Eur. Phys. J. C}\ }\textbf {\bibinfo {volume} {80}},\ \bibinfo {pages}
  {1053} (\bibinfo {year} {2020})},\ \Eprint {http://arxiv.org/abs/2009.08345}
  {arXiv:2009.08345 [hep-ph]} \BibitemShut {NoStop}%
\bibitem [{\citenamefont {Zeng}\ \emph {et~al.}(2020)\citenamefont {Zeng},
  \citenamefont {Wang}, \citenamefont {Zhang}, \citenamefont {Xie},
  \citenamefont {Wang},\ and\ \citenamefont {Chen}}]{Zeng:2020coc}%
  \BibitemOpen
  \bibfield  {author} {\bibinfo {author} {\bibfnamefont {F.}~\bibnamefont
  {Zeng}}, \bibinfo {author} {\bibfnamefont {X.-Y.}\ \bibnamefont {Wang}},
  \bibinfo {author} {\bibfnamefont {L.}~\bibnamefont {Zhang}}, \bibinfo
  {author} {\bibfnamefont {Y.-P.}\ \bibnamefont {Xie}}, \bibinfo {author}
  {\bibfnamefont {R.}~\bibnamefont {Wang}}, \ and\ \bibinfo {author}
  {\bibfnamefont {X.}~\bibnamefont {Chen}},\ }\href {\doibase
  10.1140/epjc/s10052-020-08584-6} {\bibfield  {journal} {\bibinfo  {journal}
  {Eur. Phys. J. C}\ }\textbf {\bibinfo {volume} {80}},\ \bibinfo {pages}
  {1027} (\bibinfo {year} {2020})},\ \Eprint {http://arxiv.org/abs/2008.13439}
  {arXiv:2008.13439 [hep-ph]} \BibitemShut {NoStop}%
\bibitem [{\citenamefont {Boussarie}\ and\ \citenamefont
  {Hatta}(2020)}]{Boussarie:2020vmu}%
  \BibitemOpen
  \bibfield  {author} {\bibinfo {author} {\bibfnamefont {R.}~\bibnamefont
  {Boussarie}}\ and\ \bibinfo {author} {\bibfnamefont {Y.}~\bibnamefont
  {Hatta}},\ }\href {\doibase 10.1103/PhysRevD.101.114004} {\bibfield
  {journal} {\bibinfo  {journal} {Phys. Rev. D}\ }\textbf {\bibinfo {volume}
  {101}},\ \bibinfo {pages} {114004} (\bibinfo {year} {2020})},\ \Eprint
  {http://arxiv.org/abs/2004.12715} {arXiv:2004.12715 [hep-ph]} \BibitemShut
  {NoStop}%
\bibitem [{\citenamefont {Mamo}\ and\ \citenamefont
  {Zahed}(2020)}]{Mamo:2019mka}%
  \BibitemOpen
  \bibfield  {author} {\bibinfo {author} {\bibfnamefont {K.~A.}\ \bibnamefont
  {Mamo}}\ and\ \bibinfo {author} {\bibfnamefont {I.}~\bibnamefont {Zahed}},\
  }\href {\doibase 10.1103/PhysRevD.101.086003} {\bibfield  {journal} {\bibinfo
   {journal} {Phys. Rev. D}\ }\textbf {\bibinfo {volume} {101}},\ \bibinfo
  {pages} {086003} (\bibinfo {year} {2020})},\ \Eprint
  {http://arxiv.org/abs/1910.04707} {arXiv:1910.04707 [hep-ph]} \BibitemShut
  {NoStop}%
\bibitem [{\citenamefont {Mamo}\ and\ \citenamefont
  {Zahed}(2021)}]{Mamo:2021krl}%
  \BibitemOpen
  \bibfield  {author} {\bibinfo {author} {\bibfnamefont {K.~A.}\ \bibnamefont
  {Mamo}}\ and\ \bibinfo {author} {\bibfnamefont {I.}~\bibnamefont {Zahed}},\
  }\href@noop {} {\  (\bibinfo {year} {2021})},\ \Eprint
  {http://arxiv.org/abs/2103.03186} {arXiv:2103.03186 [hep-ph]} \BibitemShut
  {NoStop}%
\bibitem [{\citenamefont {Guo}\ \emph {et~al.}(2021)\citenamefont {Guo},
  \citenamefont {Ji},\ and\ \citenamefont {Liu}}]{Guo:2021ibg}%
  \BibitemOpen
  \bibfield  {author} {\bibinfo {author} {\bibfnamefont {Y.}~\bibnamefont
  {Guo}}, \bibinfo {author} {\bibfnamefont {X.}~\bibnamefont {Ji}}, \ and\
  \bibinfo {author} {\bibfnamefont {Y.}~\bibnamefont {Liu}},\ }\href@noop {} {\
   (\bibinfo {year} {2021})},\ \Eprint {http://arxiv.org/abs/2103.11506}
  {arXiv:2103.11506 [hep-ph]} \BibitemShut {NoStop}%
\bibitem [{\citenamefont {Sun}\ \emph {et~al.}(2021)\citenamefont {Sun},
  \citenamefont {Tong},\ and\ \citenamefont {Yuan}}]{Sun:2021gmi}%
  \BibitemOpen
  \bibfield  {author} {\bibinfo {author} {\bibfnamefont {P.}~\bibnamefont
  {Sun}}, \bibinfo {author} {\bibfnamefont {X.-B.}\ \bibnamefont {Tong}}, \
  and\ \bibinfo {author} {\bibfnamefont {F.}~\bibnamefont {Yuan}},\ }\href@noop
  {} {\  (\bibinfo {year} {2021})},\ \Eprint {http://arxiv.org/abs/2103.12047}
  {arXiv:2103.12047 [hep-ph]} \BibitemShut {NoStop}%
\bibitem [{\citenamefont {Morningstar}\ and\ \citenamefont
  {Peardon}(1999)}]{Morningstar:1999rf}%
  \BibitemOpen
  \bibfield  {author} {\bibinfo {author} {\bibfnamefont {C.~J.}\ \bibnamefont
  {Morningstar}}\ and\ \bibinfo {author} {\bibfnamefont {M.~J.}\ \bibnamefont
  {Peardon}},\ }\href {\doibase 10.1103/PhysRevD.60.034509} {\bibfield
  {journal} {\bibinfo  {journal} {Phys. Rev. D}\ }\textbf {\bibinfo {volume}
  {60}},\ \bibinfo {pages} {034509} (\bibinfo {year} {1999})},\ \Eprint
  {http://arxiv.org/abs/hep-lat/9901004} {arXiv:hep-lat/9901004} \BibitemShut
  {NoStop}%
\bibitem [{\citenamefont {Chen}\ \emph {et~al.}(2006)\citenamefont {Chen} \emph
  {et~al.}}]{Chen:2005mg}%
  \BibitemOpen
  \bibfield  {author} {\bibinfo {author} {\bibfnamefont {Y.}~\bibnamefont
  {Chen}} \emph {et~al.},\ }\href {\doibase 10.1103/PhysRevD.73.014516}
  {\bibfield  {journal} {\bibinfo  {journal} {Phys. Rev. D}\ }\textbf {\bibinfo
  {volume} {73}},\ \bibinfo {pages} {014516} (\bibinfo {year} {2006})},\
  \Eprint {http://arxiv.org/abs/hep-lat/0510074} {arXiv:hep-lat/0510074}
  \BibitemShut {NoStop}%
\bibitem [{\citenamefont {Gell-Mann}\ \emph {et~al.}(1968)\citenamefont
  {Gell-Mann}, \citenamefont {Oakes},\ and\ \citenamefont
  {Renner}}]{GellMann:1968rz}%
  \BibitemOpen
  \bibfield  {author} {\bibinfo {author} {\bibfnamefont {M.}~\bibnamefont
  {Gell-Mann}}, \bibinfo {author} {\bibfnamefont {R.~J.}\ \bibnamefont
  {Oakes}}, \ and\ \bibinfo {author} {\bibfnamefont {B.}~\bibnamefont
  {Renner}},\ }\href {\doibase 10.1103/PhysRev.175.2195} {\bibfield  {journal}
  {\bibinfo  {journal} {Phys. Rev.}\ }\textbf {\bibinfo {volume} {175}},\
  \bibinfo {pages} {2195} (\bibinfo {year} {1968})}\BibitemShut {NoStop}%
\bibitem [{\citenamefont {Feynman}(1939)}]{Feynman:1939zza}%
  \BibitemOpen
  \bibfield  {author} {\bibinfo {author} {\bibfnamefont {R.~P.}\ \bibnamefont
  {Feynman}},\ }\href {\doibase 10.1103/PhysRev.56.340} {\bibfield  {journal}
  {\bibinfo  {journal} {Phys. Rev.}\ }\textbf {\bibinfo {volume} {56}},\
  \bibinfo {pages} {340} (\bibinfo {year} {1939})}\BibitemShut {NoStop}%
\bibitem [{\citenamefont {Lanik}(1984)}]{Lanik:1984fc}%
  \BibitemOpen
  \bibfield  {author} {\bibinfo {author} {\bibfnamefont {J.}~\bibnamefont
  {Lanik}},\ }\href {\doibase 10.1016/0370-2693(84)91295-4} {\bibfield
  {journal} {\bibinfo  {journal} {Phys. Lett. B}\ }\textbf {\bibinfo {volume}
  {144}},\ \bibinfo {pages} {439} (\bibinfo {year} {1984})}\BibitemShut
  {NoStop}%
\bibitem [{\citenamefont {'t~Hooft}(2003)}]{tHooft:2002pmx}%
  \BibitemOpen
  \bibfield  {author} {\bibinfo {author} {\bibfnamefont {G.}~\bibnamefont
  {'t~Hooft}},\ }\href {\doibase 10.1016/S0920-5632(03)01872-3} {\bibfield
  {journal} {\bibinfo  {journal} {Nucl. Phys. B Proc. Suppl.}\ }\textbf
  {\bibinfo {volume} {121}},\ \bibinfo {pages} {333} (\bibinfo {year}
  {2003})},\ \Eprint {http://arxiv.org/abs/hep-th/0207179}
  {arXiv:hep-th/0207179} \BibitemShut {NoStop}%
\bibitem [{\citenamefont {Biddle}\ \emph {et~al.}(2020)\citenamefont {Biddle},
  \citenamefont {Kamleh},\ and\ \citenamefont {Leinweber}}]{Biddle:2019gke}%
  \BibitemOpen
  \bibfield  {author} {\bibinfo {author} {\bibfnamefont {J.~C.}\ \bibnamefont
  {Biddle}}, \bibinfo {author} {\bibfnamefont {W.}~\bibnamefont {Kamleh}}, \
  and\ \bibinfo {author} {\bibfnamefont {D.~B.}\ \bibnamefont {Leinweber}},\
  }\href {\doibase 10.1103/PhysRevD.102.034504} {\bibfield  {journal} {\bibinfo
   {journal} {Phys. Rev. D}\ }\textbf {\bibinfo {volume} {102}},\ \bibinfo
  {pages} {034504} (\bibinfo {year} {2020})},\ \Eprint
  {http://arxiv.org/abs/1912.09531} {arXiv:1912.09531 [hep-lat]} \BibitemShut
  {NoStop}%
\end{thebibliography}%

\end{document}